\documentclass[a4paper,11pt]{article}
\pdfoutput=1 

\usepackage{jheppub} 

\usepackage[T1]{fontenc} 
\usepackage{fontawesome}
\usepackage[table, dvipsnames]{xcolor}
\usepackage[caption=false]{subfig}
\usepackage{epsfig}
\usepackage{colortbl}
\usepackage[T1]{fontenc} 
\usepackage[compat=1.1.0]{tikz-feynman}
\tikzfeynmanset{warn luatex=false}
\usepackage{adjustbox}
\usepackage{multirow}
\usepackage{slashed}
\usepackage{appendix}
\usepackage{lipsum}
\usepackage{chngcntr}
\usepackage{bigints}
\usepackage[normalem]{ulem}
\usepackage{amsmath}
\usepackage{textcomp}
\usepackage{cleveref}
\usepackage{chemarrow}
\usepackage{booktabs}
\usepackage{float}
\usepackage{amsfonts}
\usepackage{dsfont}
\PassOptionsToPackage{pdftex}{hyperref} 
\usepackage{hyperref}

\crefname{figure}{fig\,.}{figs\,.} 
\crefname{equation}{eq\,.}{eqs\,.} 
\newcommand{\stkout}[1]{\ifmmode\text{\sout{\ensuremath{#1}}}\else\sout{#1}\fi}
\definecolor{calpolypomonagreen}{rgb}{0.12, 0.3, 0.17}
\newcommand{\bea}{\begin{eqnarray}}
\newcommand{\eea}{\end{eqnarray}}

\usepackage{xcolor}

\def\Dst{{D^*}}

\def\cbar{\overline{c}}

\def\Dst{{D^*}}

\def\thD{{\theta_D}}
\def\thl{{\theta_\ell}}
\def\nn{\nonumber}

\title{\boldmath Probing Light Dark Fermions in $B \to D^{(*)}\ell X_{\rm inv}$ via Rate Distributions}

\author[a,b]{Lipika Kolay,}
\author[a]{Soumitra Nandi,}
\author[a,c]{Shantanu Sahoo}
\author[d]{and Ria Sain}

\affiliation[a]{Department of Physics, Indian Institute of Technology Guwahati,\\North Guwahati, Assam-781039, India}
\affiliation[b]{Department of Physics, Indian Institute of Technology Gandhinagar, India}
\affiliation[c]{School of Physics, University of Hyderabad, Hyderabad, Telangana-500046, India}
\affiliation[d]{Institute of Particle Physics and Key Laboratory of Quark and Lepton Physics (MOE), Central China Normal University, Wuhan, Hubei 430079, China}

\emailAdd{klipika@iitg.ac.in, soumitra.nandi@iitg.ac.in, shantanusahoo.ra@uohyd.ac.in}
\abstract{Experimental analyses of the semileptonic decays $ B \to D^{(*)} \ell \bar{\nu}$ typically rely on the assumption that the missing energy originates from a massless neutrino, as predicted by the Standard Model. However, this assumption may not hold in scenarios where the invisible final-state particle is instead massive, such as a sterile neutrino or a dark-sector fermion. In this work, we explore how the presence of a massive dark sector fermion modifies the kinematic and angular distributions of these decays. Our analysis is carried out within the framework of a general weak effective theory, and we also discuss effective and simplified models in which these interactions may arise. In addition, we study the implications of these effects for the extraction of the CKM matrix element $ |V_{cb}|$. Overall, our results show that relaxing the standard assumption of a massless neutrino can lead to observable effects and provide a framework for systematically investigating their impact on semileptonic $B$- decay distributions.}
\keywords{Invisible Decay, Light dark matter, Decay Distribution, CKM Element}

\begin{document}
\maketitle
\flushbottom

\section{Introduction}
The semileptonic decays $ B \to D^{(*)} \ell \nu $ have been the topic of growing interest in recent years. These tree-level decays with light leptons in the final state are crucial for the extraction of the CKM element $ |V_{cb}| $. New data from Belle \cite{Belle:2023bwv} and Belle-II \cite{Belle-II:2023okj} have revised earlier measurements \cite{Belle:2018ezy} and contributed to a deeper understanding of the decay kinematics and angular distributions. The differential distribution for the decay $ B \to D \ell \bar{\nu} $ was previously provided by Belle \cite{Belle:2015pkj}. Furthermore, these decays play a key role in testing the Standard Model (SM) and probing potential beyond the Standard Model (BSM) scenarios through precise determinations of associated observables.  

In these processes, the neutrinos in the final state remain invisible to detectors due to their weakly interacting nature and are therefore treated as missing energy. This invisibility poses experimental challenges, particularly in the reconstruction of decay kinematics. At the same time, it provides a powerful opportunity to test the SM and search for new physics. Current experimental analyses typically assume that the missing energy corresponds to a massless particle, consistent with the SM neutrino hypothesis. 

Traditionally identified as SM neutrinos, the invisible particles in the final state could also include low-mass sterile neutrinos or light dark matter candidates. If such particles possess BSM characteristics, they could induce measurable deviations in decay distributions or missing energy observables. Indeed, the Belle collaboration \cite{Belle:2023bwv} has analysed the missing mass squared ($m^2_{\rm miss}$) distribution, which is observed to peak around $m^2_{\rm miss}=0$ but exhibits a narrowly spread non-zero region for $m_{\rm miss}$. This observation provides a strong motivation to further explore semileptonic $B$ decays as sensitive probes of invisible particles beyond the SM. 

In this work, we explore this possibility by incorporating the contribution of a light dark sector fermion, denoted by $\chi$, to the decay rate distributions, in addition to the standard contribution from the SM neutrino. The mass generation mechanism of this dark fermion lies beyond the scope of the present analysis and is therefore not specified explicitly.

From a theoretical perspective, various models have been proposed in which $B$ mesons decay into dark sector particles or sterile neutrinos via flavour-changing neutral current (FCNC) processes \cite{Helo:2010cw, Kamenik:2011vy, Abada:2013aba, Manzari:2022iyn, Jueid:2024cge, Arcadi:2021cwg, Arcadi:2021glq, Calibbi:2025rpx}. In addition, flavour-changing charged current (FCCC) processes with right-handed neutrinos in the final state have also been explored in the literature, for example, in refs.~\cite{Cvetic:2017gkt,Kim:2019xqj, Datta:2022czw, Bernlochner:2024xiz, Becirevic:2024iyi, Das:2026uyt, Kolay:2026bjm, Sahoo:2026jmr}. In most of these studies, the invisible fermion is assumed to be a right-handed sterile neutrino.


The paper is organised as follows. In Sec.~\ref{sec:theo_framework}, we discuss the motivation for considering a finite-mass dark sector particle as the source of the missing energy. In Sec.~\ref{sec:theo_formalism}, we present the theoretical framework required to study the decays $B \to D^{(*)}\ell X_{\rm inv}$ and discuss possible new-physics scenarios that can give rise to such signatures. In Sec.~\ref{sec:analysis_results}, we present our analysis and results, including the differential decay distributions and the extraction of CKM matrix elements. Finally, we summarise our findings in Sec.~\ref{sec:summary}.

\section{Motivation and Framework}\label{sec:theo_framework}
As noted above, most existing studies have focused on scenarios involving a right-handed sterile neutrino. In contrast, in this work we analyse the semileptonic decays $ B \to D^{(*)} \mu (e) \nu $ under the hypothesis that the invisible particle could be a light fermionic dark matter candidate. In particular, we consider the decay $ B \to D^{(*)} \mu (e) \chi $, where $\chi$ denotes a light invisible fermion distinct from the SM neutrinos. The presence of such a particle modifies the missing mass distribution, with $m_{\rm miss} = m_{\chi}$. 

We adopt a model-independent effective field theory framework to describe possible BSM contributions to these decay modes. Unlike conventional approaches, we allow the invisible fermion $\chi$ to possess both left- and right-handed chiralities. To remain consistent with the experimental observations of the missing mass distribution reported in \cite{Belle:2023bwv}, we restrict our analysis to $m_{\chi} \leq 1~\text{GeV}$. Our primary goal is to investigate how the non-zero mass of the invisible particle impacts kinematic and angular observables, thereby identifying potentially important effects that are overlooked when the missing particle is assumed to be a massless SM neutrino.

The primary objective of this analysis is therefore to study the impact of a finite mass of the BSM fermion $\chi$ on the decay distributions in $B \to D^{(*)} \ell \chi$ processes. The presence of a massive invisible particle modifies both the phase-space boundaries and angular observables in a non-trivial manner. By analysing these effects, we aim to assess the sensitivity of various observables to the underlying operator structure and the mass of the dark fermion, thereby providing complementary probes of dark matter-motivated NP scenarios in flavour physics.

Within the dark matter effective theory framework, such as dark low-energy effective theory (DLEFT) or dark SM effective field theory (DSMEFT) \cite{Aebischer:2022wnl}, one encounters dimension-6 operators involving dark sector fields that can contribute to $b \to c \ell^- \bar{\chi}$ transitions at tree level. In the next section, we will present the explicit form of these operators and discuss their tree-level matching onto the effective Hamiltonian relevant for $ B \to D^{(*)} \ell^- (\bar{\chi}) $ decays. 

Moreover, several UV-complete dark matter scenarios can generate these transitions at tree level. For instance, $t$-channel dark matter models, including R-parity violating supersymmetric frameworks with bilinear interactions, have been discussed in refs. \cite{Gunion:2005rw,Dreiner:2009er,flavourfullDM_RPV,Dey:2020juy,Dreiner:2022swd,Dib:2022ppx,Wang:2023trd}. Similarly, leptoquark portal dark matter models \cite{Belfatto:2021ats} can induce effective operators analogous to those appearing in DLEFT or DSMEFT. Additional scenarios include models featuring sterile or right-handed neutrinos as dark matter candidates \cite{Kusenko:2009up,Drewes:2016upu,Escudero:2016ksa,Cho:2021yxk,Liu:2024ygz,Arcadi:2020aot}, as well as models with extended scalar sectors coupled to fermionic dark matter \cite{Bai:2014osa,Okawa:2020jea,Higuchi:2023kbt,Iguro:2022tmr}. With suitable assumptions, such frameworks can contribute to $b \to c \ell^- \bar{\chi}$ processes in the sub-GeV mass region relevant to our study.

In this work, however, we focus on a model-independent analysis based on the effective operator approach, leaving a detailed study within specific UV-complete models for future investigation. Using the effective four-fermion operators and allowing for a non-zero mass of the dark sector particle, we compute the relevant decay rate distributions in the kinematic and angular variables and the forward backward asymmetry in $B \to D \ell X_{\rm inv}$. Wherever applicable, we compare our results with available experimental data \cite{Belle:2023bwv,Belle-II:2023okj}.

 
\subsection{Effective Hamiltonian}
In this subsection, we first discuss the effective Hamiltonian relevant for the decay distributions of the processes $B \to D^{(*)} \ell X_{\rm inv}$, where $X_{\rm inv}$ can denote either a massless SM neutrino or a BSM particle $\chi$. We then discuss specific models that can generate such interactions and present the corresponding constraints on the model parameters, primarily based on cosmological bounds available in the literature.

For the analysis, we consider the following most general effective Hamiltonian for $ b \to c \, \ell \, \bar{X}_{\rm inv}$ decays, 
\begin{eqnarray}
\small
\label{eq:general_Hamiltonian_b2c}
\mathcal{H}_{\rm eff}^{b \to c \ell \nu (\chi) } = &  \frac{ 4 G_F}{\sqrt{2}} V_{cb} & \bigg[C_{V_{1L}}^{\ell,SM}  \mathcal{O}_{V_{1L}}^{\ell,SM}  + C_{V_{1L}}^{\ell}  \mathcal{O}_{V_{1L}}^{\ell} \nonumber  +  C_{V_{1R}}^{\ell}  \mathcal{O}_{V_{1R}}^{\ell}  + C_{V_{2L}}^{\ell}  \mathcal{O}_{V_{2L}}^{\ell} + C_{V_{2R}}^{\ell}  \mathcal{O}_{V_{2R}}^{\ell} + C_{S_{1L}}^{\ell}  \mathcal{O}_{S_{1L}}^{\ell} \\ && + C_{S_{1R}}^{\ell}  \mathcal{O}_{S_{1R}}^{\ell}    + C_{S_{2L}}^{\ell}  \mathcal{O}_{S_{2L}}^{\ell}  + C_{S_{2R}}^{\ell}  \mathcal{O}_{S_{2R}}^{\ell} + C_{TL}^{\ell} \mathcal{O}_{TL}^{\ell}  +  C_{TR}^{\ell} \mathcal{O}_{TR}^{\ell} \bigg]\,. 
\end{eqnarray}
In the above effective Hamiltonian, coefficients $ C^{\ell}_{K_{iL(R)}}$ associated with the operators, with $ K=V,S,T $ and $ i=1,2 $ represent the respective Wilson coefficients (WCs). The SM neutrinos are taken to be left-handed, while the dark fermion $\chi$ can have both left- and right-handed chiralities. In this work, we have mainly focused on the light leptons $\ell = (e, \, \mu)$. The corresponding four-fermion operators are given by:
\begin{align}
& \mathcal{O}_{V_{1L}}^{\ell,SM} = \left(\bar{c}\gamma_{\mu} P_{L} b \right)\left( \bar{\ell} \gamma^{\mu} P_{L}\nu_{\ell} \right) \,, \nonumber \\
& \mathcal{O}_{V_{1L(R)}}^{\ell} = \left(\bar{c}\gamma_{\mu} P_{L} b \right)\left( \bar{\ell} \gamma^{\mu} P_{L}(P_{R}) \chi \right)\,, \nonumber \\& 
\mathcal{O}_{V_{2L(R)}}^{\ell} = \left(\bar{c}\gamma_{\mu} P_{R} b \right)\left( \bar{\ell} \gamma^{\mu} P_{L}(P_{R}) \chi \right)\,,  \\& 
\mathcal{O}_{S_{1L(R)}}^{\ell} = \left(\bar{c} P_{R}  b \right)\left( \bar{\ell} P_{L} (P_{R}) \chi \right)\,, \nonumber \\& 
\mathcal{O}_{S_{2L(R)}}^{\ell} = \left(\bar{c}P_{L} b \right)\left( \bar{\ell} P_{L} (P_{R}) \chi \right) \,,  \nonumber \\& 
\mathcal{O}_{TL(R)}^{\ell} = \left(\bar{c}\sigma_{\mu\nu} P_{L} (P_{R})  b \right)\left( \bar{\ell} \sigma^{\mu \nu } P_{L} (P_{R}) \chi \right)\,. \nonumber
\end{align}

The above effective Hamiltonian is invariant under $SU(3)_c \otimes U(1)_{\mathrm{em}}$. Hence, the operator basis is referred to as the weak effective theory (WET) or low-energy effective theory basis. Tensor operators with different lepton and quark chiralities vanish identically.  We have written the operator involving the SM neutrino as $ \mathcal{O}_{V_{1L}}^{\ell,SM}$ with corresponding value of the WC $ C_{V_{1L}}^{\ell,SM} = 1$. The general Lagrangian containing SM neutrino only can be generated by setting $ \chi \rightarrow \nu $ and $ C^{\ell}_{K_{iR}} = 0$. Throughout our analysis, we will use the WCs of the above basis of eq.~\eqref{eq:general_Hamiltonian_b2c}. Note that, in this work, we have performed our analysis by considering the WCs to be real. For $\chi$ to be a viable missing energy candidate at the final state of the decay $B\to D^{(*)} \ell X_{\rm inv}$, it has to have mass in the kinematically allowed range:
\begin{equation}\label{eq:allowed_mass_chi}
m_{\chi} \leq (m_{B} - m_{D^{(*)}}-m_{\ell} ) \,.
\end{equation}
In the following paragraphs and sections, we will discuss other effective bases and NP models that can generate similar interactions of eq.~\eqref{eq:general_Hamiltonian_b2c}.
%
\paragraph{\underline{Matching with DLEFT operators}:}
The effective operators of eq.~\eqref{eq:general_Hamiltonian_b2c} can be matched with the operators of the DLEFT \cite{Aebischer:2022wnl} basis. The particle content of DLEFT includes all the SM fermions (except the top quark), photon, gluons, and the dark sector particles. Here, we are interested only in operators in which SM is extended by a dark-sector fermion. The complete set of effective operators in dim-6 relevant for our study is given in the left column of  Table~\ref{tab:DLEFT_WET_comapre}. The DLEFT basis is already defined in the mass basis; we don't need to perform any further transformation. The operators in DLEFT and LEFT are $SU(3)_{c} \times U(1)_{\rm em} $ invariant, and the basis is valid below the electroweak scale. At the tree-level, we have matched these DLEFT operators to the operators in eq.~\eqref{eq:general_Hamiltonian_b2c}. In DLEFT, we get all the operators defined in eq.~\eqref{eq:general_Hamiltonian_b2c}. The relations between WCs of the two bases are given in the right column in Table~\ref{tab:DLEFT_WET_comapre}. The tree-level matching condition between the DSMEFT and the DLEFT has been explicitly worked out in ref. \cite{Aebischer:2022wnl}. In this study, we have not separately focused on operators in DSMEFT. We only focus on DLEFT operators, which we have mentioned in the Table~\ref{tab:DLEFT_WET_comapre}.  

\begin{table}[ht]
    \centering
    \renewcommand{\arraystretch}{1.8}
    \setlength{\tabcolsep}{5pt}
    \resizebox{0.6\textwidth}{!}{%
    \rowcolors{1}{cyan!15}{lime!10}
    \begin{tabular}{|c| c |}
    \hline
    \rowcolor{Lavender!50}
    \multicolumn{2}{|c|}{Matching with DLEFT operators}   \\ 
    \hline
    \hline
    DLEFT operators &  Matching with WET  \\
    \hline
    $\left( \bar{d}_{L_p} \gamma_{\mu} u_{L_r} \right) \left( \bar{\chi}_{L} \gamma^{\mu} e_{L_s} \right) $ & $\mathcal{C}_{du\chi e}^{* \rm V,LL} = \frac{\sqrt{2}}{4 G_{F}} \frac{1}{V_{cb}}  C^{\ell}_{V_{1L}}$\\
    $\left( \bar{d}_{R_p} \gamma_{\mu} u_{R_r} \right) \left( \bar{\chi}_{L} \gamma^{\mu} e_{L_s} \right) $ & $\mathcal{C}_{du\chi e}^{* \rm V,LR} = \frac{\sqrt{2}}{4 G_{F}} \frac{1}{V_{cb}}  C^{\ell}_{V_{2L}}$\\
    $\left(\bar{d}_{R_p} \gamma_{\mu} u_{R_r}  \right) \left(\bar{\chi}_{R} \gamma^{\mu} e_{R_s}  \right) $ & $\mathcal{C}_{du\chi e}^{* \rm V,RR} = \frac{\sqrt{2}}{4 G_{F}} \frac{1}{V_{cb}}  C^{\ell}_{V_{2R}}$ \\
    $\left(\bar{d}_{L_p} \gamma_{\mu} u_{L_r}  \right) \left( \bar{\chi}_{R_{}} \gamma^{\mu} e_{R_s} \right) $ & $\mathcal{C}_{du\chi e}^{* \rm V,RL} = \frac{\sqrt{2}}{4 G_{F}} \frac{1}{V_{cb}}  C^{\ell}_{V_{2L}}$ \\
    \hline
    \hline
    $\left( \bar{d}_{R_p} u_{L_r}  \right) \left( \bar{\chi}_{R} e_{L_s} \right) $ & $\mathcal{C}_{du\chi e}^{* \rm S,LL} = \frac{\sqrt{2}}{4 G_{F}} \frac{1}{V_{cb}}  C^{\ell}_{S_{2L}}$ \\
    $\left( \bar{d}_{R_p} u_{L_r} \right) \left( \bar{\chi}_{L} e_{R_s} \right) $ & $\mathcal{C}_{du\chi e}^{* \rm S,LR} = \frac{\sqrt{2}}{4 G_{F}} \frac{1}{V_{cb}}  C^{\ell}_{S_{2R}}$ \\
    $\left( \bar{d}_{L_p} u_{R_r} \right) \left( \bar{\chi}_{R} e_{L_s} \right) $ & $\mathcal{C}_{du\chi e}^{* \rm S,RL} = \frac{\sqrt{2}}{4 G_{F}} \frac{1}{V_{cb}}  C^{\ell}_{S_{1L}}$\\
    $\left( \bar{d}_{L_p} u_{R_r}\right) \left( \bar{\chi}_{L} e_{R_s} \right) $ & $\mathcal{C}_{du\chi e}^{* \rm S,RR} = \frac{\sqrt{2}}{4 G_{F}} \frac{1}{V_{cb}}  C^{\ell}_{S_{1R}}$ \\
    \hline
    \hline
    $\left( \bar{d}_{R_p} \sigma_{\mu \nu} u_{L_r}  \right) \left( \bar{\chi}_{R} \sigma^{\mu \nu} e_{L_s} \right)$ & $\mathcal{C}_{du\chi e}^{* \rm T,L} = \frac{\sqrt{2}}{4 G_{F}} \frac{1}{V_{cb}}  C^{\ell}_{T_{L}}$ \\
    $\left( \bar{d}_{L_p} \sigma_{\mu \nu} u_{R_r}  \right) \left( \bar{\chi}_{L} \sigma^{\mu \nu} e_{R_s} \right)$ & $\mathcal{C}_{du\chi e}^{* \rm T,R} = \frac{\sqrt{2}}{4 G_{F}} \frac{1}{V_{cb}}  C^{\ell}_{T_{R}}$ \\
    \hline
    \end{tabular}
    }
    \caption{Matching of DLEFT and our WET operator basis, relevant to our analysis. The $\mathcal{C}_{i}$'s are the DLEFT WCs while $C_{i}$'s are our WET basis WCs.}
    \label{tab:DLEFT_WET_comapre}
\end{table}
\subsection{Motivation from ultraviolet-complete scenarios}
\label{sec:possible_models}

Eq.~\eqref{eq:general_Hamiltonian_b2c} presents the most general WET basis describing the decay $b \to c \,\ell \,\bar{X}_{\rm inv}$. The invisible particle, $X_{\rm inv}$, can correspond either to a massless SM neutrino or to a dark fermion with non-zero mass. As discussed earlier, a variety of dark matter scenarios can generate one or more operators appearing in the basis of eq.~\eqref{eq:general_Hamiltonian_b2c}. In this section, we discuss several representative classes of models that can give rise to such interactions, with the dark fermion playing the role of a light dark matter candidate. Before presenting these scenarios, we first clarify their role within the context of our model-independent analysis.

As described earlier, the effective Hamiltonian in eq.~\eqref{eq:general_Hamiltonian_b2c} is parametrised in terms of a general set of WCs, which encode the low-energy effects of physics beyond the Standard Model. This model-independent framework provides a convenient and systematic way to analyse possible deviations in semileptonic $B$ decays in the presence of invisible final states. By construction, it accommodates all possible Lorentz and chiral structures consistent with the underlying symmetries. In practice, however, assessing the impact of individual operators on physical observables, such as decay rates, angular distributions, and kinematic spectra, requires some guidance regarding the expected size and relative importance of the corresponding WCs. While such information is, in principle, constrained by a combination of flavour observables, electroweak precision data, and collider searches, obtaining robust, model-independent bounds on all relevant coefficients is often challenging.

A useful strategy in this regard is to consider specific ultraviolet (UV) completions in which the effective operators of eq.~\eqref{eq:general_Hamiltonian_b2c} are generated by integrating out heavy degrees of freedom. In such scenarios, the operator structures and their couplings are not arbitrary but are determined by the underlying dynamics. Moreover, these models typically induce correlated effects across multiple observables and sectors, allowing experimental data to place constraints on the underlying parameter space. Where available, bounds on mediator masses and interaction strengths derived within these frameworks can be translated into constraints on the corresponding WCs. However, it is important to emphasise that such mappings are inherently model-dependent and are not uniformly available across all operator structures or ultraviolet scenarios. In particular, for several classes of interactions relevant to our analysis, the existing literature does not provide direct or comprehensive limits on the associated low-energy coefficients.

In this work, we therefore use UV-complete models primarily as a source of qualitative guidance, rather than as a basis for systematically fixing the numerical values of all WCs. In a subset of well-studied scenarios, we identify representative regions of parameter space that are consistent with current experimental constraints and map these onto the effective theory. In other cases, where such information is not readily available, we restrict ourselves to reasonable choices motivated by the expected scale suppression and general phenomenological considerations. This approach allows us to avoid arbitrary parameter selections while maintaining sufficient flexibility to explore the phenomenological implications of different operator structures. At the same time, it preserves, where possible, a transparent connection between the model-independent description and plausible UV completions.

In the following, we outline several NP scenarios that can generate such operators, including both UV-complete constructions and simplified models featuring heavy mediators coupling the quark, lepton, and dark sectors.
 
\subsubsection{A Few Dark Matter Models}
\subsubsection*{(1) Leptoquark Models }
In this subsection, we will discuss the results of the analysis done in ref. \cite{Belfatto:2021ats}. The interaction involving a massive fermion, $ b \to c \ell \bar{\chi} $, can be generated in models where the SM is extended with a leptoquark (LQ) and the fermion $\chi$. This is possible in various LQ models incorporating both scalar and vector LQs. In what follows, we will discuss these interactions in detail.
\paragraph{\underline{Scalar Leptoquarks ($S_{1}$, $\tilde{R}_{2}$) Extension} :}
The scalar leptoquarks $S_{1} \sim (3,1)_{-1/3}$ and $\tilde{R_{2}} \sim (3, 2)_{1/6}$ can also generate similar effective interactions, with Lagrangian \cite{Bauer:2015knc,Gherardi:2020qhc,Dorsner:2016wpm}:
\begin{align}
	\mathcal{L}_{S_{1}} &= 
	\left(
	g_{\chi} \, \bar{d}_{R} \chi
	+ g_{L} \, \bar{Q}_{L} \varepsilon L^{c}_{L}
	+ g_{R} \, \bar{u}_{R} e^{c}_{R}
	\right) S_{1}
	+ \mathrm{h.c.}, \\
	\mathcal{L}_{\tilde{R}_{2}} &= 
	\left(
	g_{\chi} \, \bar{Q}_{L} \chi
	+ g_{L} \, \bar{d}_{R} L_{L} \varepsilon
	\right) \tilde{R}_{2}
	+ \mathrm{h.c.}
\end{align}
In this model, the general effective Lagrangian that will contribute to $b\to c\ell^-\bar{\chi}$ can be written as 
\begin{eqnarray}
	\mathcal{L}_{\mathrm{DM}}^{\mathrm{eff}} & = & 
	- C_{\chi R} \left( \bar{\chi} \gamma^\mu \ell_R \right) \left( \bar{d}_R \gamma_\mu u_R \right)
	+ 2 C_{\chi L} \left( \bar{\chi} L_L \right) \left( \bar{Q}_L u_R \right) \nonumber \\&&
	- C'_{\chi L} \left( \bar{L}_L \chi \right) \varepsilon \left( \bar{Q}_L d_R \right)
	- C''_{\chi L} \left( \bar{Q}_L \chi \right) \varepsilon \left( \bar{L}_L d_R \right)\,.
\end{eqnarray}
Here, the DM $\chi$ can have both chirality. The bounds on $g_{\chi}$ is not available and a comprehensive analysis of these bounds require a dedicated study of the dark matter phenomenology within a fully specified DM model which is beyond the scope of this paper.

Collider searches at the LHC impose important constraints on the Yukawa couplings $g_{L,R}^{32}$ appearing in scalar leptoquark models such as $S_1$ and $\tilde{R}_2$. While pair production searches provide largely coupling-independent lower bounds on the leptoquark mass in the TeV range, additional constraints arise from high-$p_T$ dilepton tails and single leptoquark production, both of which are sensitive to $t$-channel leptoquark exchange and scale with the Yukawa couplings. Recent recast analyses that consistently combine these effects demonstrate that LHC data place significant bounds on the leptoquark-quark-lepton Yukawa interactions over a wide mass range~\cite{Angelescu:2021lln}. In particular, this study indicates that, for leptoquark masses in the range $M_{\rm LQ} \sim 1 \mathrm{TeV}$, the Yukawa couplings are typically constrained to be at most of moderate size, with representative bounds of $g_{L,R}^{32} \sim \mathcal{O}(1)$. Larger values are increasingly disfavoured due to their impact on single production channels and their enhancement of deviations in the high-energy tails of dilepton distributions. It is important to note, however, that these limits are inherently model-dependent, with sensitivity to the assumed flavour structure and chirality of the couplings. Nevertheless, they provide a useful benchmark for the collider sensitivity to scalar leptoquark interactions.

On the other hand, the coupling $g_\chi$ governs the interaction of the leptoquark with the dark-sector fermion, and the leptoquark acts as a portal between the SM and the dark sector. In this analysis, our focus is on the light dark matter regime. It is worth noting that collider bounds are comparatively less restrictive in the region $m_\chi \sim 1~\mathrm{GeV}$, as the sensitivity of monojet searches becomes largely independent of the dark matter mass once $m_\chi \ll \mathcal{O}(100~\mathrm{GeV})$. In this light dark matter limit, collider constraints primarily set bounds on the mediator mass, while the coupling $g_\chi$ is instead controlled predominantly by direct and indirect detection considerations.

For spin-0 mediated $t$-channel interactions, the same coupling $g_\chi$ governs both the annihilation cross section relevant for thermal freeze-out and the elastic scattering cross section probed in direct detection experiments. In the regime where the mediator mass lies at the TeV scale, $M_{\mathrm{med}} \sim \mathcal{O}(1\text{-}2~\mathrm{TeV})$, the annihilation cross section is strongly suppressed, scaling approximately as $\langle \sigma v \rangle \propto g_\chi^4\, m_\chi^2 / M_{\mathrm{med}}^4$. For light dark matter masses $m_\chi \sim \mathcal{O}(1~\mathrm{GeV})$, this leads to inefficient annihilation and consequently an overabundant relic density unless relatively large values of $g_\chi$ are invoked. For instance, the analyses in refs.~\cite{Mohan:2019zrk,Arina:2023msd} indicate that, for $m_{\chi} \sim \mathcal{O}(1~\mathrm{GeV})$ and mediator masses around $\mathcal{O}(1~\mathrm{TeV})$, satisfying both relic density and direct detection constraints typically requires $g_\chi \gtrsim 5$.

Given the allowed ranges of $g_\chi$ and $g_{L,R}^{32}$ for a leptoquark mass of order $1~\mathrm{TeV}$, one can estimate the corresponding WCs of the effective operators using the mapping relation
\begin{equation}
	C_{S_{iL(R)}}^{\ell} = \frac{\sqrt{2}}{4 G_F V_{cb}} C_{\chi L(R)} 
	= \frac{\sqrt{2}}{4 G_F V_{cb}} \frac{g_{\chi} g_{L(R)}^{32}}{M_{LQ}^2}.
\end{equation}
For representative values $g_{\chi} \sim g_{L(R)}^{32} \sim \mathcal{O}(1)$, this relation naturally yields $C_{S_{iL(R)}}^{\ell}$ of order unity.

However, for light dark matter and TeV-scale mediators, achieving a simultaneous explanation of the relic abundance and satisfying direct detection constraints is highly non-trivial. In particular, the values of $g_\chi$ required for efficient annihilation tend to approach or exceed the perturbative regime. Such large couplings are usually disfavoured by direct detection limits in the case of scalar interactions, while in pseudoscalar scenarios they remain insufficient to reproduce the observed relic density due to suppressed annihilation rates. Consequently, only a very small region of parameter space remains viable. This tension is a generic feature of $t$-channel portal models with light dark matter and heavy mediators. It therefore motivates extensions of the minimal framework, for instance through the introduction of additional annihilation channels, in order to consistently accommodate both relic density and direct detection constraints in the light dark matter regime.

\paragraph{\underline{Multi-component Dark Matter Scenario} :}
While most studies assume a single dark matter component, this minimal picture may be too restrictive given the present lack of definitive signals. In many well-motivated extensions of the SM, the dark sector naturally contains more than one stable particle, leading to a multicomponent dark matter scenario. In such frameworks, the observed relic density is shared among different species, which can ease experimental constraints and allow for richer thermal and phenomenological histories. This motivates considering multicomponent setups as a simple and realistic extension of the standard dark matter paradigm. As discussed by Belfatto \textit{et al.}~\cite{Belfatto:2021ats}, in the presence of more than one DM component, the effective interaction Lagrangian introduced earlier is modified and can be written as
\begin{equation}
	\mathcal{L}_{\mathrm{eff}} \supset 
	C_{\mathrm{DM}_1} (\bar{\chi}_1 \ell)(\bar{d} u)
	+ C_{\mathrm{DM}_2} (\bar{\chi}_2 \ell)(\bar{d} u)
	+ C_{\mathrm{DM}_{12}} (\bar{\chi}_1 \chi_2)(\bar{q} q)\,.
\end{equation}
In this scenario as well, instead of being stabilized by an imposed external symmetry, the DM is treated as a long-lived particle (LLP). 
By making the coupling $C_{\rm DM_{1}}$ very small, the DM can be made sufficiently stable while simultaneously achieving the observed relic density through the freeze-out mechanism. 
This is accomplished with the following DM couplings:
\begin{eqnarray}
	C_{\rm DM_{2}} \sim C_{\rm DM_{12}} \sim G_{F}\,.
\end{eqnarray}
If we map these couplings to our basis of eq.~\eqref{eq:general_Hamiltonian_b2c}, it converts to:
\begin{equation}
	\left( \mathcal{C}_{S_{1L}} + \mathcal{C}_{S_{1R}} + \mathcal{C}_{S_{2L}} + \mathcal{C}_{S_{2R}}  \right)  \sim 8.48 
\end{equation}

A comprehensive analysis of these bounds would require a dedicated study of the dark matter phenomenology within a fully specified model framework. Such an investigation is beyond the scope of the present work. Consequently, while representative ranges for $g_\chi$ can be identified in specific realisations, we refrain from imposing generic numerical limits and instead treat this coupling as a free parameter, subject only to broad consistency considerations.

\paragraph{\underline{Vector Leptoquark ($\mathcal{U}$) Extension} :}
If the SM is extended via a vector LQ $\mathcal{U}^{\mu}$, with SM charges $\mathcal{U} \sim (3, 1 )_{2/3}$, the possible interaction will look like \cite{Buttazzo:2017ixm,Fajfer:2015ycq}:
\begin{equation}\label{eq:LQ_vector_U}
	\mathcal{L}_{\mathcal{U}} =
	\left(
	g_{\chi}^i \, \bar{u}^i_{R} \gamma^{\mu} \chi
	+ g_{R}^{ij} \, \bar{d}^i_{R} \gamma^{\mu} \ell^j_{R}
	+ g_{L}^{ij} \, \bar{q}^i_{L} \gamma^{\mu} L^j_{L}
	\right) \mathcal{U}_{\mu}
	+ \mathrm{h.c.}
\end{equation}
Such interactions can arise in models with a $SU(4)$ Pati-Salam symmetry \cite{Pati:1974yy, Barbieri:2015yvd, Barbieri:2017tuq, DiLuzio:2017vat}.

This interaction can generate the effective Lagrangian interaction via integrating out the heavy LQ field and doing the Fierz transformation:
\begin{equation}\label{eq:LQ_eff_vector_U}
	\mathcal{L}^{\mathrm{eff}}_{\mathcal{U}} =
	- C_{\chi R} \, (\bar{\chi} \gamma^{\mu} \ell_{R}) (\bar{d}_{R} \gamma_{\mu} u_{R})
	+ 2 C_{\chi L} \, (\bar{\chi} L_{L}) (\bar{q}_{L} u_{R})
	+ \cdots
\end{equation}
The above WCs of the effective Lagrangian will have a relation with the LQ model parameters as
\begin{equation}
	C_{\chi R} = \frac{g_{\chi} g_{R}}{M_{\mathcal{U}}^2}, 
	\quad
	C_{\chi L} = \frac{g_{\chi} g_{L}}{M_{\mathcal{U}}^2}\,.
\end{equation}

Important bounds on the coupling $g_L^{32}$ arise from high-energy collider searches at the LHC. These constraints originate primarily from $t$-channel exchange of the vector leptoquark in high-$p_T$ processes such as dilepton and lepton-neutrino production, as well as from single leptoquark production channels. In particular, high-$p_T$ measurements of $pp \to \ell\ell$ and $pp \to \ell\nu$ cross sections are sensitive to the effective four-fermion contact interactions induced by $t$-channel leptoquark exchange, thereby constraining the combination $(g_L^{32})^2/M_{\mathcal U}^2$~\cite{Bessaa:2014jya,Angelescu:2021lln,Bhaskar:2021pml}. These analyses of dilepton production probe contact interaction scales in the multi-TeV regime. Interpreting these results in the context of vector leptoquark scenarios suggests that, for representative leptoquark masses $M_{\mathcal U} \sim \mathcal{O}(1~\mathrm{TeV})$, couplings of order unity are not excluded by present collider data. More precisely, the high-$p_T$ tails of dilepton distributions typically allow $g_{L,R}^{32} \sim \mathcal{O}(1)$ for $M_{\mathcal U} \gtrsim 1~\mathrm{TeV}$, although the exact bounds remain analysis- and model-dependent. These considerations provide a useful, albeit approximate, indication of the range of couplings compatible with current high-energy collider constraints.

A similar conclusion holds for vector and axial-vector interactions, where the combined relic density and direct detection constraints typically favour larger values of the coupling $g_\chi$ for leptoquark masses around the TeV scale. Such values can readily generate effective WCs of $\mathcal{O}(1)$, consistent with the requirements of low-energy observables. However, as discussed in the scalar-mediated case, these regions are often in tension with perturbativity bounds, since the required couplings approach or exceed the perturbative regime. This interplay between relic abundance, direct detection, and perturbativity therefore significantly restricts the viable parameter space of the minimal single-component dark matter framework.

This tension can be alleviated in multicomponent dark matter scenarios, where the dark matter candidate considered here contributes only a fraction of the total relic abundance. In such setups, both relic density and direct detection constraints are effectively relaxed, allowing for larger values of $g_\chi$ without conflicting with experimental bounds or perturbativity requirements. Consequently, multicomponent dark matter provides a well-motivated framework in which TeV-scale mediator models with $\mathcal{O}(1)$ couplings can remain phenomenologically viable and yield sizable WCs.

We can analogously construct an effective Lagrangian involving vector and axial-vector operators at low energies. In this case, multicomponent dark matter scenarios can naturally accommodate vector operators with WCs of comparable magnitude to those in the scalar sector. The presence of multiple dark matter species relaxes the tight correlation between relic density and interaction strength: one component can remain sufficiently long-lived due to suppressed couplings, while another component can possess interactions of $\mathcal{O}(G_F)$ that govern thermal freeze-out.

From a theoretical perspective, vector and axial-vector operators are particularly well-suited to generate sizeable WCs. This originates from their chirality structure: vector and axial-vector currents are chirality conserving, connecting fermions of the same chirality (LL or RR). In contrast, scalar operators necessarily connect opposite chiralities (LR or RL), and are therefore associated with chirality-flipping transitions. In relativistic theories, such chirality flips are typically linked to mass insertions, which in ultraviolet completions translate into Yukawa-like suppressions. Vector and axial-vector interactions do not suffer from this limitation and can therefore support unsuppressed contributions. As a result, they can generate WCs of similar or even larger magnitude than scalar operators in certain regions of parameter space.

Overall, the inclusion of vector and axial-vector operators, together with a multicomponent dark matter framework, provides a theoretically consistent and phenomenologically viable extension of the minimal $t$-channel setup, allowing for $\mathcal{O}(1)$ WCs while satisfying cosmological and experimental constraints. While a complete construction of such extended models is beyond the scope of the present work, our analysis demonstrates that sizeable values of the effective WCs remain viable and can be consistently realised within a phenomenologically well-motivated framework.


\subsubsection*{(2) Lepton Portal Dark Matter Model's Bounds }
Another possible way to obtain such a low-mass fermionic dark matter candidate that generates these kinds of interactions is through lepton-portal dark matter models, as discussed in \cite{Okawa:2020jea, Iguro:2022tmr, Higuchi:2023kbt}. In this framework, the realization of light fermionic dark matter is studied in the context of a two-Higgs-doublet model. The SM is extended by introducing two Higgs doublets, $\Phi$ and $\Phi_{\nu}$, together with a fermionic dark matter particle $\Psi$. The scalar doublets take the forms:
\begin{equation}
	\Phi = \begin{pmatrix} G^{+} \\ \frac{1}{\sqrt{2}}(v + h + iG^{0}) \end{pmatrix}, \quad \Phi_{\nu} = \begin{pmatrix} H^{+} \\ \frac{1}{\sqrt{2}}(H + iA) \end{pmatrix}.
\end{equation}
The interaction of the DMs, which helps it to get relic, is given by \cite{Okawa:2020jea}:
\begin{equation}
	-\mathcal{L}_{\ell} = y_{\nu}^{i} \left[ \frac{1}{\sqrt{2}} \bar{\nu}_{L}^{i} (H - iA) \psi_{R} - \bar{e}_{L}^{i} H^{-} \psi_{R} \right] + \text{h.c.},
\end{equation}
 From the analysis of lepton portal dark matter in the works \cite{Iguro:2022tmr,Okawa:2020jea, Higuchi:2023kbt}, we note that for the charged Higgs mass $M_{H^\pm}=220$ GeV, the allowed values of $y_{\nu}^{\ell}$ which satisfies present DM and collider constraints could be as high as 1, which is even more relaxed for higher values of $M_{H^{\pm}}$ and could be greater than 1. In this model, we do not have any contribution in the charged current processes $b\to c \ell^-\bar{X}_{\rm inv}$ decays, since the charged higgs does not interact with the quarks. Neither does it have any contribution via loop-level processes. In the following paragraph, we will discuss an extended version of this model, which could contribute to $b\to c \ell^-\bar{X}_{\rm inv}$ decays at the tree level. Furthermore, we will be able to utilise the bounds given in \cite{Iguro:2022tmr,Okawa:2020jea, Higuchi:2023kbt} to obtain the allowed values of the relevant WCs of a few operators in eq.~\eqref{eq:general_Hamiltonian_b2c}. 

\paragraph{\underline{Connecting with the quark sector}:} 
In the scenario discussed in \cite{Okawa:2020jea}, DM does not interact with SM quarks, as the analyses are carried out in a lepton portal framework. To realize the interaction in eq.~\eqref{eq:general_Hamiltonian_b2c}, this setup can be extended in two ways. The first option is to allow the doublet $\Phi_{\nu}$ to couple directly to quark doublets. In this case, DM stability is determined by kinematics: for heavy quarks, stability is automatic, while for decays into lighter quarks, sufficiently small couplings can be chosen so that DM behaves as an LLP. The second option is to introduce a third scalar doublet that mixes with $\Phi_{\nu}$, thereby enabling DM to couple to heavy quarks through a charged scalar mediator. In either case, DM stability must be ensured by kinematics and suitably suppressed couplings to heavy and light quarks.

We consider a three Higgs doublet model (3HDM) with the doublets $\Phi$, $\Phi_1$ and $\Phi_2$, respectively. Here, $\Phi$ is the SM Higgs doublet and $\Phi_1$ is equivalent to $\Phi_\nu$ of ref.~\cite{Okawa:2020jea} and couples to leptons and fermionic DM and is odd under $Z_2$ symmetry. The doublet $\Phi_2$ interacts with the quark sector. Additionaly, we assume that $\Phi_1$ and $\Phi_2$ mixes via soft breaking term of strength $\mu_{12}^2$ and $\Phi_1$ do not mixes with the $\Phi$. If we focus only on the Yukawa terms associated with $\Phi_1$ and $\Phi_2$, a possible form of the corresponding Yukawa Lagrangian is given by:
\begin{eqnarray}
	\mathcal{L}_{Y} = Y_{u}^{i} \bar{q}_{L} (i \sigma_{2} \Phi_{2}) u_{R} + Y_{d}^{i}  \bar{q}_{L} \Phi_{2} d_{R} + Y_{\ell}^{i} \bar{\ell}_{L} \Phi_{2} e_{R} + + Y_{\nu}^{i} \bar{\ell}_{L} \Phi_{1} \Psi_{R} + \rm h.c.
\end{eqnarray} 
Here, $\Psi$ is the low-mass dark matter. 
The scalar doublets, $\Phi_{1}$ and $\Phi_{2}$ can mix and give the fields of mass eigenstates. The doublets will have the general structure:
\begin{equation}
	\Phi_{i} = \begin{pmatrix}
		\phi_{i}^{+} \\
		\frac{H_{i} + i A_{i}}{\sqrt{2}}
	\end{pmatrix}
\end{equation}
After mixing, the mass eigenstate fields are:
\begin{eqnarray}
	&& H_{1}^{+} = \cos\theta \, \phi_{1}^{+} + \sin \theta \, \phi_{2}^{+} \,, \nonumber \\
	&& H_{2}^{+} = -\sin\theta \, \phi_{2}^{+} + \cos \theta \, \phi_{2}^{+} \,.
\end{eqnarray}
The couplings can be chosen to be non-degenerate across different quark sectors. To prevent DM decay into light quark states, the corresponding couplings can be set to zero. Our mass range of interest for DM is $m_{\psi} \leq 1$ GeV, so non-zero couplings to the bottom and charm quarks are allowed to generate non-zero WCs in eq.~\eqref{eq:general_Hamiltonian_b2c}. The Lagrangian describing the interaction of the charged scalar with quarks is then given by:
\begin{eqnarray}
	& \mathcal{L}_{\rm cc} & =  \bar{u} \left[-Y_{u} \cos \theta \, P_{L} + Y_{d} \cos\theta \,P_{R}   \right] d \, H_{1}^{+} \nonumber \\
	&& +\bar{u} \left[Y_{u} \sin \theta \, P_{L} - Y_{d} \sin\theta \,P_{R}   \right] d \, H_{2}^{+}\,. 
\end{eqnarray}
Now, if the masses of $H_{1}^{+}$ and $H_{2}^{+}$ are nearly equal (but not exactly), they can both be represented as $H^{+}$, and the Lagrangian then reduces to:
\begin{eqnarray}
	&\mathcal{L}_{\rm cc}& = \bar{u} \left[ - Y_{u} (\cos \theta - \sin \theta ) P_{L} + Y_{d} (\cos \theta - \sin \theta) P_{R} \right]d H^{+}\,, \nonumber \\
	&& = \bar{u} \left[ Y_{u}^{\prime} P_{L} + Y_{d}^{\prime}P_{R} \right] d H^{+} \,. 
\end{eqnarray}
The model setup is discussed in detail in appendix \ref{appndx:model_3HDM}.
The analysis of ref.~\cite{Bernal:2023aai} directly determined the values of $Y_{u}^{\prime}$ and $Y_{d}^{\prime}$ from various observables. Among these, the most relevant observable for our case is the decay $H^{+} \to c \bar{b}$. The authors performed their analysis for $M_{H^{+}} = 130$ GeV. The main constraint arises from the decay $H^{+} \to c \bar{b}$, which is derived from the bound on the branching ratio of the following process:
\begin{eqnarray}\label{eq:t2bH2cb}
	\mathcal{B} (t \to H^{+} b ) \times \mathcal{B} (H^{+} \to c \bar{b})\,.
\end{eqnarray}
The works of \cite{Akeroyd:2018axd} and \cite{Akeroyd:2022ouy} provide bounds on their model parameters $|X|$ and $|Y|$, which can, in turn, be translated into constraints on our model parameters $|Y_{u}^{\prime}|$ and $|Y_{d}^{\prime}|$. It is important to note that, unlike their setup, our couplings are not related through a single parameter across all quark generations. We have the freedom to relax MFV assumptions and set certain couplings to be small in order to evade FCNC bounds. Moreover, since we are not interested in the lepton sector couplings (i.e., the DM-lepton interaction), $Y_{\ell}$ can also be taken to be negligibly small.

We have used the updated bounds from \cite{Bernal:2023aai}. The authors find that the constraints of eq.~\eqref{eq:t2bH2cb} are satisfied for $\hat{Y}_{A}^{U, D} < 1$. Using this, we obtain bounds on our couplings as $Y_{u}^{\prime} \sim Y_{d}^{\prime} \sim V_{cb}$, calculated for $M_{H^{\pm}} = 130$ GeV. The DM analysis in \cite{Iguro:2022tmr} is valid for $M_{H^{\pm}} = 220$ GeV. Although the quark-side analysis was performed for $130$ GeV, we use this bound as a reference for $M_{H^{\pm}} = 220$ GeV. While this is only approximate, it provides a reasonable estimate for combining with the DM constraints.

These couplings can be related to the scalar WCs as
\begin{equation}
	\frac{4 G_{F}}{\sqrt{2}} \, V_{cb} \, C_{S_{1(2)R}} \equiv \frac{y_{\nu} \, Y_{u(d)}^{\prime}}{M_{H^{\pm}}^2}\,,
\end{equation}
which leads to an approximate bound on the scalar WCs:
\begin{equation}
	C_{S_{iR}} \lesssim 0.63 \,.
\end{equation}

Similar types of interactions arise in other lepton-portal dark matter models, as well as in $t$-channel dark matter scenarios. In those cases, the relic density, direct detection bounds, the lepton anomalous magnetic moment $(g-2)$, and limits from lepton-flavour-violating processes can be simultaneously satisfied in the high-mass region of the DM parameter space. In our case, the relic density and direct detection bounds can be accommodated while remaining consistent with low-energy constraints, owing to the presence of an additional annihilation channel mediated by the neutral Higgs component of the $\Phi_{\nu}$ doublet. This yields a viable DM scenario that is compatible with all current experimental bounds.

However, it is important to emphasise that such estimates are inherently model-dependent and can vary significantly with the operator structure, flavour assumptions, and the choice of observables included in the analysis. Moreover, for many ultraviolet completions or less-explored operator combinations, robust and systematically derived bounds on the associated couplings are not readily available in the literature.

In light of this, we do not attempt to assign universal numerical ranges to all WCs. Instead, wherever possible, we refer to existing phenomenological studies and global analyses to motivate representative values, while in other cases we restrict ourselves to qualitatively reasonable parameter choices guided by the underlying scale suppression and consistency with current experimental constraints.

\section{Methodology}\label{sec:theo_formalism}
\subsection{Differential Decay Rate Distributions}
In this section, we will discuss the differential decay distributions of $B \to D^{(*)} \ell \bar{X}_{\rm inv}$ decays. We pointed out that apart from the SM neutrinos ($\nu$), $X_{\rm inv}$ represents a light mass dark-sector fermion $\chi$. Neither neutrinos nor $\chi$ will be detected experimentally, and they will have missing energy signatures. However, since $\chi$ has a light mass, the phase space available to the decays $B \to D^{(*)} \ell \bar{\nu} $ and $B \to D^{(*)} \ell \bar{\chi} $ will be different. In SM, the di-lepton invariant mass square $ q^2 $ has the following range: 
\begin{equation}\label{eq:q2_limit_SM}
	q^2 \Rightarrow\left[ m_{\ell}^2 ,~ (m_{B} -m_{D^{(*)}})^2 \right] \,.
\end{equation}
This range will be modified for a massive particle, like $\chi$, in the final state and the modified range will be
\begin{equation}\label{eq:q2_limit_NP}
	q^2 \Rightarrow\left[ \left(m_{\ell} + m_{\chi} \right)^2 ,~ (m_{B} -m_{D^{(*)}})^2 \right]\,,
\end{equation}
with $ m_{\chi} $ being the mass of the dark-fermion. Therefore, the contributions from the SM neutrinos and from the massive dark-sector fermions will be added at the decay width level. This is because, although both decays have experimentally identical signatures, they represent distinct processes, and the phase space available to them will be different. As pointed out above, the integration limit of \( q^2 \) will differ from that of the SM. 

Following the discussion above, we can write the decay width for $B \to D^{(*)} \ell \bar{X}_{\rm inv}$ as  
\begin{equation}\label{eq:total_decay_width}
	\Gamma(B \to D^{(*)} \ell \bar{X}_{\rm inv}) = \Gamma(B \to D^{(*)} \ell \bar{\nu})|_{SM} + \Gamma(B \to D^{(*)} \ell \bar{\chi} )|_{NP}\,.
\end{equation}
The detailed mathematical expressions for the decay widths $\Gamma(B \to D^{(*)} \ell \bar{\nu})|_{SM}$ and the respective decay distributions are available in the literature, for example, see the refs.~\cite{Sakaki:2013bfa,Becirevic:2019tpx}. 

The differential decay width of the process $ B \to D \ell \chi $ can be written as: 
\begin{eqnarray}
	\frac{d^2 \Gamma}{d q^2  \cos \theta_{\ell}} = \frac{G_{F}^2}{4}\frac{|V_{cb}|^2}{32 m_{B}^2} \frac{1}{(2 \pi)^3}  \frac{\lambda^{1/2}(q^2, m_{\ell}^2, m_{\chi}^2) |\vec{p}_{D}|}{q^2}  \left| \mathcal{M} (B \to D \ell \bar{\chi} )\right|^2\,,
\end{eqnarray}
with $|\vec{p}_{D}| = \frac{\lambda^{1/2}(m_{B}^2, m_{D}^2, q^2)}{2m_{B}}$. Here, $ |\mathcal{M}| $ is the matrix amplitude, summed over the lepton and dark fermion's spin. 
The K{\"a}llen function is defined as $\lambda(\alpha,\beta,\gamma) = \alpha^2+\beta^2+\gamma^2-2\alpha\beta - 2\beta\gamma-2\alpha\gamma$. In these differential decay distributions, $ q^2$ and $\cos \theta_{\ell}$ are the di-lepton invariant mass square and the angle between the leptons and the direction to the $B$ meson in the virtual $W$-boson rest frame, respectively.
We obtain the decay width $\Gamma(B \to D \ell \bar{\chi})$ after integrating the two-fold decay rate distribution $\frac{d^2 \Gamma(B \to D \ell \bar{\chi})}{d q^2  \cos \theta_{\ell}}$.  

The differential decay width for the process $\Gamma(B \to D^* \ell \bar{\chi})$ can be obtained by integrating the four-fold angular distribution
$\frac{d^4 \Gamma(B \to D^{(*)} \ell \bar{\chi})}{d q^2 , d\cos \theta_D , d\cos \theta_\ell , d\phi}$, which is defined in the subsequent subsection. For a massive $\chi$, we present the most general expression for this four-fold decay distribution, derived from the effective interaction introduced in eq.~\eqref{eq:general_Hamiltonian_b2c}. The corresponding SM result can then be recovered by retaining only the contribution from the operator $\mathcal{O}_{V_{1L}}^{\ell,SM}$ and taking the limit of a massless $\chi$. 

\subsection{Four-Fold Decay Distribution}
\begin{figure}[t]
	\centering
	\includegraphics[width=0.7\linewidth]{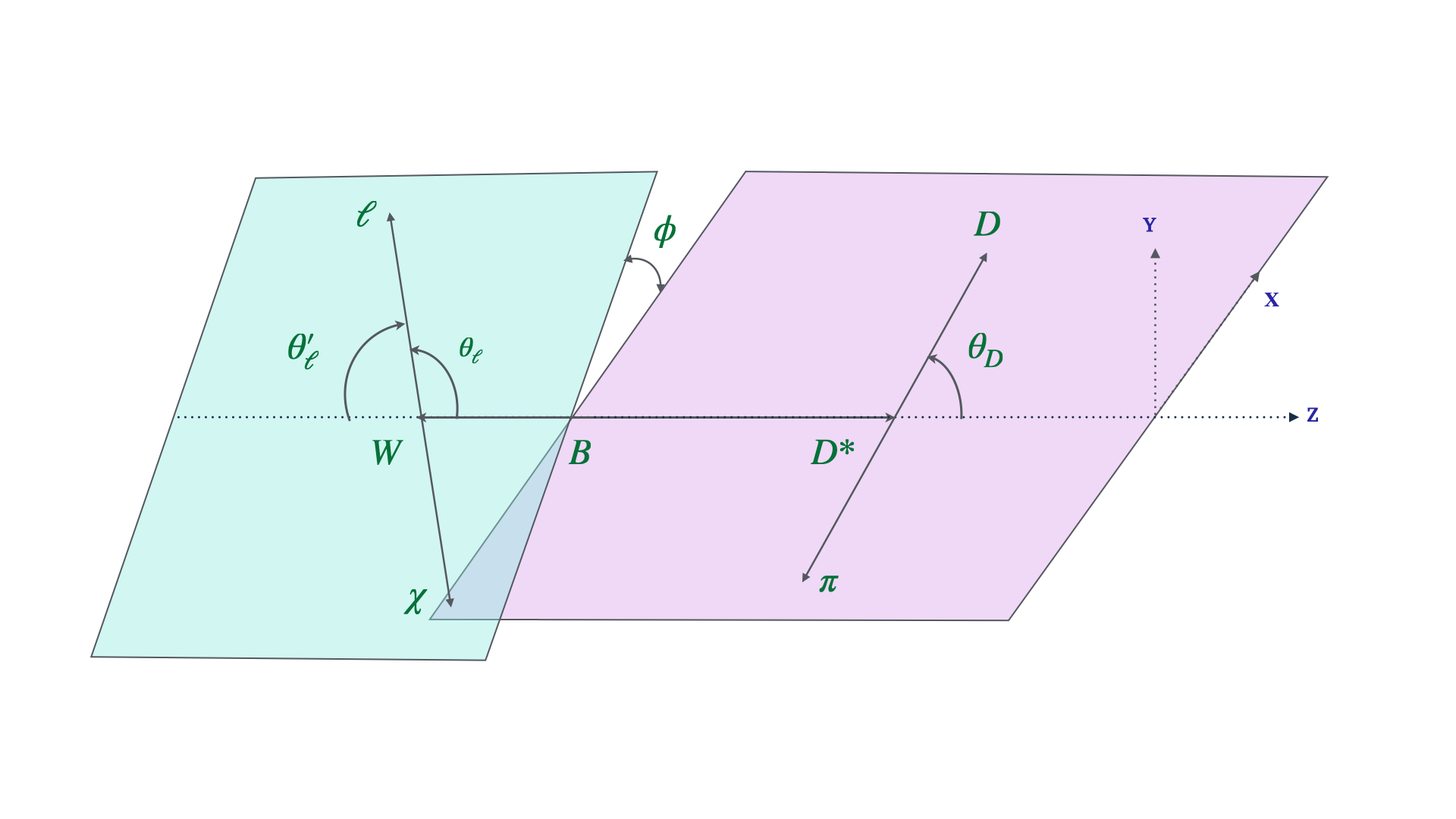}
	\caption{Schematic diagram of the decay plane of $B \to D^{*} (\to D \, \pi) \, \ell \, \bar{\chi}$.}
	\label{fig:decay_plane}
\end{figure}

We consider $ B \to D^{*}\ell \bar{\chi} $ followed by the decay $D^*(p_D, \epsilon) \to D (p_D) \pi(p_\pi)$. For the kinematics, we adopt the convention for angles and momenta as in fig.~\ref{fig:decay_plane}, with the lepton pair momentum $q=(p_B-p_{D^*})$. 
The differential decay width of the process $ B \to D^{*}(D\pi) \ell \bar{\chi} $ can be written as:
{\footnotesize \begin{equation}\label{eq:fivefold_decaywidth}
		\frac{d}{ dm_{D\pi}^2}\left(\frac{d^4 \Gamma }{d q^2 \, d\cos \theta_{D} \,  d\cos \theta_{\ell} \, d\phi} \right) = \frac{\lambda^{1/2}(m_B^2, m_{D^*}^2, q^2)}{256 (2\pi)^6 m_{B}^3} \frac{\lambda^{1/2}(q^2, m_{\ell}^2, m_{\nu}^2)}{q^2} \frac{|\vec{p}_{D}|}{m_{D\pi}} \sum_{\lambda_{\ell}, \lambda} \left| \mathcal{M}^{\lambda_{D^*}}_{\lambda_{\ell},\lambda_{\chi}}(B\to D \pi \ell \bar{\chi}) \right|^2.
	\end{equation}
}
In the above equation $\mathcal{M}^{\lambda_{D^*}}_{\lambda_{\ell},\lambda_{\chi}}(B\to D \pi \ell \bar{\chi})$ represents the four-body helicity amplitude and the parameters $\lambda_\ell$, $\lambda_{D^*}$ and $\lambda_{\chi}$, respectively, represent the helicities of the charged lepton, $D^*$ meson and dark sector fermion $\chi$. Corresponding to the decay plane in fig.~\ref{fig:decay_plane}, the kinematic variables are defined as
\begin{itemize}
	\item $q^2$: di-lepton invariant mass square, which we could express in terms of $w = \frac{m_{B}^2 + m^2_{D^{(*)}}-q^2}{2 m_{B}\, m_{D^{(*)}} }$. 
	\item $\theta_{D}$: The angle between the $D$ meson and the direction opposite to the $D^{*}$ meson in the $D^{*}$ rest frame.
	\item $\theta_{\ell}$: The angle between the leptons and the direction to the $B$ meson in the virtual $W$-boson rest frame. 
	\item $\phi$: The azimuthal angle between the two decay planes of $\ell-\chi$
	and $D -\pi$ in the $B$-meson rest frame.
\end{itemize}
The kinematical variables will vary within the following ranges:  
\begin{eqnarray}
	\cos \theta_{\ell}\, : \,  \{ -1, +1 \} \, , ~~ \cos \theta_{D^{*}} \, : \, \{ -1, +1 \}  \,, ~~\phi \, : \, \{ 0, 2 \pi \}\,.
\end{eqnarray}
and the allowed ranges of $q^2$ for the SM neutrinos and for $\chi$ are given in eqs.~\eqref{eq:q2_limit_SM} and \eqref{eq:q2_limit_NP}, respectively. 
We can write the amplitude for the process $\mathcal{M}_{\lambda_{\ell},\lambda}^{\lambda_{D^*}}(B\to D \pi \ell \bar{\chi})$ as:  
\begin{equation}\label{eq:matrix_amp_four_body}
	\mathcal{M}^{\lambda_{D^*}}_{\lambda_{\ell},\lambda_{\chi}}(B\to D \pi \ell \bar{\chi}) = \sum_{\lambda_{D^{*}}} \mathcal{M}^{\lambda_{D^*}}_{ \lambda_{D}}(D^* \to D\pi) \ \frac{i}{(m_{D\pi}^2 - m_{D^*}^2) + i m_{D^*} \Gamma_{D^*} } \  
	\mathcal{M}^{\lambda_{D^*}}_{\lambda_{\ell},\lambda_{\chi}}(B\to D^{*}\ell \bar{\chi})  
	\,.
\end{equation}
The square of this amplitude is given by
\small{
	\begin{equation}\label{eq:B2Dst_diffDecaywidth}
		\left|  \mathcal{M}^{\lambda_{D^*}}_{\lambda_{\ell},\lambda_{\chi}} (B\to D \pi \ell \bar{\chi}) \right|^2 =   \frac{\pi}{m_{D^{*}} \Gamma_{D^*}} \delta(m_{D\pi}^2 - m_{D^*}^2) \left| \sum_{\lambda_{D^{*}}} \mathcal{M}^{\lambda_{D^*}}_{\lambda_{\ell},\lambda_{\chi}}(B\to D^{*}\ell \bar{\chi})   \mathcal{M}^{\lambda_{D^*}}_{ \lambda_{D}}(D^* \to D \pi)  \right|^2.
	\end{equation} 
}
The origin of the Dirac-delta function in this amplitude square is the square of the modulus of the Breit-Wigner propagator of the $D^*$ meson. In the present case, the decay width $ \Gamma_{D^*} << m_{D^{*}} $, in narrow width approximation, we can write this as 
\begin{equation}
	\frac{1}{(m_{D\pi}^2 - m_{D^*}^2)^2 + m_{D^*}^2 \Gamma_{D^*}^2} \quad \xrightarrow{\text{\small$\Gamma_{D^*} << m_{D^{*}}$} }{}\quad  \frac{\pi}{m_{D^*} \Gamma_{D^*}} \delta(m_{D\pi}^2 - m_{D^*}^2)\,,
\end{equation}  
After integrating eq.~\eqref{eq:fivefold_decaywidth} with respect to $ dm_{D\pi}^2, $, we obtain the following four-fold angular decay distribution:
\begin{align}\label{eq:fourfold_decaywidth}
	\frac{d^4 \Gamma }{d q^2 \, d\cos \theta_{D} \,  d\cos \theta_{\ell} \, d\phi} &= \frac{\lambda^{1/2}(m_B^2, m_{D^*}^2, q^2)}{256 (2\pi)^6 m_{B}^3} \frac{\lambda^{1/2}(q^2, m_{\ell}^2, m_{\chi}^2)}{q^2}   |\vec{p}_{D}|  \frac{\pi}{m^2_{D^{*}} \Gamma_{D^*}} \times \nonumber \\ 
	&\qquad \sum_{\lambda, \, \lambda_{\ell}, \, \lambda_D}  \left| \sum_{\lambda_{D^{*}}} \mathcal{M}^{\lambda_{D^*}}_{\lambda_{\ell},\lambda_{\chi}}(B\to D^{*}\ell \bar{\chi})  \mathcal{M}^{\lambda_{D^*}}_{ \lambda_{D}}(D^* \to D\pi)  \right|^2.
\end{align}
To obtain the above equation, we have used the following integration
\begin{equation}
	\int \frac{1}{m_{D^*} m_{D\pi}}\delta(m_{D\pi}^2 - m_{D^*}^2) dm_{D\pi}^2 \rightarrow \frac{1}{m_{D^*} ^2}\,.
\end{equation}
The eq.~\eqref{eq:fourfold_decaywidth} can be reduced to 
\begin{equation}\label{eq:diffrate1}
	\frac{d^4 \Gamma}{dq^2\, d \cos\thl\,	d\cos\thD\, d\phi} = \mathcal{N} \sum_{\lambda_{\ell}, \lambda} \sum_{ \lambda_{D^*}}
	\Bigl|\mathcal{M}^{\lambda_{D^*}}_{\lambda_{\ell},\lambda_{\chi}} \, \mathcal{M}_{\lambda_{D^*}}^{ \lambda_{D}}\Bigr|^2.
\end{equation}
We have obtained the above expression of eq.~\eqref{eq:diffrate1} using the narrow width approximation for the Breit-Wigner propagator. The helicity amplitudes  $\mathcal{M}^{\lambda_{D^*}}_{ \lambda_{D}}$ for $D^* \to D \pi^+$ decay are discussed in the sec.~\ref{subsec:secdecay}. With the different pieces collected together We have the following expression for $\mathcal{N}$:  
\begin{equation} 	
	\mathcal{N} =  \frac{1}{2^{15} \pi^5 m_B^3 m_{D^*}^3 \Gamma_{D^*} } \; \frac{ \sqrt{\lambda(p_{B}^2, p_{D^*}^2, q^2)}
		\sqrt{\lambda(p_{D^*}^2, p_{D}^2, p_{\pi}^2)}
		\sqrt{\lambda(q^2,p_\ell^2,p_X^2)}} {q^2}  \, \,\,  ,
\end{equation}
where, $\lambda(p_{B}^2, p_{D^*}^2, q^2)= Q_+ Q_-, \, \lambda(p_{D^*}^2, p_{D}^2, p_{\pi}^2)= r_+ r_-$, with $Q_{\pm}$=$(m_{B} \pm m_{D^*})^2-q^2$ and $r_{\pm}$=$(m_{D^*} \pm m_{D})^2-m_{\pi}^2$.
\subsubsection{Secondary Decay: $D^*\to D \pi$}
\label{subsec:secdecay}
The $D^* \to D \pi$ decay amplitude can be written as \cite{Becirevic:2019tpx}
\begin{equation}
	\mathcal{M}_{\lambda_{D^*}}^{\lambda_D} = g_{D^* D\pi} (\epsilon_{\mu} \cdot p_{D}^{\mu}),
\end{equation}
with $|p_D|=\frac{\sqrt{\lambda(p_{D^*}^2, p_{D}^2, p_{\pi}^2)}}{2 m_{D^*}}$ the D three-momentum in the $D^*$ rest frame and the strong interaction strength $g_{{D^* D \pi}} $ can be extracted from the total width of the decay which is given as follows:
\begin{align}
	\Gamma(D^{*} \to D \pi) &= C ~\frac{|\vec{p}_D|}{24 \pi m_{D^*}^2} |\mathcal{M}(D^* \to D \pi)(m,n)|^2 \nn \,, \\
	&= C \frac{|\vec{p}_D|}{24 \pi m_{D^*}^2} g^2_{D^*D\pi}(\epsilon_{\mu}(m) \cdot p_{D}^{\mu})(\epsilon_{\nu}(n) \cdot p_{D}^{\nu})^*\,,
\end{align}
with $ C = 1,\frac{1}{2} $ for the decay to $ D^* $  to $ D^{\pm}\pi^{\mp} $ and $ D^0 \pi^0 $ respectively.
The polarization vector for the $D^*$ meson in the $D^*$ rest frame can be expressed as
\begin{equation}
	\epsilon_{\mu}(\pm)=\frac{1}{\sqrt{2}}(0,\pm 1, +i, 0), \, \, \,\,\, \epsilon_{\mu}(0)=(0,0,0,-1)\,,
\end{equation}
and the three-momentum for the $D$ meson is expressed as
\begin{equation}
	p_D^{\mu}=(E_D ,\, |p_D| \sin \thD ,\, 0 , \, |p_D| \cos \thD)\,,
\end{equation}
Specifying the $D^*$ polarization indices, we can write we can
\begin{align}
	\label{sec:secdecayamp}
	\Gamma(D^{*} \to D \pi) &= C \, \sum_{m,n} \frac{|\vec{p}_D|}{24 \pi m_{D^*}^2} |g^2_{D^*D\pi} F(m,n)|\,,
\end{align}
where,
\begin{equation}
	F(m,n) = (\epsilon_{\mu}(m) \cdot p_{D}^{\mu})(\epsilon_{\nu}(n) \cdot p_{D}^{\nu})^* \,,
\end{equation}
Using four momentum and polarization we have
\begin{equation}
	F(m,n)=\frac{|{\vec p}_D|^2}{2} \left(\begin{array}{ccc}
		\sin^2 \thD  &  -\sin^2 \thD & -\frac{1}{\sqrt{2}} \sin 2\thD \\
		- \sin^2 \thD  &  \sin^2 \thD & \frac{1}{\sqrt{2}} \sin 2 \thD \\
		- \frac{1}{\sqrt{2}} \sin 2\thD & \frac{1}{\sqrt{2}} \sin 2\thD & 
		2 \cos^2 \thD
	\end{array}\right),
\end{equation}
With rows and columns corresponding to values of $m=n= (+1, -1, \, 0)$ from top to bottom and left to right. Then the total decay width is evaluated as:
\begin{equation}
	\Gamma(D^{*} \to D \pi) = C \, \frac{|\vec{p}_D|^3 g^2_{D^*D\pi}}{24 \pi m_{D^*}^2}.
\end{equation}
Here, we have shown the helicity formalism for the two-body decay $D^* \to D \pi$. This framework is essential to construct the four-fold differential decay distribution of the process $B\to D^*(\to D \pi) \ell X$. In particular, we have demonstrated how the secondary polar angle $\theta_D$ and the azimuthal angle $\phi$ naturally emerge in the four-fold decay distribution. In the following subsection, we will discuss the calculation of the helicity amplitude $ M^{\lambda_{D^*}}_{\lambda_{\ell},\lambda_{\chi}}$ relevant to the decay rate distributions discussed above. 

\subsection{Total Decay Amplitude and Transition Matrix Elements}
Assuming a factorization between the hadronic and leptonic components, we can express the three-body helicity amplitude \( M_{\lambda_{\ell}\lambda_{\chi} }^{\lambda_{D^*}}(B\to M\ell \bar{\chi}) \) (with $M = D$ or $D^*$) in eq.~\eqref{eq:fourfold_decaywidth} as

\begin{align}\label{eq:amplitude_gen}
M_{\lambda_{\ell},\lambda_{\chi}}^{\lambda_{M}}(B\to M\ell \bar{\chi}) = 
\frac{G_{F}}{\sqrt{2}} V_{cb} \sum_{i=1}^{2} \bigg[ 
C^{\ell}_{V_{i}\,L(R)}\sum_{\lambda} \, \eta_{\lambda} H_{V_{i},\lambda}^{\lambda_{M}} L^{V,L(R)}_{\lambda_{\ell},\lambda_{\chi},\lambda} 
& + C^{\ell}_{T_{i}\,L(R)} \sum_{\lambda,\lambda'} \eta_{\lambda} \eta_{\lambda'} H_{T_{i},\lambda, \lambda '}^{\lambda_{M}} L^{T,L(R)}_{\lambda_{\ell},\lambda_{\chi},\lambda, \lambda'} \nonumber \\
& + C^{\ell}_{S_{i}\,L(R)} H_{S_{i}}^{\lambda_{M}} L^{S,L(R)}_{\lambda_{\ell},\lambda_{\chi}} \bigg]\,.
\end{align}
To obtain this amplitude, we have used the effective Hamiltonian defined in eq.~\eqref{eq:general_Hamiltonian_b2c} and the $C_{I_j}$s represent the WCs corresponding to different operators. In the above equation, $H$ and $L$ parametrise the hadronic and leptonic amplitudes. Here, $ \lambda,\, \lambda' = (0,\pm 1, t) $,  are the virtual gauge boson helicity, while  $ \eta_{\lambda} $ being the metric factor. 

The hadronic amplitudes in eq.~\eqref{eq:amplitude_gen} are defined as \cite{Sakaki:2013bfa}: 
\begin{align}\label{eq:hadronic_amplitudes}
	H^{\lambda_{M}}_{{V_{1(2)}}, \lambda } & =  \epsilon_{\mu}^{*} \langle M (\lambda_{M}) | \bar{c}\gamma^{\mu}(1 \mp \gamma_5)b|\bar{B} \rangle \,, \nonumber \\
	H^{\lambda_{M}}_{{S_{1(2)}}, \lambda } & =  \langle M (\lambda_{M}) |\bar{c} (1\mp \gamma_5)b | \bar{B}\rangle \,, \nonumber \\
	H^{\lambda_{M}}_{T_{1(2)} , \lambda \lambda'} &= H^{\lambda_{M}}_{T_{1(2)}, \lambda ' \lambda }  =  \epsilon_{\mu}^{*}(\lambda) \epsilon_{\nu}^{*}(\lambda')  \langle M (\lambda_{M}) |\bar{c} \sigma^{\mu \nu }(1\mp \gamma_5)b | \bar{B}\rangle\,, 
\end{align}
where $\lambda_{M}$ and $\lambda$ denote the helicities of the final state meson and the virtual intermediate boson. We set $ \lambda_{M} = s $ to represent the $D$ meson and for the $D^{*}$ meson, the helicities are $ \lambda_{M} = (0, \pm 1) $, respectively, and $ \lambda = (0, \pm 1, t) $ in the $ B  $ rest frame. Here, $ \epsilon^*_{\mu} $'$s$ are the polarisation vectors associated with the W boson in the $B$ rest frame, are written as:
 \begin{equation}\label{eq:pol_wboson1}
 \epsilon^*_{\mu}(t)= \frac{1}{\sqrt{q^2}}(q_0,0,0,|q|); \quad \epsilon^*_{\mu}(\pm ) = \frac{1}{\sqrt{2}}( 0, \pm 1, i, 0) ; \quad \epsilon^*_{\mu}(0) = \frac{1}{\sqrt{q^2}}(|q|,0,0,q_0); 
 \end{equation}
where, $q_0=(m_B^2+q^2-m_{D^*}^2)/2 m_B$. Here, $ \epsilon^{\mu}(k,\lambda)$'$s$ are the polarization vectors associated with the $D^*$ meson in the $B$ rest frame are written as:
 \begin{equation}\label{eq:pol_Dstmeson1}
\epsilon^{*{\mu}}(k,\pm ) = \frac{1}{\sqrt{2}}( 0,\, \mp 1, \, i, \, 0) \quad ; \quad \quad \epsilon^{*_{\mu}}(k,0) = \frac{1}{m_{D^*}}(|q|, \, 0,\, 0, \, m_B-q_0) \quad.
 \end{equation}
The kinematical details are presented in appendix \ref{appndxsec:leptonic_helicity}. The metric tensor is defined by $g_{\mu \nu}=(+, -, -, -)$.  We have expressed all these hadronic amplitudes in terms of the hadronic form factors. We have presented the detailed mathematical expressions of all the hadronic helicity amplitudes in the sec.~\ref{appndxsec:hadronic_amp} in the appendix.

Similarly, we define the leptonic amplitudes for charged lepton and the dark-fermion $ \chi $ as follows: 
\begin{subequations}\label{eq:leptonic_helicity_amp}
	\begin{eqnarray}
	L^{V, L(R)}_{\lambda_{\ell},\lambda_{\chi},\lambda} (q^2,\cos \theta_{\ell}) & = &  \epsilon^{\mu} (\lambda) \langle\ell(p_{\ell},\lambda_{\ell}),\bar{\chi}(p_{\ell},\lambda_{\chi}) | \bar{\ell} \gamma_{\mu}(1 \pm \gamma_5)\chi |0\rangle \nonumber \\
	& = & \epsilon^{\mu}(\lambda) \bar{u}_{\ell} \gamma_{\mu } (1\pm \gamma_5) v_{\bar{\chi}}\,, \\
	L^{S, L(R)}_{\lambda_{\ell},\lambda_{\chi}} (q^2,\cos \theta_{\ell}) & = &  \langle \ell(p_{\ell} ,\lambda_{\ell}) ,\bar{\chi}(p_{\ell},\lambda_{\chi}) | \bar{\ell} (1\pm\gamma_5)\chi |0\rangle = \bar{u}_{\ell} (1\pm \gamma_5) v_{\bar{\chi}}\,,  \\
	L^{T, L(R)}_{\lambda_{\ell},\lambda_{\chi},\lambda \lambda'}  (q^2,\cos \theta_{\ell}) & = & - i \epsilon^{\mu}(\lambda) \epsilon^{\mu}(\lambda') \langle \ell(p_{\ell} ,\lambda_{\ell}) ,\bar{\chi}(p_{\ell},\lambda_{\chi}) | \bar{\ell} \sigma_{\mu \nu} (1\pm\gamma_5)\chi |0\rangle  \nonumber \\
	& = & - i \epsilon^{\mu}(\lambda) \epsilon^{\mu}(\lambda') \bar{u}_{\ell} \sigma_{\mu \nu} v_{\bar{\chi}} \,. 
	\end{eqnarray}
\end{subequations}
 Here, $ \epsilon^{\mu} $'$s$ are the polarisation vectors associated with the virtual gauge boson in the $ W^{*} $ rest frame, and are written as:
\begin{equation}\label{eq:pol_wboson2}
 \epsilon^{\mu}(t)= (1,0,0,0); \quad \epsilon^{\mu}(\pm ) = \mp \frac{1}{\sqrt{2}}( 0, 1,\mp i,0) ; \quad \epsilon^{\mu}(0) = (0,0,0,-1); 
 \end{equation}
The energy and momentum of the lepton and the massive fermion will be modified as:
\begin{equation}
E_{\ell} = \frac{m_{\ell}^2 - m_{\chi}^2 + q^2}{2 \sqrt{q^2}}, \quad E_{\chi} = \frac{ q^2- m_{\ell}^2 + m_{\chi}^2 }{2 \sqrt{q^2}}, \quad |\vec{p}_{\ell}| = \frac{\lambda^{1/2} (q^2, m_{\ell}^2, m_{\chi}^2) }{2 \sqrt{q^2}}\,.
\end{equation} 
The detailed mathematical expressions for the leptonic helicity amplitudes are shown in  appendix~\ref{appndxsec:leptonic_helicity} with the leptonic helicity spinor required to calculate the leptonic helicity amplitudes. 

\section{Analysis and Results}\label{sec:analysis_results}
In the previous sections, we have discussed the formalism to study the differential decay width distribution of the decays $B \to D^{(*)} \ell \bar{\nu}(\bar{\chi})$ in the presence of a massive dark-sector fermion in addition to the massless SM neutrinos.
In this section, we will present the results of our analysis. We divide our analysis into the following two parts:
\begin{itemize}
	\item First, we will study the effects of the new BSM effective operators given in eq.~\eqref{eq:general_Hamiltonian_b2c} in the differential rate distributions and angular coefficient distributions of $B \to D^{(*)} \ell \bar{\chi}$ decays. These rate distributions will be sensitive to the mass $m_{\chi}$ and the new WCs. For certain allowed benchmark points (BPs) of these parameters, we will test the sensitivity of these rate distributions with respect to corresponding SM predictions for $B \to D^{(*)} \ell \bar{\nu}$ decays. 
	
	\item Furthermore, we can use the available data to extract the CKM element $|V_{cb}|$ alongside the new WCs simultaneously. This detailed study will help us constrain the new WCs and test the impact of these new BSM contributions on the extracted value of $|V_{cb}|$. 
    Extracting the CKM element and the corresponding WCs from the current experimental data does not allow us to include all available observables. In particular, we cannot use the binned distributions of $\frac{d \Gamma}{d\cos \theta_{D}}$, $\frac{d \Gamma}{d\phi}$, and $\frac{d \Gamma}{d \cos \theta_{\ell}}$ for $B\to D^*\ell^-X_{\rm inv}$. However, we can use the data on $\frac{\Delta\Gamma}{\Delta w}$, defined above for the $B \to D^{(*)} \ell X_{\rm inv}$ decays, obtained by integrating over small $q^2$ or $w$ bins within the physically allowed kinematic ranges.
    
\end{itemize} 
The helicity or transversity amplitudes of the hadronic currents depend on the form factors, which are functions of the transfer momentum \( q^2 \) or its equivalent variable \( w \). Hence, to study the decay rate distributions, we need proper knowledge of the form factors associated with the decays. In this regard, the inputs from the lattice collaborations are useful \cite{MILC:2015uhg, Na:2015kha, FermilabLattice:2021cdg, Aoki:2023qpa}. In the following paragraphs, we discuss the shapes of the $B\to D$ and $B\to D^*$ form factors, which we have obtained using the lattice data in the given references and the inputs on $B\to D^*$ form factors at $q^2=0$ obtained in the light cone sum rule (LCSR) approach \cite{Gubernari:2018wyi}.

\subsection{$q^2$-Shapes of the Form Factors}\label{subsec:q2shapes}
To determine the $q^2$ shape of the form factors, a suitable parametrization is necessary. Following Boyd-Grinstein-Lebed (BGL) \cite{Boyd:1994tt,Boyd:1997kz} parameterizations, we can express $B\to M$ form factors $ \mathcal{F}_{i}$ as a series expansion in $z$, such as 
\begin{eqnarray}
	\mathcal{F}_i (z) = \frac{1}{P_i (z) \phi_i (z)} \sum_{j=0}^{N} a_{j}^{\mathcal{F}_i} z^j,
	\label{eq:FF-BGL}
\end{eqnarray}
where $z$ is related to the recoil angle $w$ as:
\begin{eqnarray}\label{eq:z_definition}
	z = \frac{\sqrt{w+1}-\sqrt{2}}{\sqrt{w+1}+\sqrt{2}}\,.
\end{eqnarray}
The recoil variable is related to the momentum transfer $q^2$ as $q^2 = m_B^2 + m_{M}^2 - 2 m_B m_{M} w$.
The functions $P_i (z)$, called the Blaschke factors, are given by
\begin{eqnarray}
	P_i(z) = \prod_p \frac{z-z_{p_{i}}}{1 - z z_{p_{i}}},
	\label{eq:Blaschke-fact}
\end{eqnarray}
which are used to eliminate the poles at $z=z_p$ where,
\begin{equation}\label{eq:Zp}
	z_p = \frac{\sqrt{(m_B + m_M)^2 - m_P^2} - \sqrt{4 m_B m_M}}{\sqrt{(m_B + m_M)^2 - m_P^2} + \sqrt{4 m_B m_M}}.
\end{equation}
Here $m_P$ denotes the pole masses below the pair production threshold; for the details, please see the ref. \cite{Bigi:2017jbd}. In the following discussions, we will present the expressions for the outer functions $\phi_i (z)$ for $B\to D^{(*)}$ form factors. Furthermore, we will also mention the pole masses associated with the BGL expansion of the associated form factors. These outer functions are analytic functions of $q^2$ and are taken from \cite{Boyd:1997kz}.

\begin{table}[t]
	\begin{center}
		\begin{tabular}{|c|c|}
			\hline
			Form factor involved & $B_c^{(*)}$ pole masses (GeV)\\
			\hline
			$f_+$ and $g$ & 6.32847, 6.91947, 7.030\\
			$f$ and $F_1$ &  6.73847, 6.750, 7.145, 7.150\\
			$F_2$ & 6.27447, 6.8712, 7.250\\
			$f_0$ & 6.70347, 7.122\\
			\hline
		\end{tabular}
		\caption{Pole masses used in the $B \to D^{(*)}$ modes.}
		\label{tab:poleBD}
	\end{center}
\end{table} 
\paragraph{\underline{\bf Extraction of $B\to D$ form factors}:}
Following the above definitions, we can express $B\to D$ form factors $ \mathcal{F}_{i} = \{ f_{+}(z), f_{0}(z)\}$ as a series expansion in $z$. The recoil variable is related to the momentum transfer $q^2$ as $q^2 = m_B^2 + m_{D}^2 - 2 m_B m_{D} w$. The functions $P_i (z)$ will be as given in eq.~\eqref{eq:Blaschke-fact} with the replacement $m_M \to m_D$ in the definition of $z_p$ given in eq.~\eqref{eq:Zp}.
The pole masses  $m_P$ associated with $f_{+,0}(z)$ can be seen from Table \ref{tab:poleBD}.
The respective outer functions $\phi_i (z)$ are given as 
 \begin{subequations}\label{eq:outer_function_B2D}
 	\begin{eqnarray}
 		\phi_{f_{+}} & = & \frac{8 r^2}{m_{B}} \sqrt{\frac{\eta_{I}}{3 \pi \tilde{\chi}_{1}^{T} (0)}} \, \, \frac{(1+z^2 (1-z)^{1/2})}{\left[(1+r) (1-z) + 2 \sqrt{r} (1+z) \right]^4}\,, \\
 		\phi_{f_{0}} & = & r(1-r^2) \, \sqrt{\frac{8 \eta_{I}}{\pi \tilde{\chi}_{1^{-}}^{L}(0) }} \,\, \frac{(1-z^2) (1-z)^{1/2}}{\left[ (1+r) (1-z) + 2 \sqrt{r} (1+z) \right]^{4}} \,.
 	\end{eqnarray}
 \end{subequations}
 where $r = m_{D}/m_B$. The coefficients of the BGL parametrisation are constrained by unitary constraints: 
 \begin{equation}\label{eq:unitrary_B2D}
 	\sum_{n=0}^{\infty} \left( a_{j}^{i}\right)^2 <1\,. 
 \end{equation}
 In order to cancel the divergence at $ q^2 = 0$, the following condition needs to be imposed: 
 \begin{equation}\label{eq:FF_B2D_constraint}
 	f_{+} (0) = f_{0}(0)\,.
 \end{equation}

In refs.~\cite{MILC:2015uhg,Na:2015kha}, the Fermilab-MILC and HPQCD collaborations have provided data on $f_{+,0}(q^2)$ at a few values of $q^2\ne 0$. In this work, we combine the results of both these lattice collaborations. The inputs for the form factors at different $q^2$ points and their respective correlations can be seen from Table~1 of ref.~\cite{Jaiswal:2017rve}. We have adopted the methodology and the other relevant inputs related to BGL parametrization of the form factors from refs.~\cite{Jaiswal:2017rve,Ray:2023xjn}. For further discussion of unitarity bounds and form-factor parametrisations, see Refs.~\cite{Gopal:2024mgb, Gubernari:2026sqc}.

In this analysis, we have considered the BGL series expansion defined in eq.~\eqref{eq:FF-BGL} up to $N=2$. Furthermore, we have utilised the QCD and unitarity constraints in eqs.~\eqref{eq:FF_B2D_constraint} and \eqref{eq:unitrary_B2D}, respectively. Using the lattice inputs we have discussed above, we have extracted the BGL coefficients and presented the respective results in Table \ref{tab:B2D_combined_FF_fit}. With these results, we have obtained the $q^2$ shapes of the form factors and use these results further to extract the $w$ or $q^2$ distributions of $B\to D\ell^-\bar{X}_{\rm inv}$ decays, which we discuss in the next subsection. 
 
\begin{table}[t]
	\centering
	\renewcommand{\arraystretch}{1.4}
        \setlength{\tabcolsep}{6pt}
        \resizebox{0.35\textwidth}{!}{%
	\rowcolors{1}{cyan!15}{lime!10}
	\begin{tabular}{|c|c|}
		\hline
		Parameters & Fit Value \\
		\hline\hline
		$a_{0}^{+}$  	  &  $\text{0.01573(12)}$  \\
		$a_{1}^{+}$  	  &  $\text{-0.0386(35)}$  \\
		$a_{2}^{+}$  	  &  $\text{-0.0875(222)}$  \\
		$a_{0}^{0}$ 	  &  $\text{0.07735(58)}$  \\
		$a_{1}^{0}$  	  &  $\text{-0.213(16)}$  \\
		\hline 
	\end{tabular}
    }
	\caption{Fit results of the form factors coefficients of $ B \to D $  transition from the combined data of MILC \cite{MILC:2015uhg} and HPQCD\cite{Na:2015kha}. }
    \label{tab:B2D_combined_FF_fit}
\end{table}

\paragraph{\underline{\bf Extraction of $B\to D^*$ form factors}:}
Following eq.~\eqref{eq:FF-BGL}, we express the four form factors $\mathcal{F}_i = \{  f(z)$, $g(z)$, $\mathcal{F}_1 (z), \mathcal{F}_2 (z)\} $ relevant for $B \to D^{*} $ decay as a series expansion in $z$. In this decay, the lepton invariant mass squared is defined as $q^2 = m_B^2 + m_{D^*}^2 - 2 m_B m_{D^*} w $. The functions $P_i (z)$s are defined in eq.~\eqref{eq:Blaschke-fact}, where $ z_{p} $ defined in eq.~\eqref{eq:Zp} with $ m_{M} \to m_{D^{*}} $\,. The pole masses  $m_P$ associated with $\{  f(z)$, $g(z)$, $\mathcal{F}_1 (z)$ and $\mathcal{F}_2 (z)\} $ can be seen from Table~\ref{tab:poleBD}. The outer functions $\phi_i (z)$ are given as follows~\cite{Ray:2023xjn}:
\begin{subequations}\label{eq:outer-func}
	\begin{eqnarray}
		\phi_f &=& \frac{4r}{m_B^2} \sqrt{\frac{n_I}{6\pi \chi_{1^+}^T (0)}} \frac{(1+z)(1-z)^{3/2}}{\left[(1+r)(1-z) + 2\sqrt{r}(1+z)\right]^4}\,,  \\
		\phi_g &=& 16r^2 \sqrt{\frac{n_I}{3\pi \tilde{\chi}_{1^-}^T (0)}} \frac{(1+z)^2(1-z)^{-1/2}}{\left[(1+r)(1-z) + 2\sqrt{r}(1+z)\right]^4}\,,\\
		\phi_{\mathcal{F}_1} &=& \frac{4r}{m_B^3} \sqrt{\frac{n_I}{6\pi \chi_{1^+}^T (0)}} \frac{(1+z)(1-z)^{5/2}}{\left[(1+r)(1-z) + 2\sqrt{r}(1+z)\right]^5}\,,  \\
		\phi_{\mathcal{F}_2} &=& 8\sqrt{2}r^2 \sqrt{\frac{n_I}{\pi \tilde{\chi}_{1^+}^L (0)}} \frac{(1+z)^2 (1-z)^{-1/2}}{\left[(1+r)(1-z) + 2\sqrt{r}(1+z)\right]^4}\,.
	\end{eqnarray}
\end{subequations}
where $r = m_{D^*}/m_B$ and the other inputs can be found in \cite{Bigi:2017jbd}. Therefore, for $N = 2$, there are twelve coefficients, $a_{j}^{\mathcal{F}_i} $ for the four form factors. These coefficients satisfy the following weak unitary constraints \cite{Boyd:1997kz}:
\begin{eqnarray}
	\sum_{j=0}^{N} (a_j^{g})^2 < 1,~~\sum_{j=0}^{N} (a_j^{f})^2 + (a_j^{\mathcal{F}_1})^2  < 1, ~~\sum_{j=0}^{N} (a_j^{\mathcal{F}_2})^2  < 1.
	\label{eq:FF-unity-constr}
\end{eqnarray}
Additionally, these form factors satisfy the following QCD constraints at both zero and maximum recoil: 
\begin{subequations}
	\begin{eqnarray}
		\mathcal{F}_1 (1) &=& m_B (1-r) f(1),\\
		\mathcal{F}_2 (w_{max}) &=& \frac{1+r}{m_B^2 (1+w_{max}) (1-r)r} \mathcal{F}_1 (w_{max}).
		\label{eq:FF-kin-constr}
	\end{eqnarray}
\end{subequations}

We consider these constraints in our analysis, which will relate the BGL coefficients and reduce the number of independent BGL coefficients. In our analysis, we have considered the masses of the leptons. Hence, only 10 independent form factor coefficients are required to fit the theory to the data. For the numerical analysis presented here, we perform maximum-likelihood parameter estimation using OpTex \cite{sunando_patra_2019_3404311}, a Mathematica-based package. 
 
For the analysis, we have used the inputs on the form factors from the Fermilab-MILC collaboration~\cite{FermilabLattice:2021cdg}, JLQCD~\cite{Aoki:2023qpa}, and HPQCD collaboration~\cite{Harrison:2023dzh}. We have also used the inputs from LCSR \cite{Gubernari:2018wyi} at $q^2 = 0$. We did a combined analysis with these available inputs and extracted the relevant BGL coefficients, which we have presented in Table~\ref{tab:B2Dst_combined_FF_fit}.
We used the combined analysis results to obtain the decay rate and angular distributions, which we discuss below.

  \begin{table}[t!]
 	\centering
 	\renewcommand{\arraystretch}{1.2}
 	\setlength{\tabcolsep}{3.5pt}
 	\resizebox{0.3\textwidth}{!}{%
 		\rowcolors{1}{cyan!15}{lime!10}
 		\begin{tabular}{|cc|}
 			\hline
 			Parameter & Fit Value \\
 			\hline\hline
 			$a_{0}^{g}$  &  $\text{0.03117(84)}$  \\
 			$a_{1}^g$    &  $\text{-0.054(33)}$  \\
 			$a_{2}^{g}$  &  $\text{-0.70(115)}$  \\
            $a_{0}^f$    &  $\text{0.01206(13)}$  \\
 			$a_{1}^f$    &  $\text{0.0121(83)}$  \\
 			$a_{2}^f$    &  $\text{-0.077(339)}$  \\
 			$a_{1}^{\mathcal{F}_1}$  &  $\text{-0.0041(20)}$  \\
            $a_{2}^{\mathcal{F}_1}$  &  $\text{-0.0051(708)}$  \\
 			$a_{0}^{\mathcal{F}_2}$  &  $\text{0.0484(10)}$  \\
 			$a_{1}^{\mathcal{F}_2}$  &  $\text{-0.219(49)}$ \\
            \hline
            d.o.f & 22 \\
 	        p-value & 0.71\\
 			\hline
 	\end{tabular}}
 	\caption{Fitted results of the BGL coefficients associated with the form factor for the $B \to D^{*}$ transition were obtained from the combined data of the JLQCD~\cite{Aoki:2023qpa}, MILC~\cite{FermilabLattice:2021cdg}, and HPQCD~\cite{Harrison:2023dzh} collaborations at various $q^2$ values, along with LCSR~\cite{Gubernari:2018wyi} data at $q^2=0$.}\label{tab:B2Dst_combined_FF_fit}
 \end{table}
\subsection{Available Experimental Measurements}\label{sec:methodology}
\paragraph{\underline{Observables of $ B \to D \ell X_{\rm inv} $}:}
The data on the differential rate for $B \to D \ell \bar{\nu}$ $(\ell = \mu, e)$ is provided by the Belle collaboration \cite{Belle:2015pkj} in which the collaboration has given the data only on the bin-integrated observable: $ \frac{\Delta \Gamma}{\Delta w} $, where 
\begin{equation}\label{eq:obsdelG}
\Delta \Gamma = \int_{w_{i}}^{w_{j}}\frac{d \Gamma}{d w} d w \,,
\end{equation}
and $ w_{j}-w_{i} = \Delta w$, the difference between two consecutive bins.  In this work, we will calculate these observables in the SM and with the set of effective operators we have discussed in eq.~\eqref{eq:general_Hamiltonian_b2c}. The comparison with the SM will help us understand whether we will observe any deviations with respect to the SM, given the choices of new physics parameters like the mass $m_\chi$ or the new WCs. Furthermore, given the mass $m_{\chi}$, within the allowed range of $q^2$ or $w$, we can compare our estimate with the data mentioned in the above reference. Note that the Belle collaboration has used the same data to extract the CKM element $V_{cb}$, assuming rate distributions purely from the SM. In the presence of new operators, we can check the extracted value of $V_{cb}$, off-course, for this we need to choose only the bins allowed by the changes in the kinematics due to a massive $\chi$.  
\paragraph{\underline{Observables of $ B \to D^{*} \ell X_{\rm inv} $}:}
For the decay $ B \to D^{*} \ell \bar{\nu}_{\ell}$, the most recent result is provided by Belle-II, both on normalised partial decay width \cite{Belle:2023bwv} and un-normalised partial decay width \cite{Belle-II:2023okj}. In normalised decay width, the data on bin-wise partial branching ratios $ \frac{\Delta \Gamma}{\Gamma} $ are provided, where $ \Gamma $ is the total decay width and $ \Delta \Gamma $ is defined as
\begin{equation}\label{eq:Obs_Belle_normalised}
	\Delta \Gamma = \int_{x_i}^{x_j} \frac{d \Gamma}{dX}dX\,,
\end{equation}
$ X $ stands for the kinematical variables $ w, \cos \theta_{D}, \phi $ and $ \cos \theta_{\ell}$. And $ x_{i}, x_j$ define the integration limits of different bins obtained for the relevant kinematical variable $X$. On the other hand, Belle-II \cite{Belle:2023bwv} gives the data on $ \Delta \Gamma $, which is integrated in the bins of the parameters $ w, \cos \theta_{D}, \phi $ and $\cos\theta_{\ell}$ i.e the observable defined in \eqref{eq:Obs_Belle_normalised}. 

In all these observables in different bins of the different kinematical variables, we will estimate the contributions of the new effective operators of eq.~\eqref{eq:general_Hamiltonian_b2c} and check their respective sensitivities. Given the allowed benchmark scenarios, we will look for possible deviations in NP scenarios concerning the SM contribution, if any. Note that in the experimental analyses, the differential distributions, like $\frac{d \Gamma}{d\cos \theta_{D}}$, $\frac{d \Gamma}{d\phi}$ and $\frac{d \Gamma}{d \cos \theta_{\ell}}$ are after integration over $q^2$ or $w$. The integration range of $q^2$ or $w$ is defined assuming the presence of a left-handed massless neutrino in the final state; hence, the respective integration range is given in eq.~\eqref{eq:q2_limit_SM}. To check the impact of the new effective operators on the extraction of $V_{cb}$ we can not use these three angular distributions, however, we can still use the data $\frac{d \Gamma(B\to D^*\ell\nu)}{dq^2}$ or $\frac{d \Gamma(B\to D^*\ell\nu)}{dw}$ within the allowed ranges of $q^2$ (eq.~\eqref{eq:q2_limit_NP}).

Furthermore, in~\cite{Belle:2015pkj}, Belle has provided the distribution of missing energy for different bins of $w$ from the decay $B \to D \ell \bar{\nu}$. For the lower bin of $w$, the distribution peaks near $~0.5 \, $ $\rm GeV^{2}$, whereas for higher bins, the peak tends to zero. For the decay $B \to D^{*} \ell \bar{\nu}$, Belle~\cite{Belle:2023bwv} shows the distribution of $m_{\rm miss}^2$, the vicinity of the peak lies near $\rm -0.5\, GeV^2 < m_{miss}^2 < 0.5 \, GeV^{2} \, \rm $.  

\subsection{Decay Distribution}\label{sec:B2Dst_unnormalised}

\begin{figure}[t]
	\centering
	\vspace{-0.5cm}
	\subfloat[]{\includegraphics[scale=0.175]{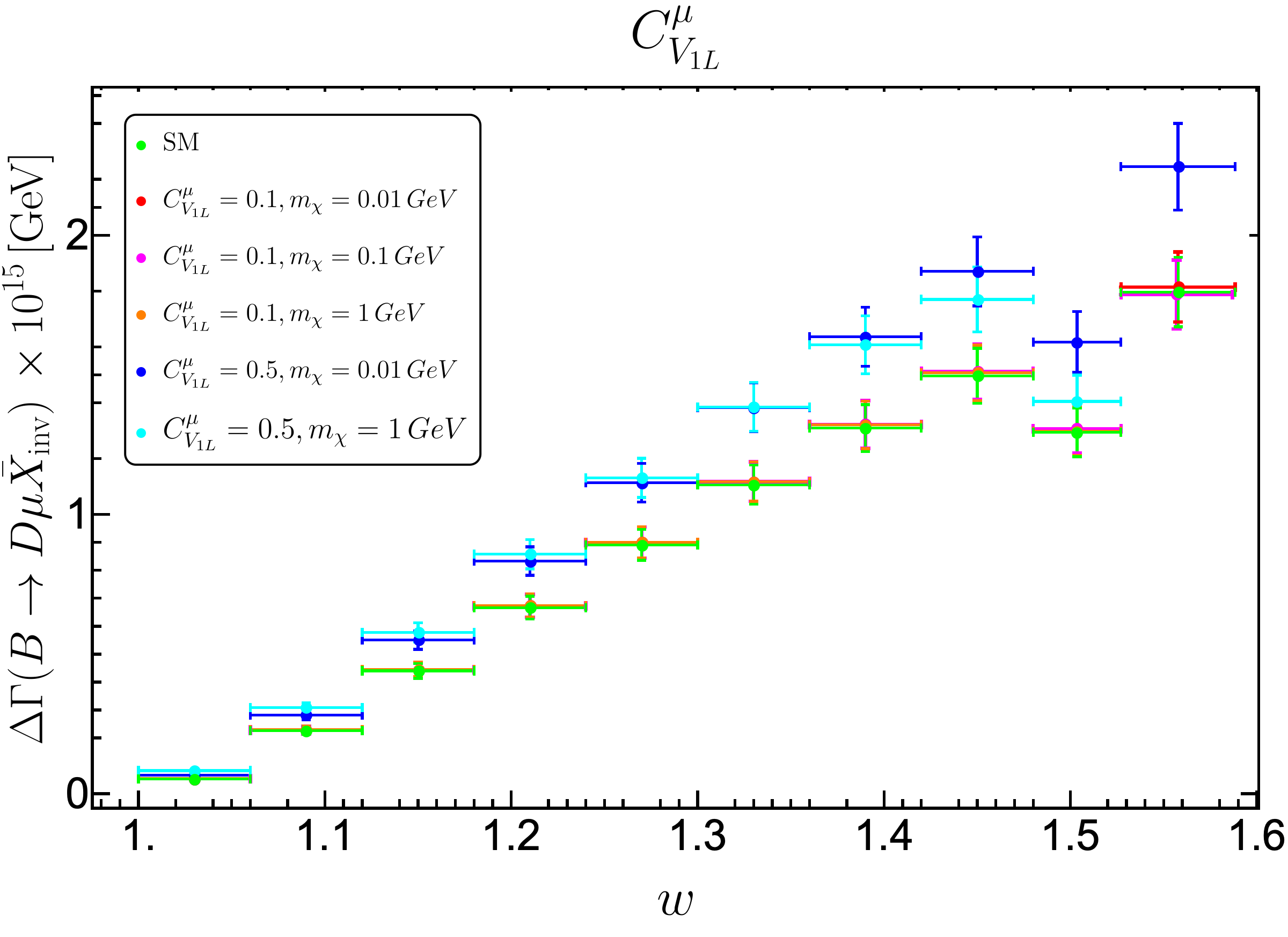}}\hspace{0.0001cm}
	\subfloat[]{\includegraphics[scale=0.175]{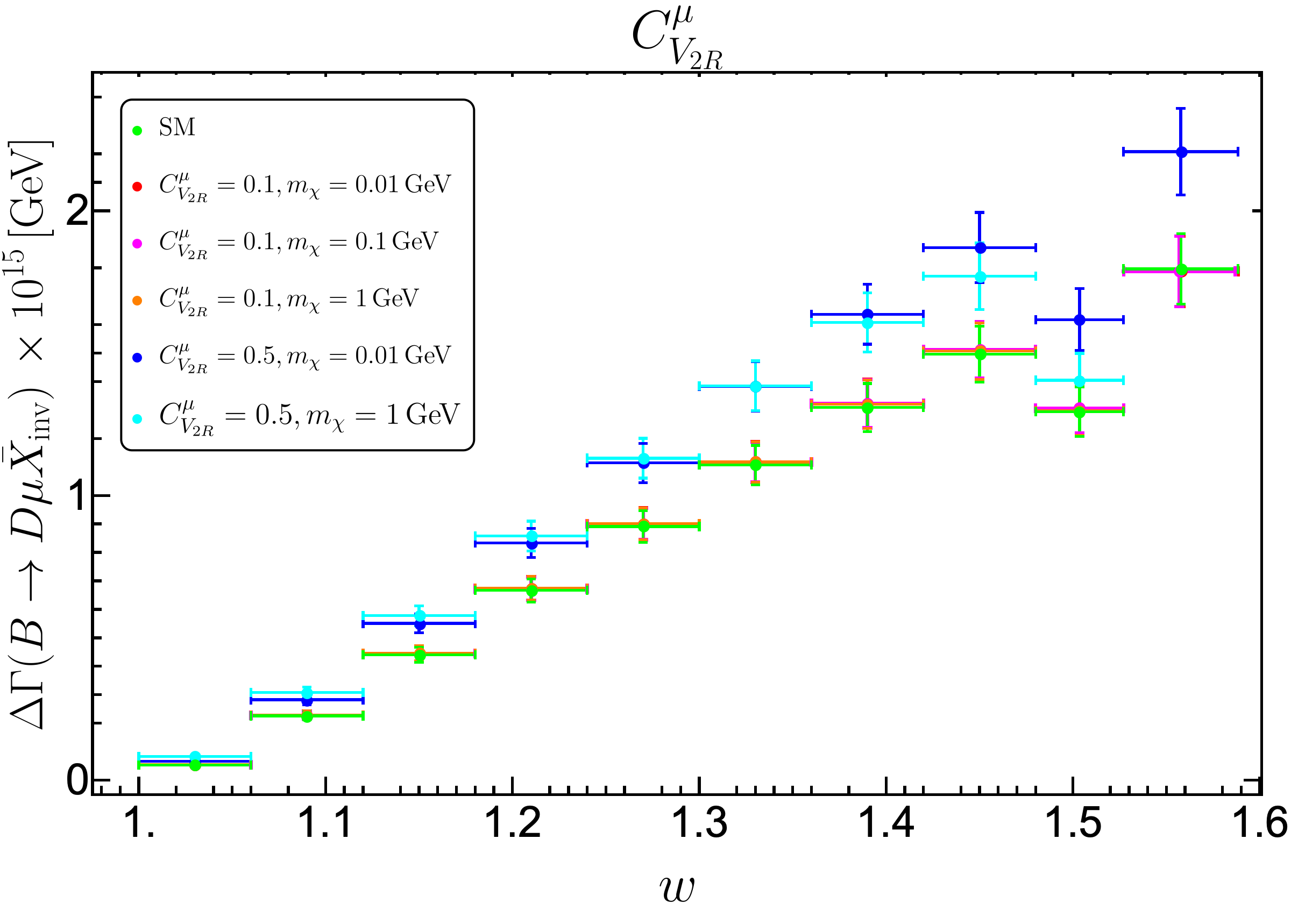}}\hspace{0.0001cm}	
	\subfloat[]{\includegraphics[scale=0.175]{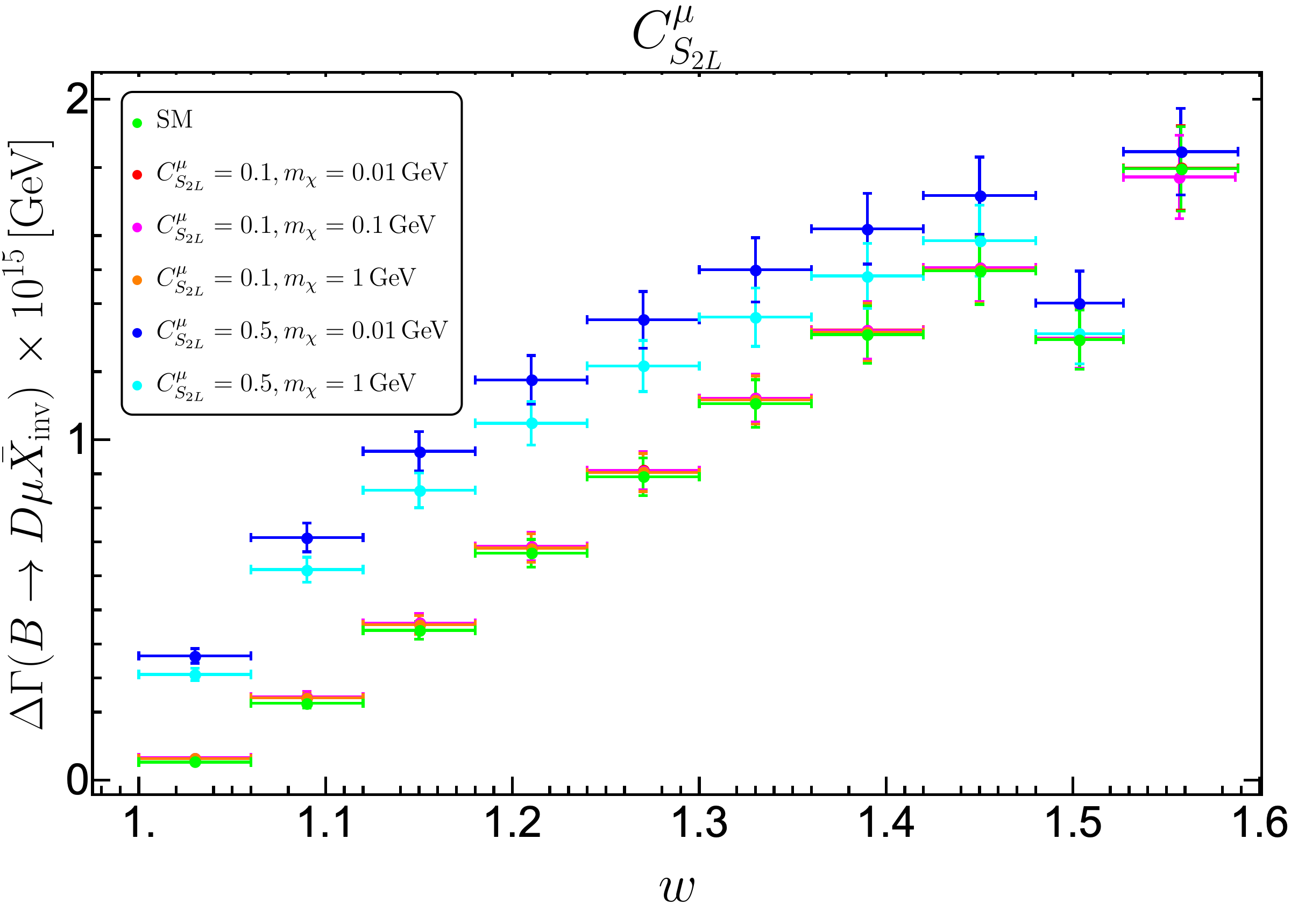}}\hspace{0.0001cm}
	\subfloat[]{\includegraphics[scale=0.175]{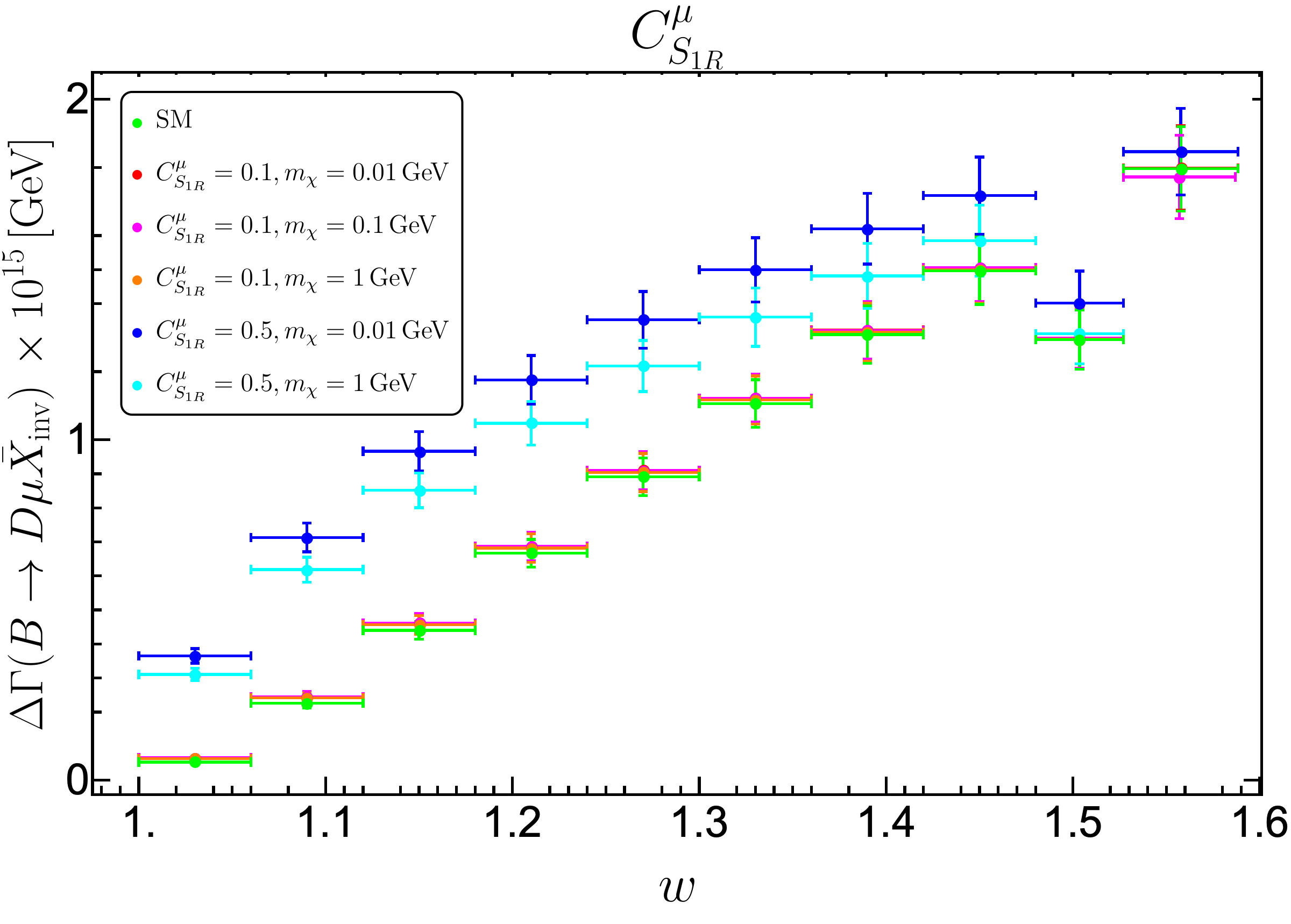}}\\
	\subfloat[]{\includegraphics[scale=0.175]{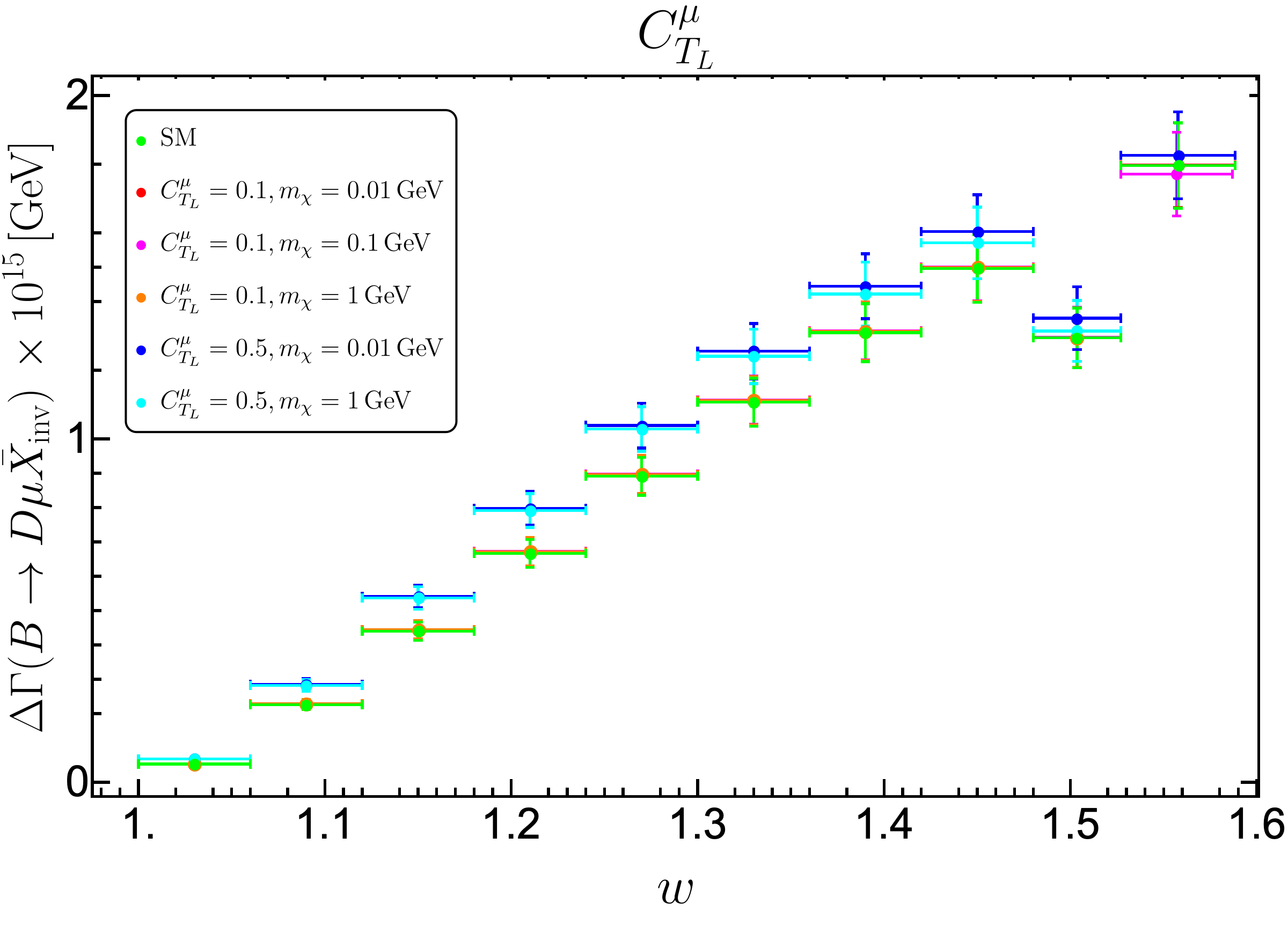}}\hspace{0.0001cm}
	\subfloat[]{\includegraphics[scale=0.175]{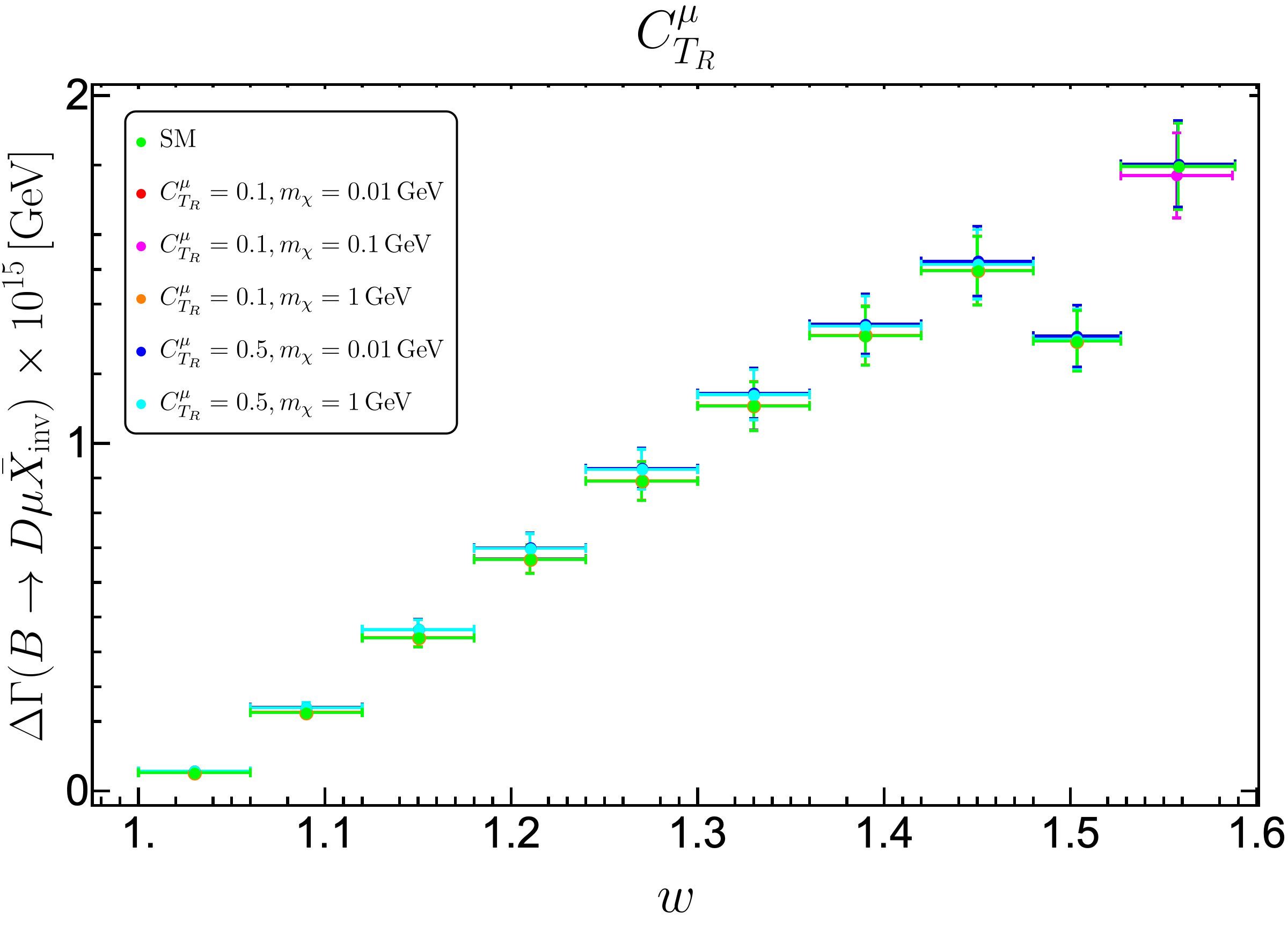}}
	\caption{Variation of $ \Delta \Gamma $ for the decay $B \to D \, \ell \, X_{\rm inv}$, integrated in bins of $w$ for different NP operator scenario. Each NP plot is done for $ m_{\chi} =(0.01, 0.1, 1)$ GeV for each NP operators with two different values $ C_{i}=0.1$ and $ 0.5 $. In each plot, the green colour points are the SM distribution. }\label{fig:B2D_unnormalised_omega_NP}
\end{figure}

In this subsection, we present our estimate of all the different bin-wise integrated decay rate distributions associated with $B\to D^{(*)}\ell^-\bar{\nu}$ (SM) and $B\to D^{(*)}\ell^-\bar{\chi}$ decays using the extracted form factors presented in sec.~\ref{subsec:q2shapes}. We have compared the SM results with the respective rates obtained in the scenarios, including the contributions from massive dark-sector fermions for each distribution. First, we will discuss the observables associated with $B\to D$ decays, followed by the bin-wise integrated decay rate distributions in $B\to D^*$ decays.

\subsubsection{Decay distribution in $B\to D\ell^-\bar{X}_{\rm inv}$:}
In this paragraph, we will discuss our estimate of the rate distribution of $B \to D \ell^- \bar{\chi}$ decay, for different benchmark values of the new physics WCs and the mass of the dark sector particle $\chi$ and look for any possible deviations with respect to the corresponding SM predictions. For purely SM type interaction, where the final state particle is a massless neutrino, for the decay $B \to D \mu \bar{\nu}$, we get the allowed range for the kinematic variable $w$ as: $ =  [1.00-1.59] $. The allowed kinematic range will be reduced as a massive particle is introduced into the final state, depending on its mass. For different masses of the final state particle $\chi$, the allowed range for $w $ is obtained and given below: 
\begin{subequations}\label{eq:B2D_allowed_omega}
	\begin{eqnarray}
		&m_{\chi} = 0.01\, \text{ GeV},  \,w \, : \,  [1.00-1.588]\,, \\
		&m_{\chi} = 0.1\, \text{ GeV},  \,w \, : \,  [1.00-1.587]\,, \\
		&m_{\chi} =1 \,\text{ GeV}, \, w  \, : \, [1.00-1.527]\,.
	\end{eqnarray}
\end{subequations}
It is obvious from eq.~\eqref{eq:B2D_allowed_omega}, that the maximum allowed value of $w$ decreases in the presence of the massive final state particle $\chi$, compared to the massless SM like final state particle $\nu$, i.e, $w_{\chi}^\text{max} < w_{\text{SM}}^\text{max}$. 

We have estimated $\Delta\Gamma$ using eq.~\eqref{eq:obsdelG} from the differential distribution $\Delta\Gamma/\Delta w$ after integrating in small $w$-bins. To do so, we need the $q^2 \, (w)$ shapes of the form factors $F_{+,0}(q^2)$. In sec.~\ref{subsec:q2shapes}, we have extracted the BGL coefficients which we require to get $F_{+,0}(q^2)$. Additionally, to obtain $\Delta\Gamma$ we need input on the CKM element $|V_{cb}|$ for which we have used \cite{Belle-II:2023okj,ParticleDataGroup:2024cfk}:
\begin{equation}\label{eq:CKM_and_lifetime}
	|V_{cb}| = (40.57 \pm 0.31 \pm 0.95 \pm 0.58) \times 10^{-3}.
\end{equation}


\paragraph{\underline{$ \Delta \Gamma  $- integrated in bins of $w$}: }
Fig.~\ref{fig:B2D_unnormalised_omega_NP} shows the distribution of the differential decay width \( \Delta \Gamma \), integrated in bins of \( w \), for the SM and for $B\to D\ell^-\bar{X}_{\rm inv}$ decays in different benchmark NP scenarios. Each plot contains the SM distribution and the distributions correspond to a specific WC, with different values of the dark fermion mass \( m_{\chi} \), like \( m_{\chi} = 0.01, \, 0.1 \) and \( 1 \) GeV. Following the discussions in the earlier section, we have considered the values of WCs \( C_{ij} = 0.1 \) and \( 0.5 \), for each value of $m_{\chi}$. The cases with the benchmark values \( C_{ij}\) and \( m_{\chi} \) (in GeV) intermediate to the values mentioned above are omitted to keep the plots clear, as one can easily infer that behaviour from the given plots. As mentioned earlier, for a massive dark sector fermion in the final state, the allowed range of \( w \) becomes smaller than that in the SM. Thus, we have shown only the kinematically allowed bins for each $m_{\chi}$ value. Hence, for $m_{\chi} = 0.01$ GeV, there will be one extra bin compared to those for $m_{\chi} = 1$ GeV. In these plots, we present the estimated 1$\sigma$ error bars in all the bins. 
Note that in the NP scenarios, the additional bin is not available for the massive fermion and therefore follows the SM distribution; it is not shown separately.
In both the SM and NP scenarios, the dominant source of uncertainty arises from the form factors. For the unnormalised decay distribution, the uncertainty associated with the CKM element further increases the bin-wise uncertainty. 
The plots are shown only for WCs: $\mathcal{C}_{V_{1L}}^{\ell}$, $\mathcal{C}_{V_{2R}}^{\ell}$, $\mathcal{C}_{S_{2L}}^{\ell}$, $\mathcal{C}_{S_{1R}}^{\ell}$ and $\mathcal{C}_{T_{L,R}}^{\ell}$. In this case, the distributions are similar for left and right-handed lepton and quark-currents, which is because the hadronic matrix elements and the matrix element squared of the leptonic current are the same in both these cases. The plots for $\mathcal{C}_{V_{2L}}^{\ell}$ and $\mathcal{C}_{V_{1R}}^{\ell}$ will be similar to those in the scenarios $\mathcal{C}_{V_{1R}}^{\ell}$ and $\mathcal{C}_{V_{2L}}^{\ell}$, respectively. A similar observation is there for scalar pseudoscalar current where the distributions are similar for $\mathcal{C}_{S_{1L}}^{\ell}$, $\mathcal{C}_{S_{2R}}^{\ell}$, $\mathcal{C}_{S_{2R}}^{\ell}$ and $\mathcal{C}_{S_{1R}}^{\ell}$. Hence, we have not shown those plots separately. For the $\mathcal{C}_{T_{L}}^{\ell}$ and $\mathcal{C}_{T_{R}}^{\ell}$, the distributions are different and we note that the deviations in the left-handed quark and lepton currents but not for the right handed lepton and quark currents, this is because the hadronic and leptonic current amplitudes will contribute differently in the two cases. 

In the items below, we will summarise a few other interesting observations: 
\begin{itemize}
	\item Apart from $\mathcal{O}_{T_{R}}^{\ell}$, in all the rest of the one operator scenarios we note deviations in the decay rate $\Delta\Gamma(B\to D\ell^-\bar{X}_{\rm inv})$ in a certain number of $w$-bins as compared to their respective SM predictions. Of course, the degree of discrepancies is much larger in the scenarios $\mathcal{O}_{S_{1R(L)}}^{\ell}$ and $\mathcal{O}_{S_{2L(R)}}^{\ell}$ as compared to the other operators. 
	
	\item The discrepancies between the SM and the NP scenarios increase with the lowering of the $m_{\chi}$ values.
	
	\item We observe that the larger deviations occur for higher values of the WCs, which is expected since NP contributions are added at the decay width level. We see from the plots, in our chosen benchmark scenarios, the maximum deviations we observe for $\mathcal{C}_{i_{1R(L)}}^{\ell} \approx 0.5$. 
	
	\item For the scenarios $\mathcal{O}_{V_{1L(R)}}^{\ell}$,  $\mathcal{O}_{V_{2L(R)}}^{\ell}$ and $\mathcal{O}_{T_{L}}^{\ell}$, the deviations are around 2$\sigma$ level even for $\mathcal{C}_{i_{1R(L)}}^{\ell} = 0.5$. However, for $\mathcal{O}_{S_{1R(L)}}^{\ell}$ and $\mathcal{O}_{S_{2L(R)}}^{\ell}$ in the four lower $w$-bins the discrepancies are much larger than 5$\sigma$ for these values of the respective WCs. Even if we consider $\mathcal{C}_{i_{1R(L)}}^{\ell} \approx 0.2$, we note deviations in the first two $w$ bins which are 5$\sigma$ or a little above. With the higher values of $w$, the errors in the predictions are increasing, hence, the degree of discrepancies is decreasing.  

    \item The differential decay width distributions, integrated over bins of $w$, for the decay $B \to D\mu X_{\rm inv}$ are shown above. In the vector and scalar operator scenarios, the left- and right-chiral cases lead to nearly identical distributions, making it challenging to distinguish between the two chiralities using this observable. In contrast, the tensor operator scenarios produce distinctly different distributions for $C_{T_L}^{\mu}$ and $C_{T_R}^{\mu}$, demonstrating that this observable can effectively discriminate between the two tensor chiralities, i.e., between left- and right-handed lepton currents. If deviations in the $q^2$ distribution similar to those shown above are observed, which cannot be attributed to the presence of right-handed neutrinos, they may instead point to the existence of left-handed invisible fermions. 

\end{itemize}

\paragraph{\underline{Angular Observable $ A_{FB}$}: }
The observable $\Delta \Gamma (B \to D \ell X_{\rm inv})$ is obtained after integrating the angular variable $\theta_{\ell}$ over the full region. To find any sensitivity of the angular distribution with the NP, the observable $A^\ell_{FB}(q^2)$ is important in this sector. The dependence on $\cos\theta_\ell$ can be understood in the forward-backwards asymmetry, which is defined as:
\begin{align}
    \mathcal{A}^\ell_{FB}(q^2)=&\frac{\int^{1}_{0} \frac{d^2 \Gamma}{d q^2 \, d \cos\theta_\ell}-\int^{0}_{-1} \frac{d^2 \Gamma}{d q^2 \, d \cos\theta_\ell}}{\frac{d \Gamma}{d q^2}}\,.
\end{align}
\begin{figure}[t]
	\centering
	\vspace{-0.4cm}
	\subfloat[]{\includegraphics[width=0.48\linewidth]{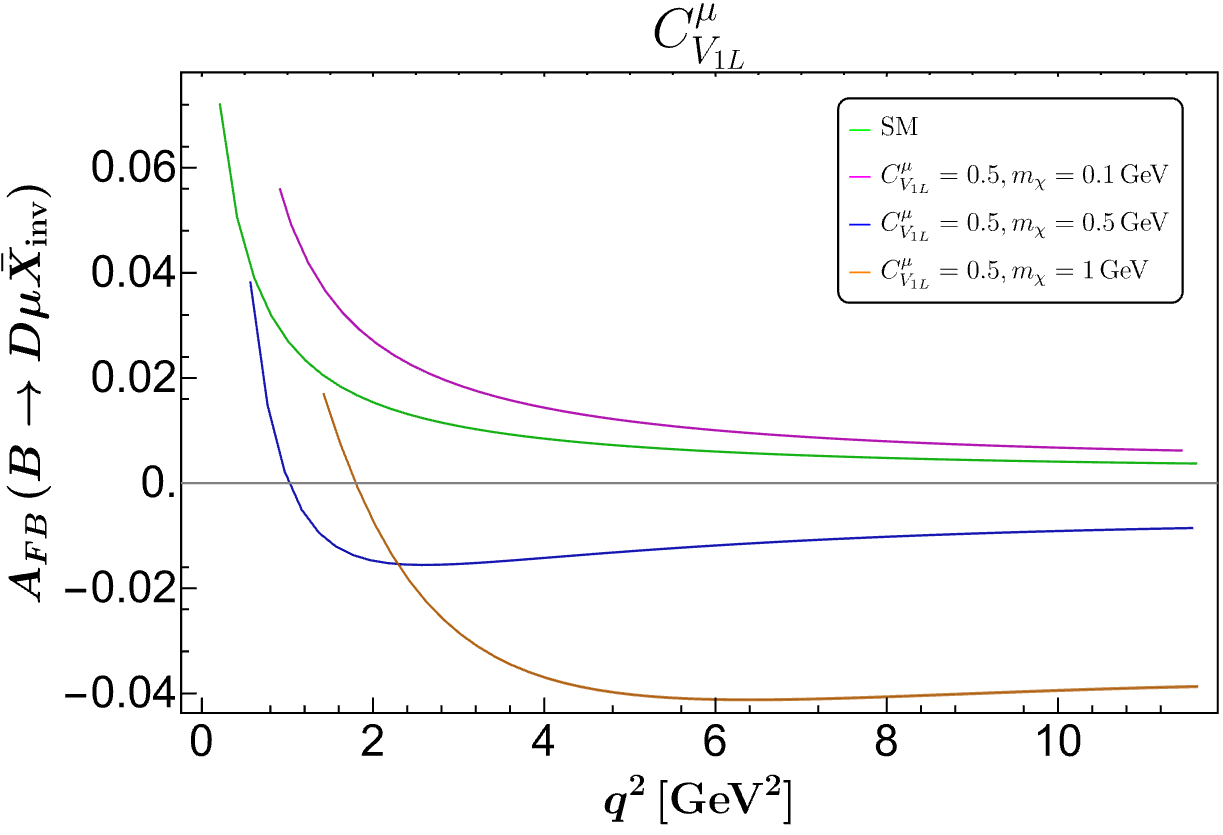}\label{fig:B2D_AFB1}}~~~~~~
	\subfloat[]{\includegraphics[width=0.48\linewidth]{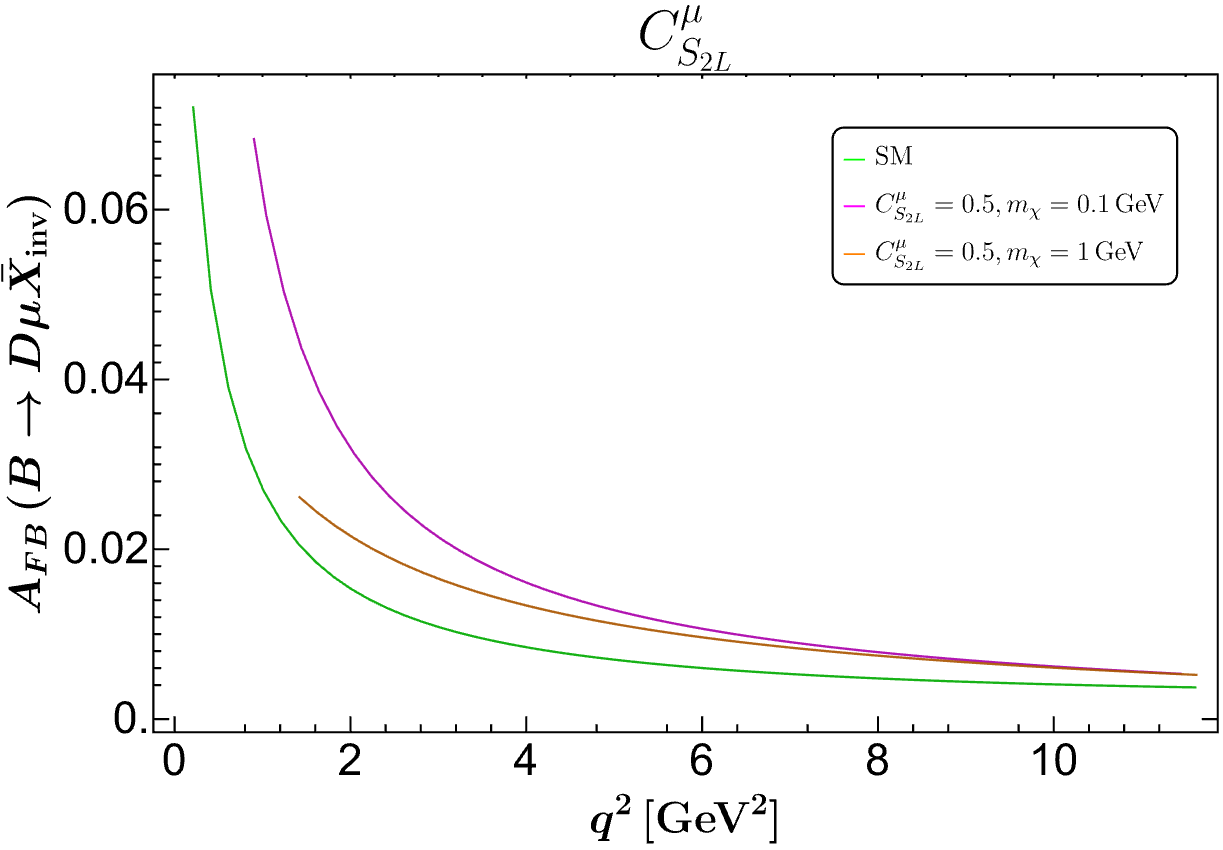}\label{fig:B2D_AFB2}}\vspace{-0.3cm}\hspace{0.0001cm}
	\subfloat[]{\includegraphics[width=0.48\linewidth]{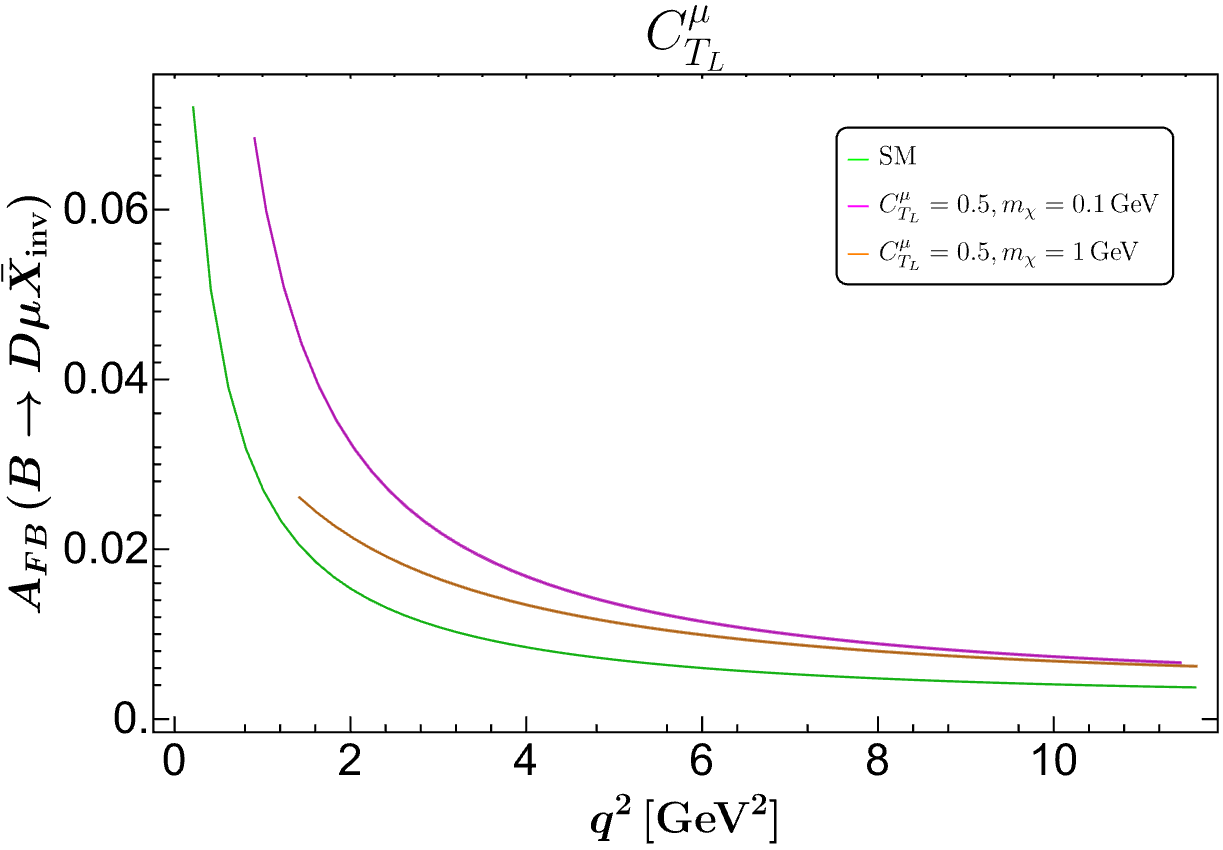}\label{fig:B2D_AFB3}}~~~~~~
	\subfloat[]{\includegraphics[width=0.48\linewidth]{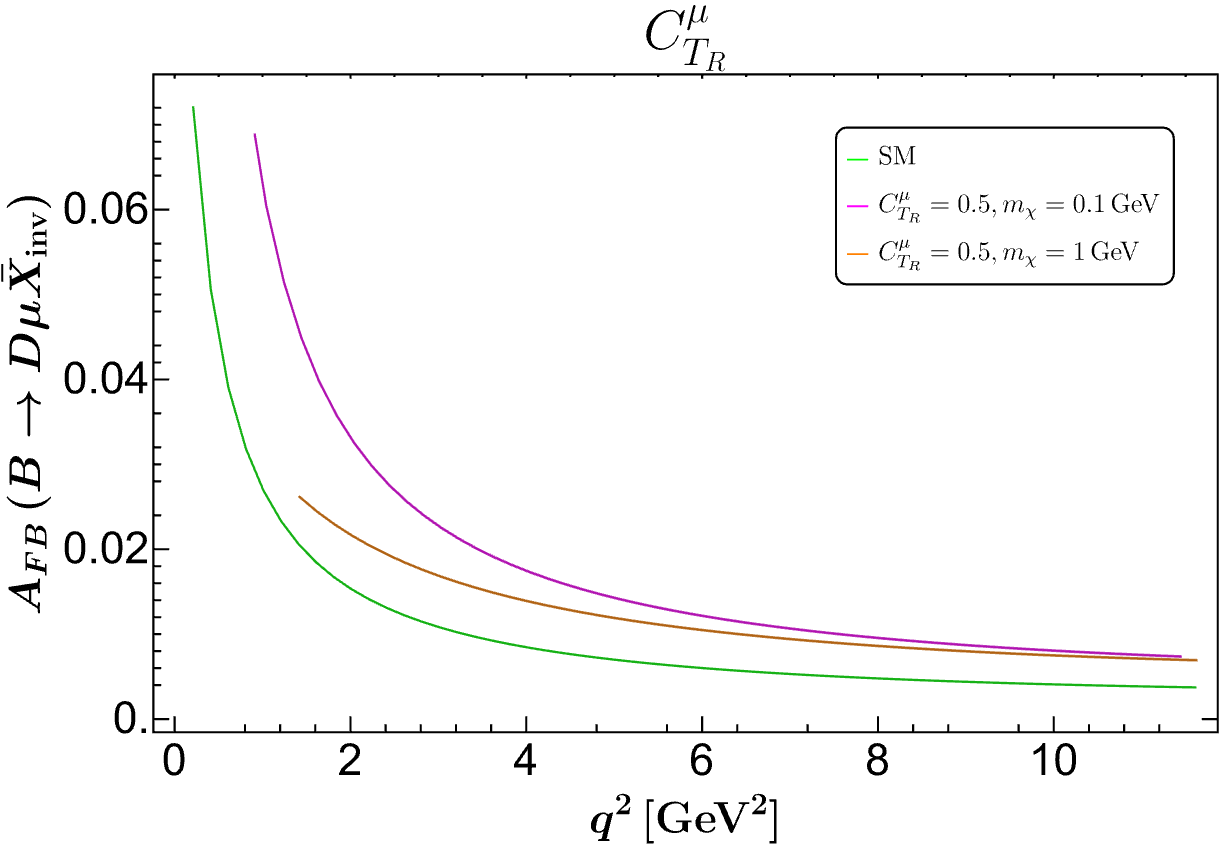}\label{fig:B2D_AFB4}}\vspace{-0.4cm}
	\caption{Variation of the angular observable $A_{FB}$ for the decay $B \to D \ell X_{\rm inv}$, for different masses of the fermion $\chi$ and different values of the WCs, as specified in the plots. The green colour corresponds to the SM distribution. } \label{fig:B2D_AFB}
\end{figure}

Fig.~\ref{fig:B2D_AFB} shows the $q^2$ dependence of the forward-backward asymmetry, $A^\ell_{FB}(q^2)$. In each panel, the green band represents the SM prediction. The uncertainties associated with these distributions are too small to be visible on the scale of the plots. Fig.~\ref{fig:B2D_AFB1} illustrates the effect of the left-handed vector operator with $C_{V_{1L}}^\mu = 0.5$ for three benchmark dark matter masses, $m_\chi = 0.1$, $0.5$, and $1$ GeV. In the SM, $A_{FB}$ remains positive throughout the entire kinematic range. In contrast, the NP contribution can significantly modify its behaviour. Although $m_{\chi}=0.1$ GeV keeps $A^\ell_{FB}(q^2)$ in the positive region whereas for $m_\chi = 0.5, 1$ GeV, a zero-crossing occurs near the low $q^2 $ region of the distribution. The remaining vector operators, $C_{V_{1R}}^\mu$, $C_{V_{2L}}^\mu$, and $C_{V_{2R}}^\mu$, exhibit qualitatively similar distributions. The other panels of fig.~\ref{fig:B2D_AFB} correspond to the WCs $C_{S_{2L}}^\mu$ and $C_{T_{L,R}}^\mu$, each taken to be $0.5$. In all these cases, $A^\ell_{FB}(q^2)$ remains positive throughout the allowed $q^2$ range. For scalar interactions, all four scalar WCs lead to identical distributions. In the case of tensor interactions, the overall $q^2$ dependence is similar for both tensor operators, although the numerical values differ slightly. While the scalar and tensor operators cause similar distributions, the vector operators lead to distinctly different behaviour, allowing them to be clearly distinguished from the scalar and tensor scenarios. 
For a fixed value of the WC, $A^\ell_{FB}(q^2)$ exhibits a clear dependence on the dark fermion mass, allowing the different benchmark values of $m_\chi$ to be distinguished in all the scenarios considered. This represents an improvement over the differential decay rate distribution of $B \to D \ell X_{\rm inv}$, where the vector and tensor operator scenarios show little sensitivity to $m_\chi$. Thus, $A^\ell_{FB}(q^2)$ emerges as a sensitive probe of the presence of the dark fermion for all operator scenarios discussed here in this analysis.

\begin{figure}[t]
	\centering
	\vspace{-0.4cm}
	\subfloat[]{\includegraphics[width=0.48\linewidth]{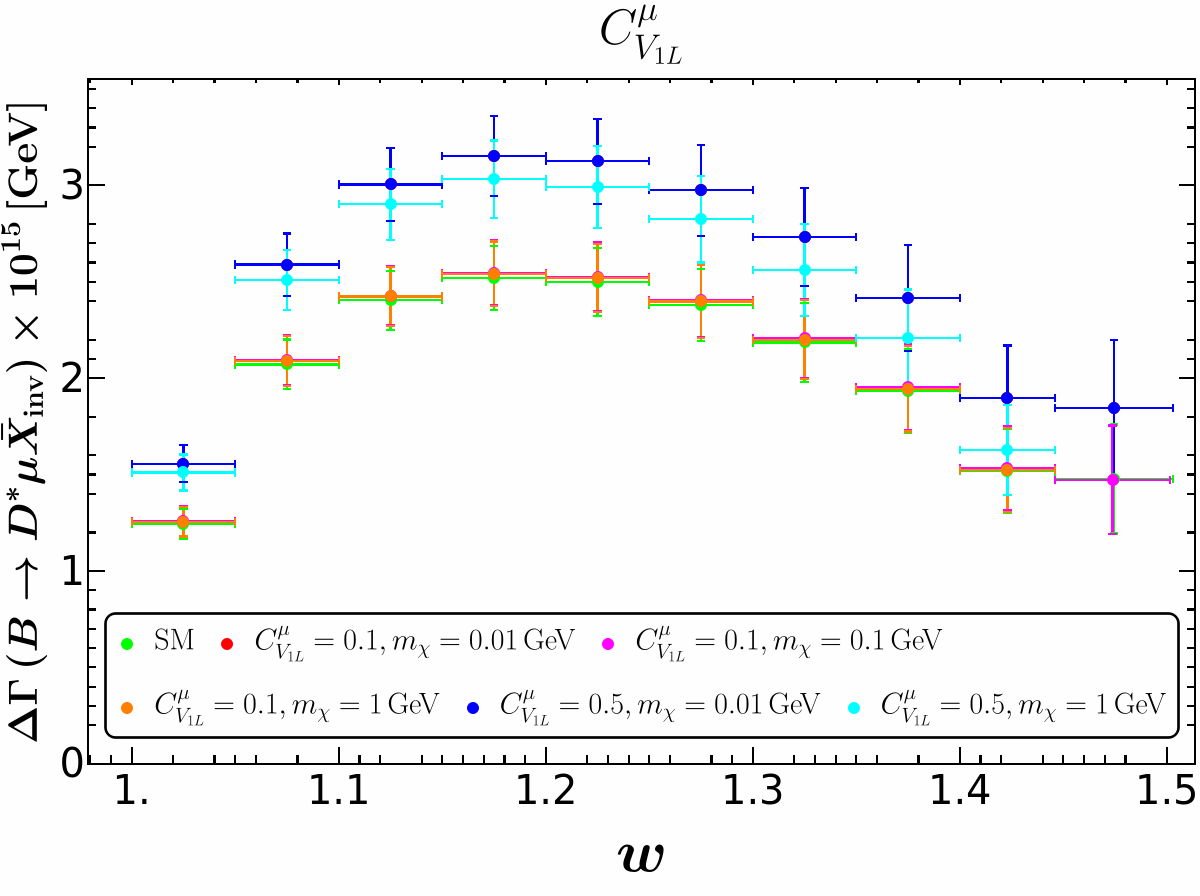}}\hspace{0.0001cm}
	\subfloat[]{\includegraphics[width=0.48\linewidth]{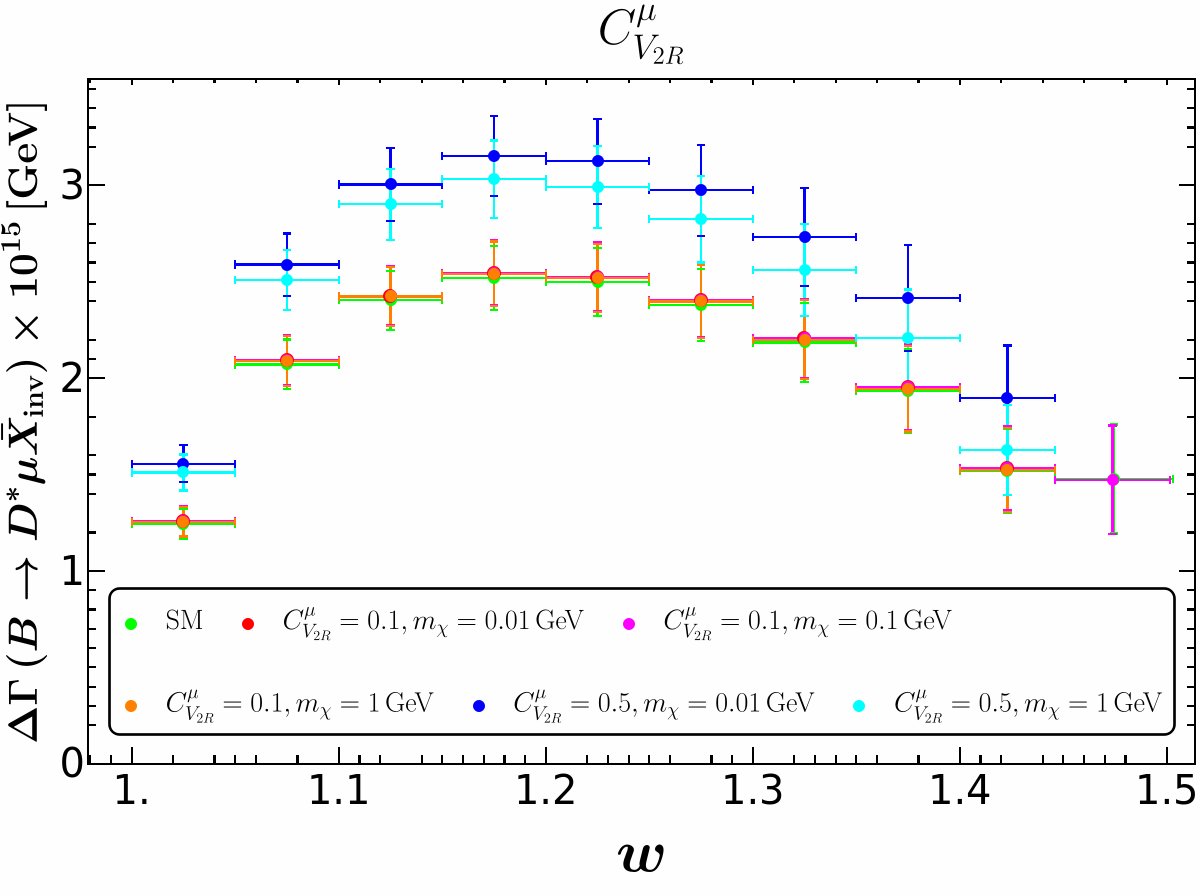}}\vspace{-0.4cm}\hspace{0.0001cm} \\
	\subfloat[]{\includegraphics[width=0.48\linewidth]{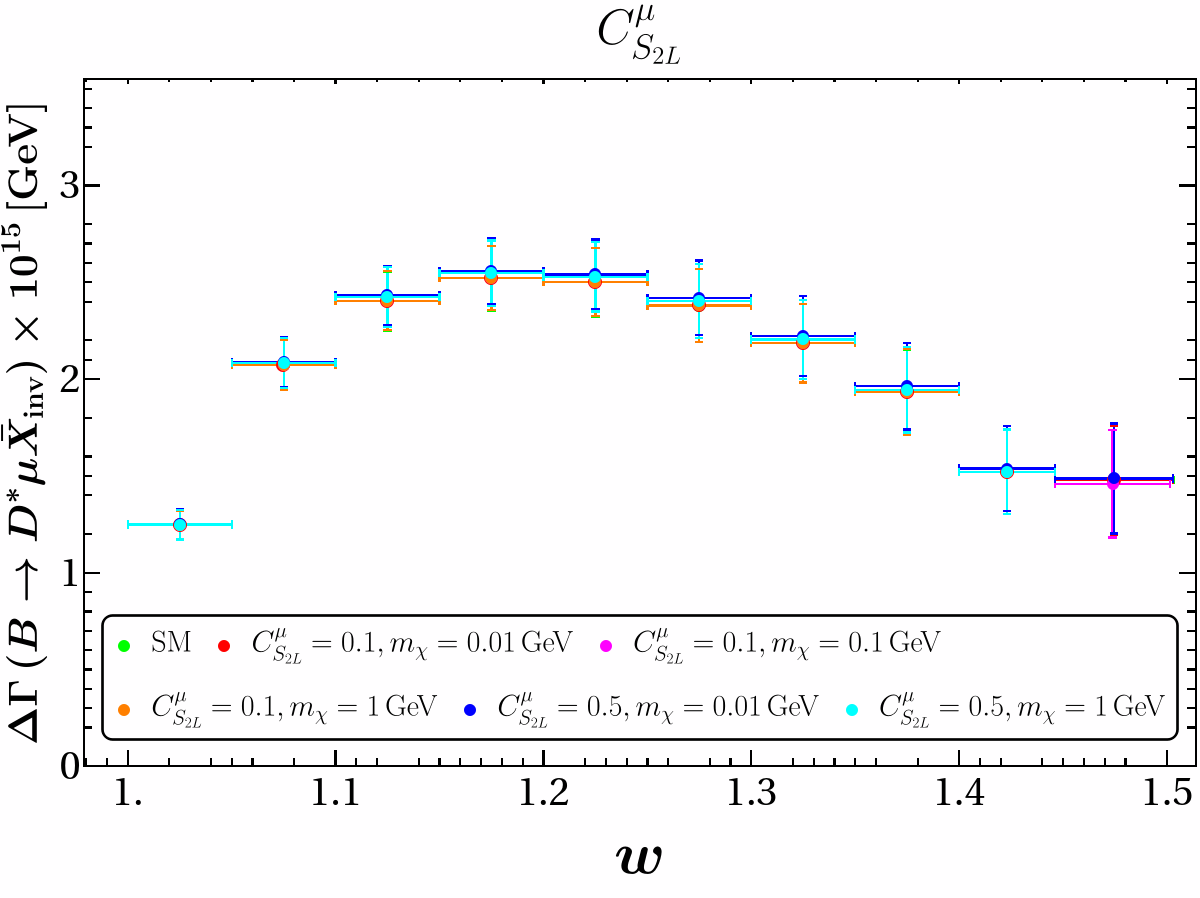}}\hspace{0.0001cm} 
	\subfloat[]{\includegraphics[width=0.48\linewidth]{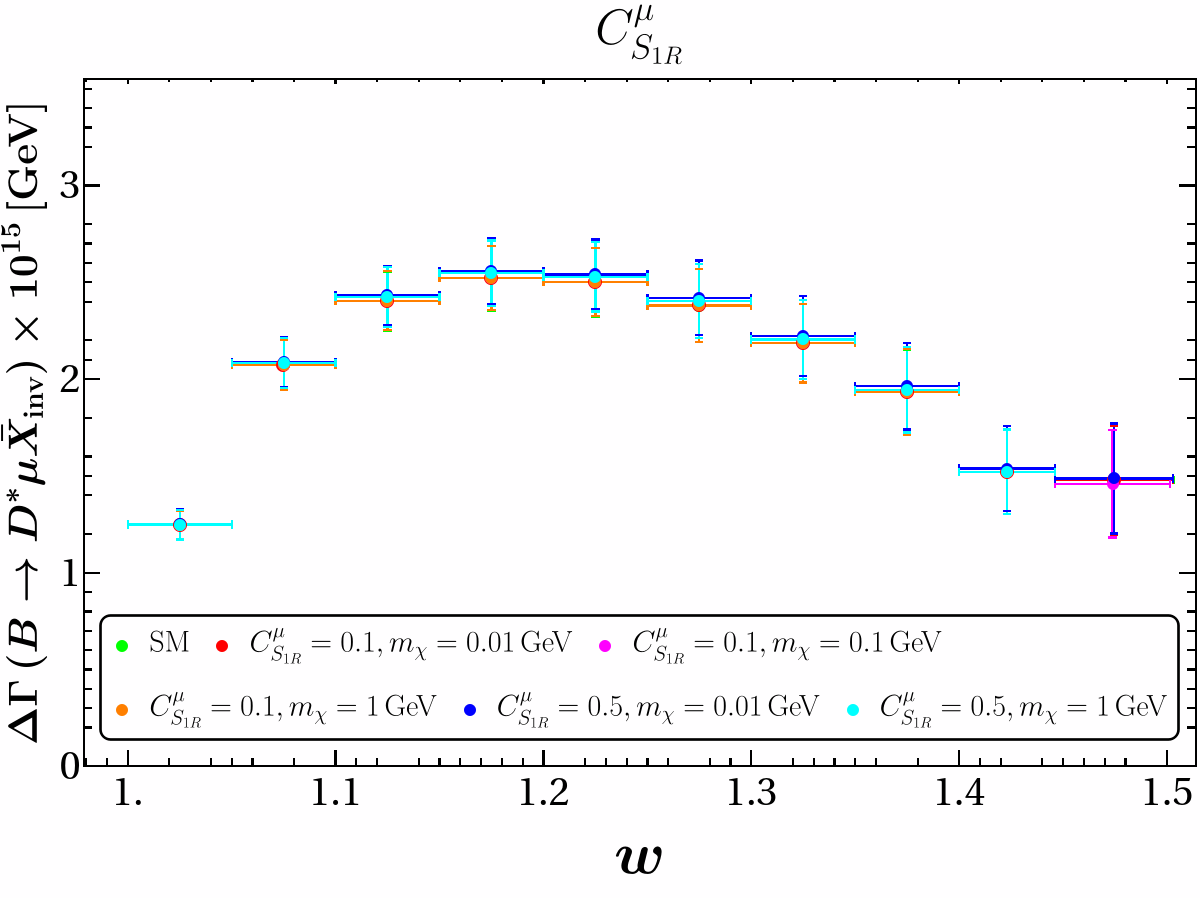}}\hspace{0.0001cm}  \\
	\subfloat[]{\includegraphics[width=0.48\linewidth]{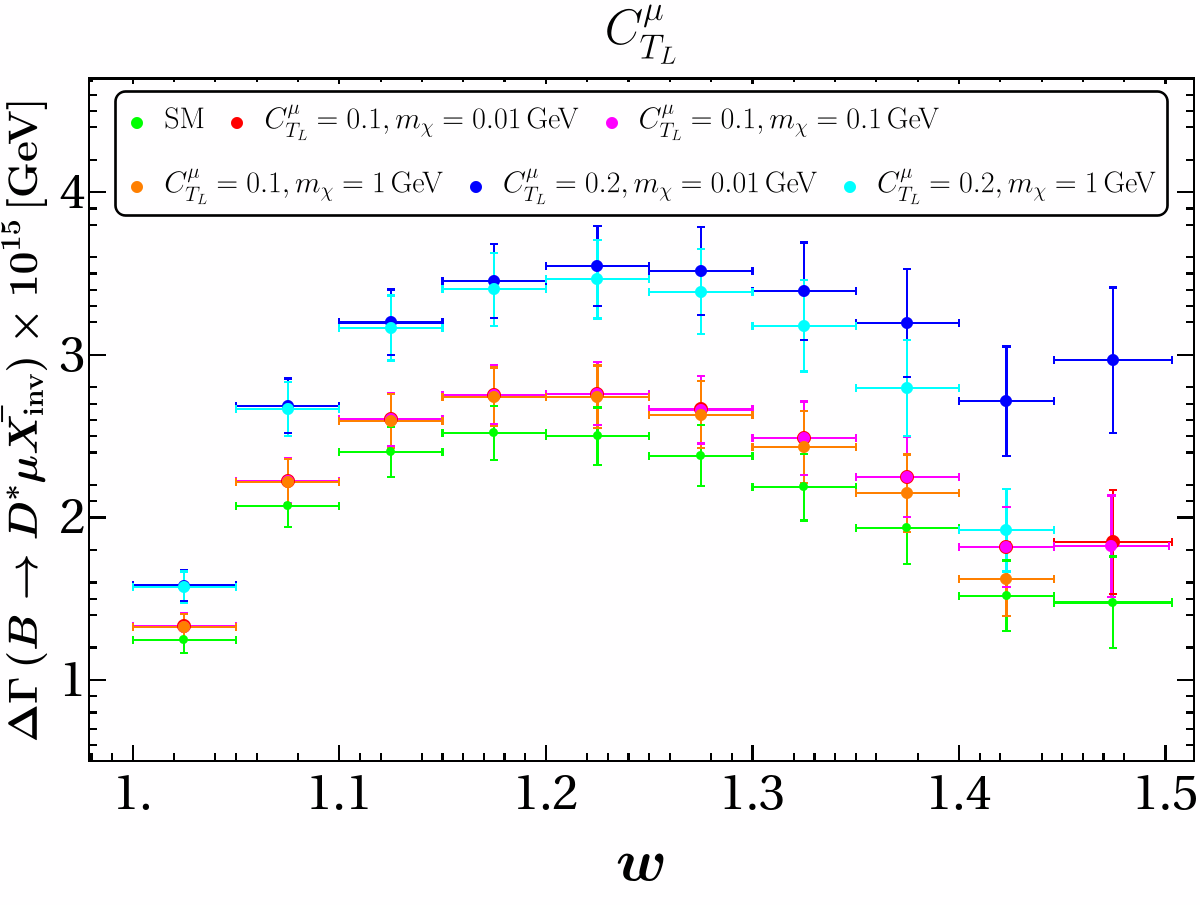}}\hspace{0.1cm}
	\subfloat[]{\includegraphics[width=0.48\linewidth]{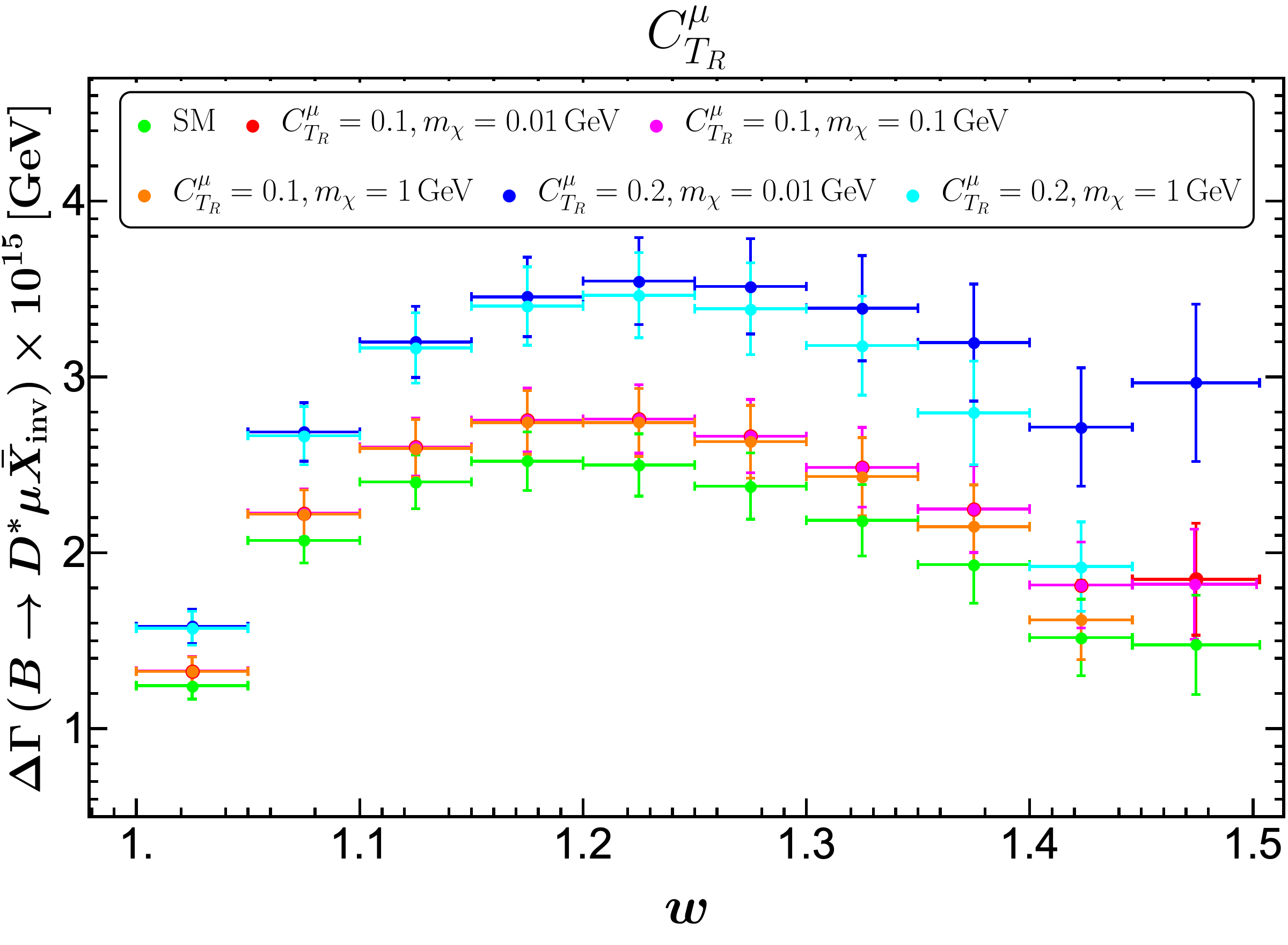}}\vspace{-0.2cm}
	\caption{Variation of $\Delta \Gamma $, for the decay $B \to D^{*} \, \ell \, X_{\rm inv}$ integrated over different bins of $w$ for different NP scenarios. We consider two BPs corresponding to each WCs, where $m_{\chi}$ is varied for three different masses: $(0.01, 0.1, 1)~{\rm GeV}$. In each plot, the green points represent the SM distribution. } \label{fig:B2Dst_unnorm_omega_SM_NP}
\end{figure}

\begin{figure}[t]
	\centering
	\vspace{-0.4cm}
	\subfloat[]{\includegraphics[width=0.48\linewidth]{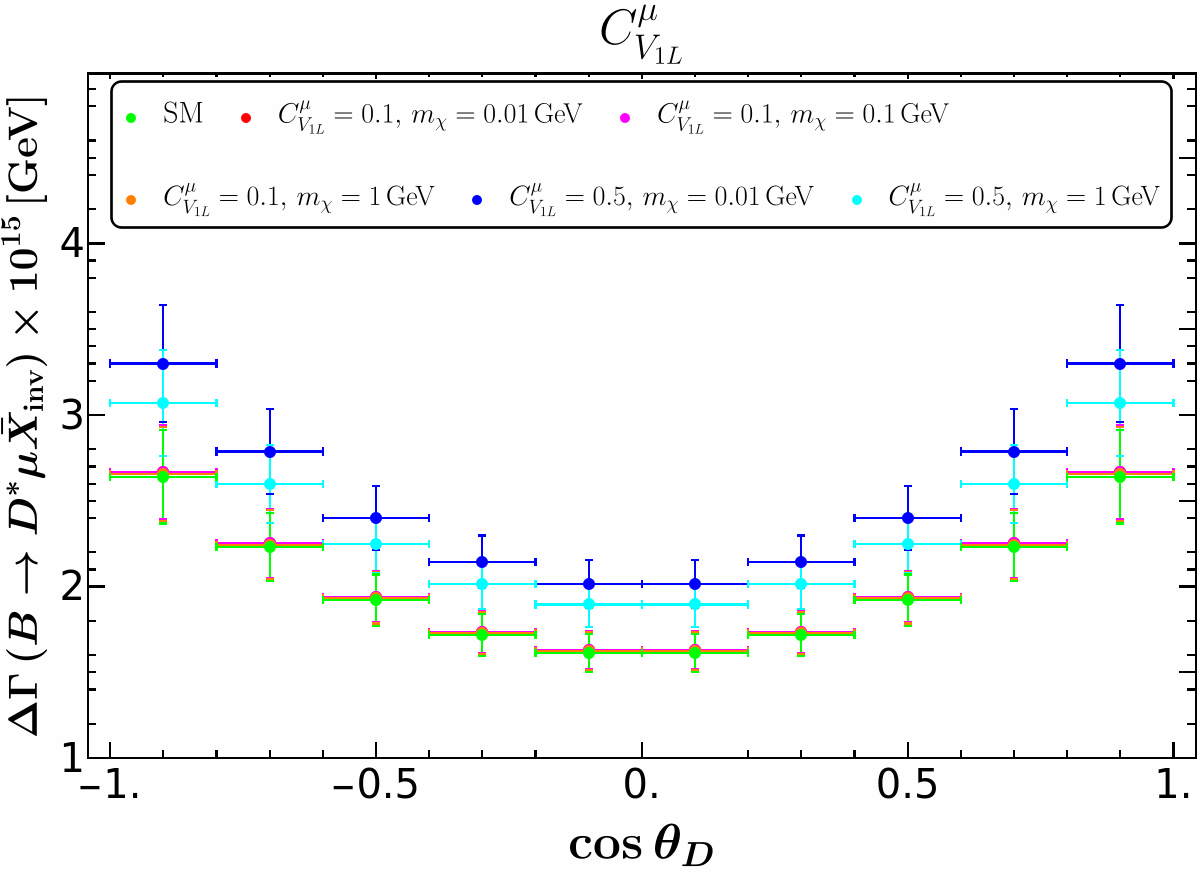}}~~~~~~
	\subfloat[]{\includegraphics[width=0.48\linewidth]{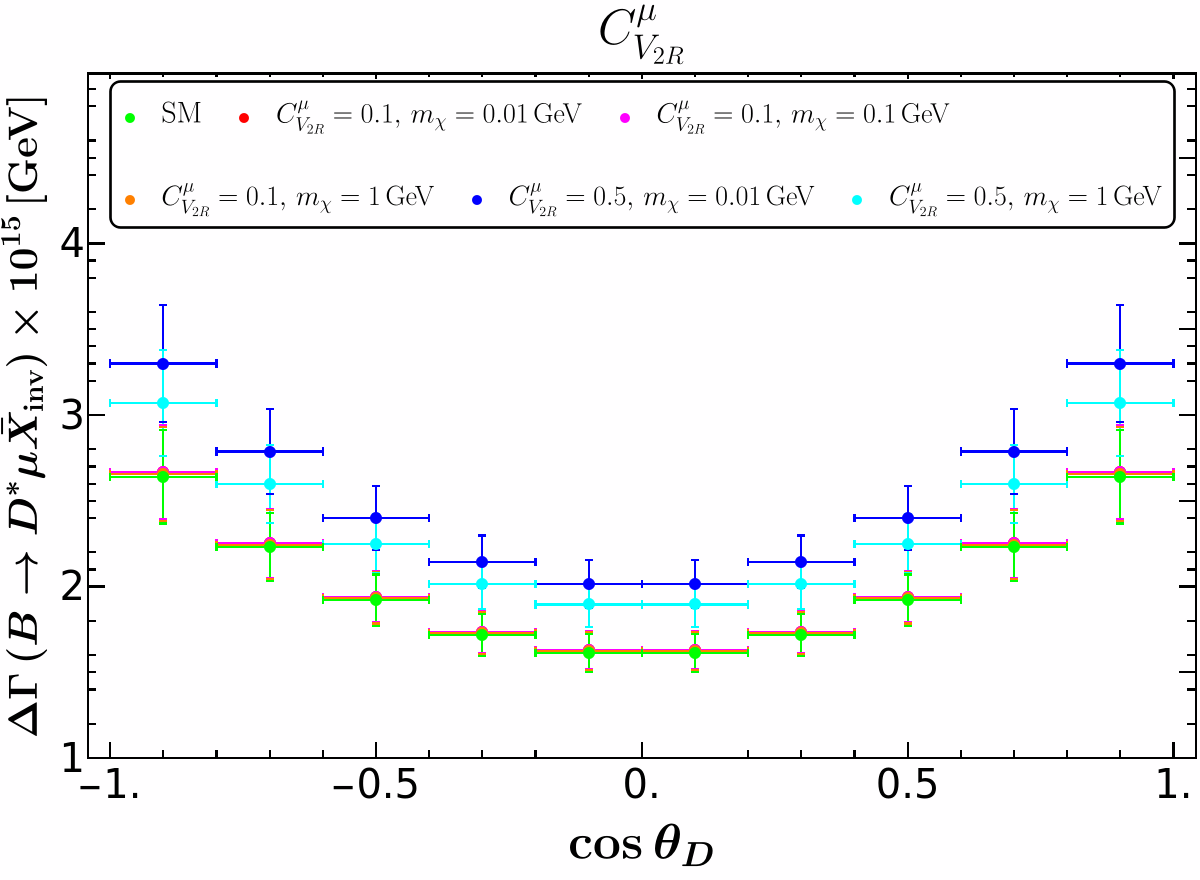}}\vspace{-0.3cm}\hspace{0.0001cm}
	\subfloat[]{\includegraphics[width=0.48\linewidth]{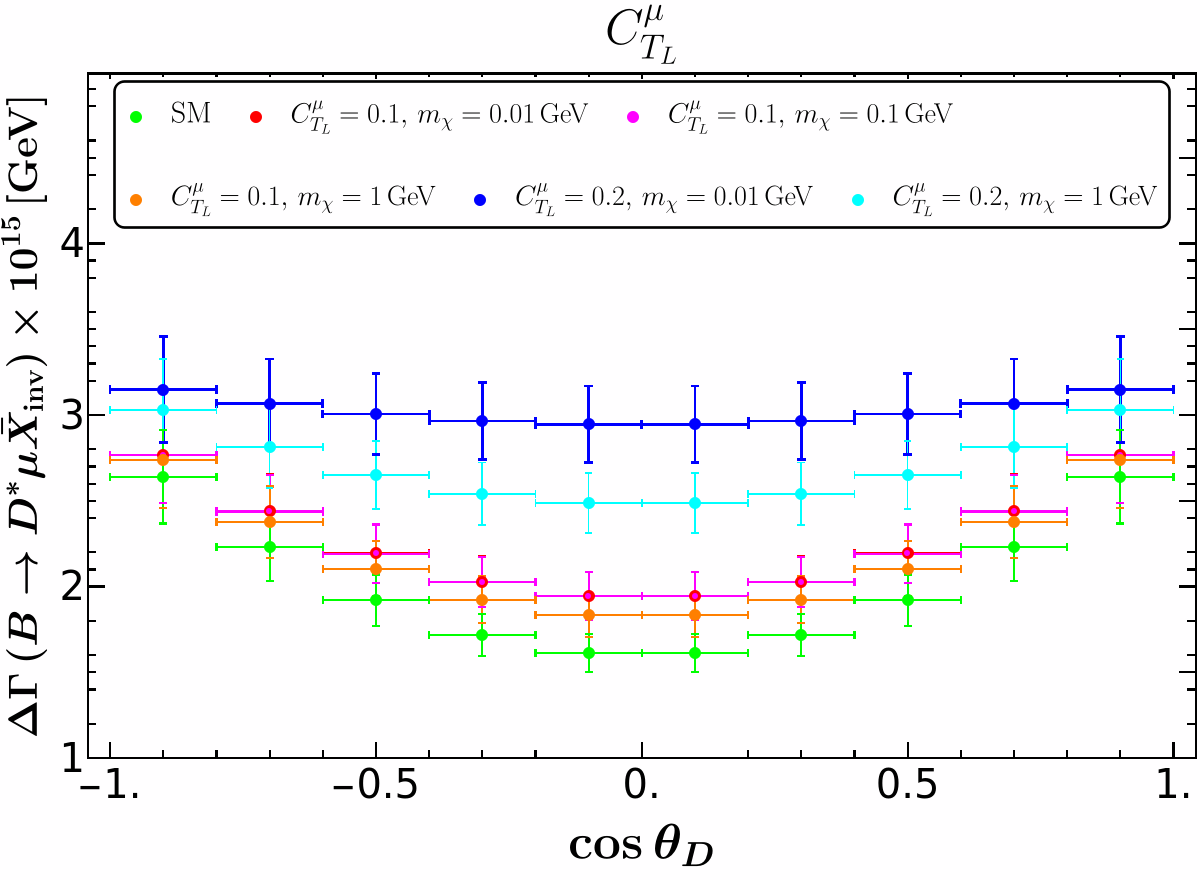}}~~~~~~
	\subfloat[]{\includegraphics[width=0.48\linewidth]{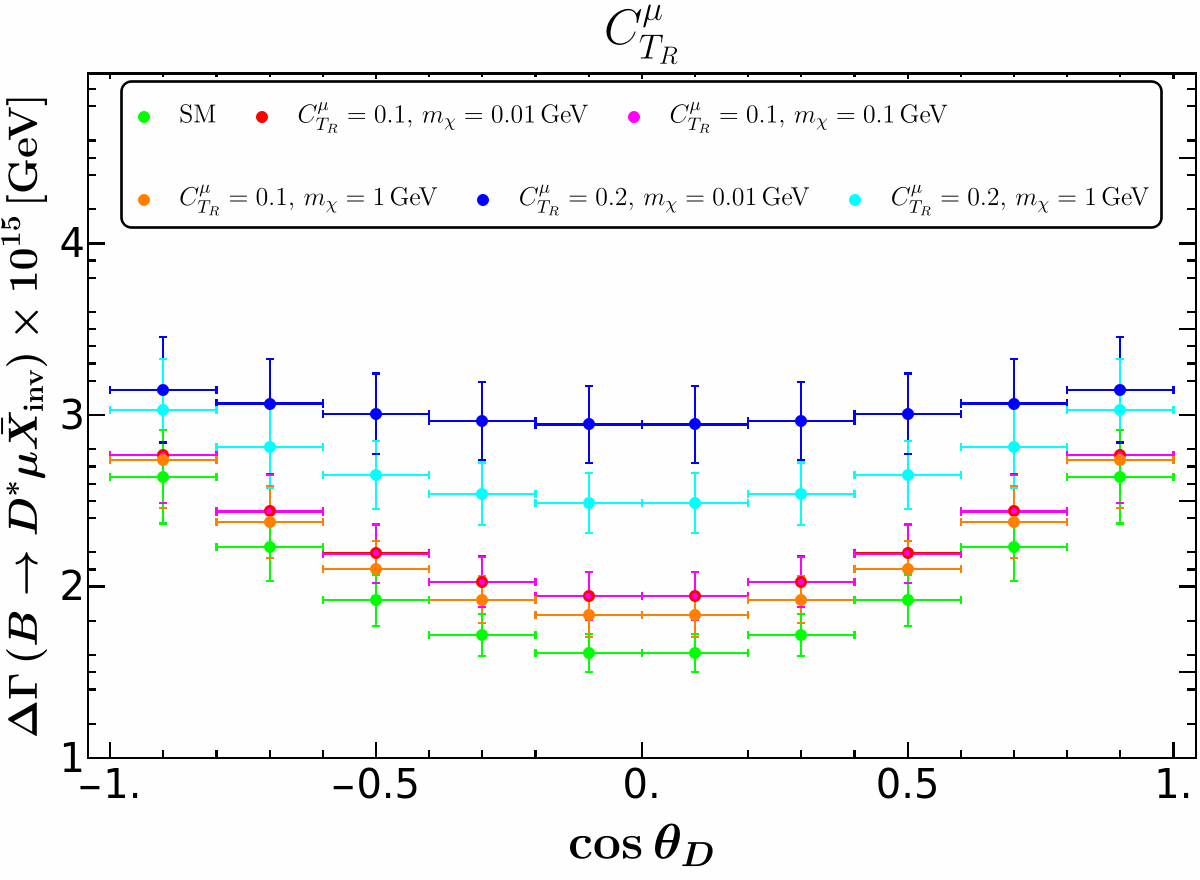}}\vspace{-0.4cm}
	\caption{Variation of $ \Delta \Gamma$, for the decay $B \to D^{*} (D \pi) \mu X_{\rm inv}$ integrated in different bins of $\cos \theta_{D}$, for SM and different NP scenarios. Each NP plot is done for $ m_{\chi} = (0.01\,, 0.1\,, 1)~\rm GeV $ for each NP operators. The green points in each plot show the SM distribution.} \label{fig:B2Dst_unnorm_cosD_SM_NP}
\end{figure}

\subsubsection{Decay Distribution in $B\to D^*\ell^-X_{\rm inv}$:}

In this subsection, we will discuss and compare our estimates of the decay rate distributions of $B \to D^{*} \ell^-\bar{\nu}$ decays and $B \to D^{*} \ell^-\bar{X}_{\rm inv}$ decays. Unlike the previous case, here, in addition, we will have the angular distribution $\frac{d\Gamma}{d\cos\theta_{D}}, \frac{d\Gamma}{d\cos\theta_{\ell}}, \frac{d\Gamma}{d\phi}$ along with the kinetic distribution $\frac{d \Gamma }{dw}$. Here, we will discuss the SM distribution of $\Delta\Gamma(B\to D^*\ell^-\bar{\nu})$ with $\Delta\Gamma(B\to D^*\ell^-\bar{X}_{\rm inv})$ (eq.~\eqref{eq:total_decay_width}) for different NP benchmark scenarios. We obtain $\Delta\Gamma$ from eq.~\eqref{eq:Obs_Belle_normalised} from the $q^2$ and other different angular distributions. For the decay of $B\to D^{*} \mu \nu$, in the SM, the available bin-length for $w$ is $ [1.00-1.51]$. In the presence of $\chi$, 
the lower limit of the recoil parameter $w$ will be modified from $m_{\ell}^2 \to (m_{\ell}+m_{\chi})^2$. Like the previous case, we have studied the NP scenarios for three different masses of the heavy fermion, i.e., $m_{\chi} = (0.01, 0.1, 1)$ GeV. For each case, available bins in $w$ will be different and given below:
\begin{subequations}\label{eq:B2Dst_allowed_omega}
	\begin{align}
		&m_{\chi} = 0.01 \, \text{GeV}, \ \ \ \ w : [1.00-1.503] \,, \\
		&m_{\chi} = 0.1 \,\text{GeV}, \ \ \ \ w : [1.00-1.501] \,,  \\
		&m_{\chi} = 1 \,\text{GeV}, \ \ \ \ w : [1.00-1.446] \,.
	\end{align}
\end{subequations}
Our goal is to look for the possible shift in the distribution $\Delta\Gamma(B\to D^*\ell^-\bar{X}_{\rm inv})$ from the respective SM predictions in different NP benchmark scenarios. In the following, we will present the main results obtained from the different kinematic and angular distributions.   

\begin{figure}[t]
	\centering
	\subfloat[]{\includegraphics[width=0.48\linewidth]{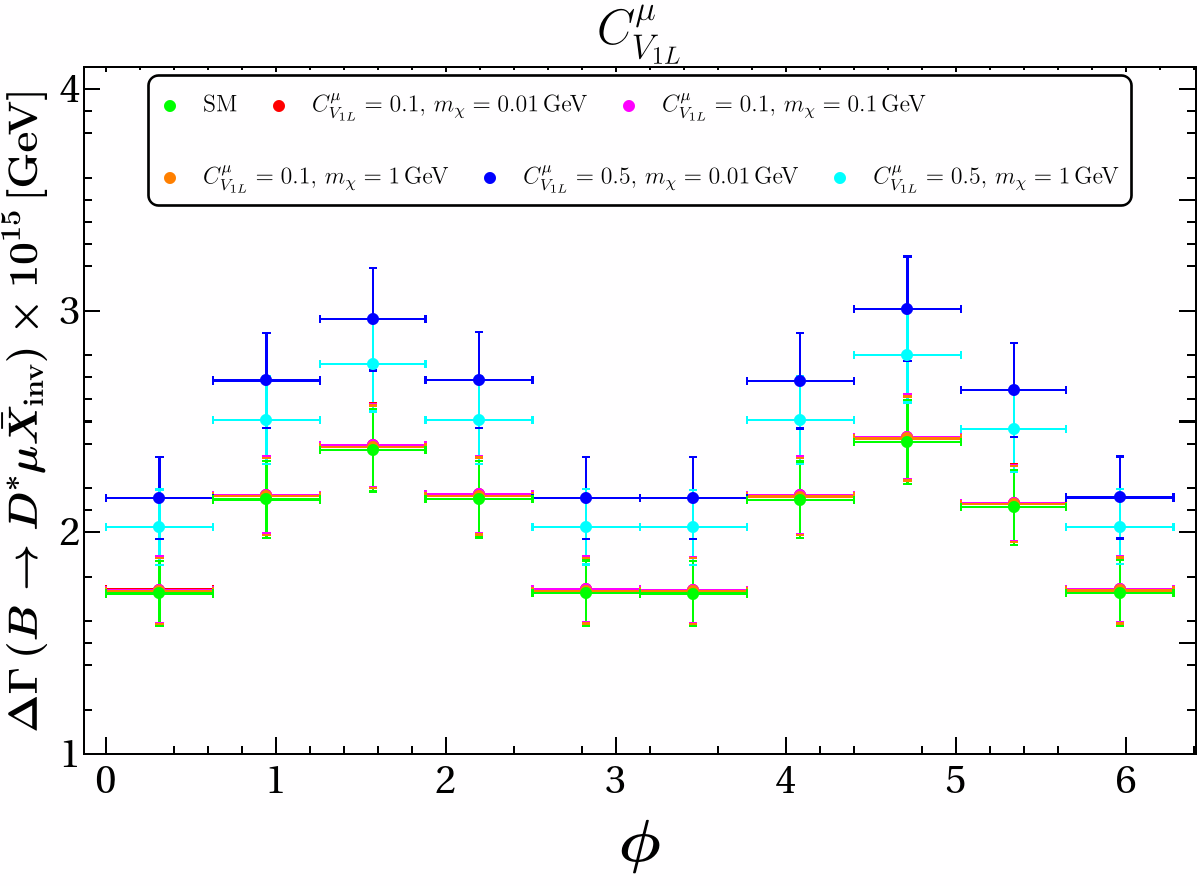}}~~~~~~
	\subfloat[]{\includegraphics[width=0.48\linewidth]{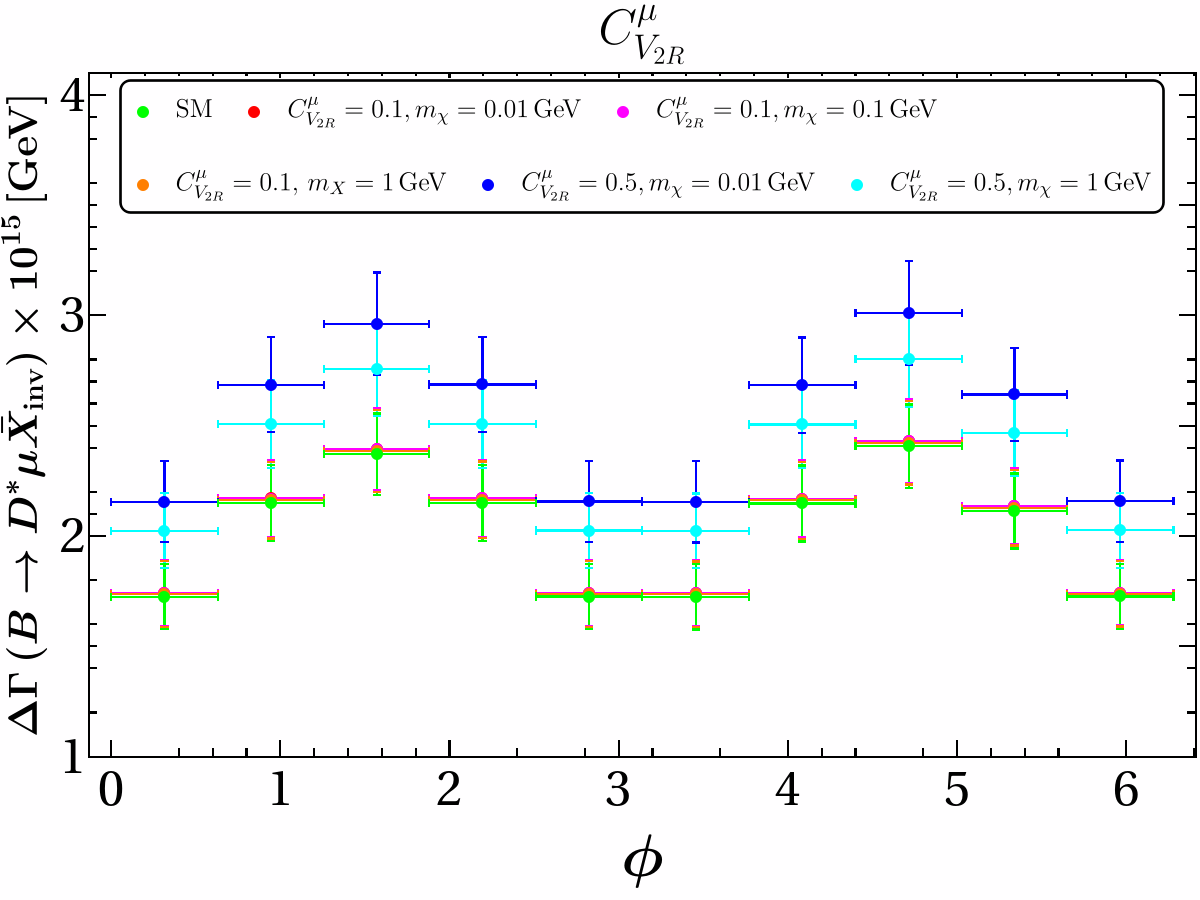}}\vspace{-0.4cm}\hspace{0.0001cm}
	\subfloat[]{\includegraphics[width=0.48\linewidth]{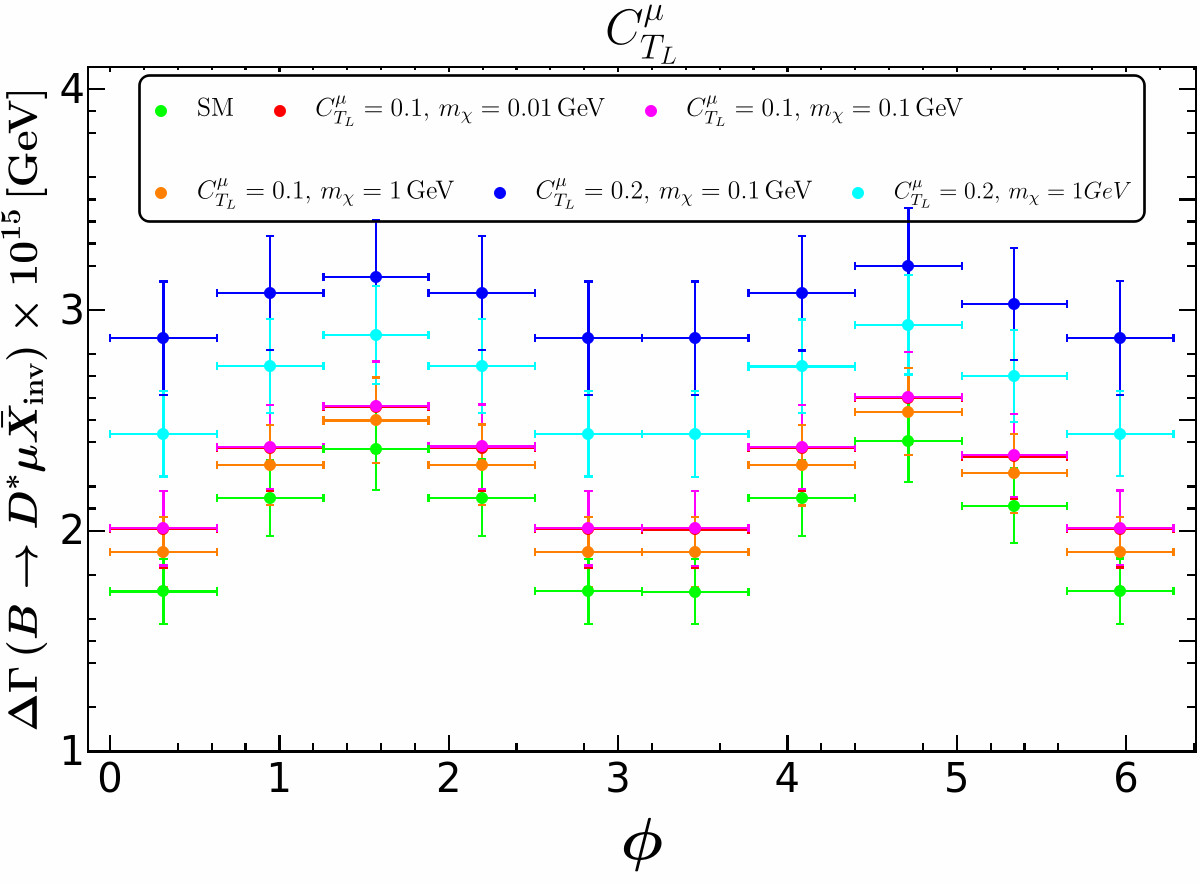}}~~~~~~~
	\subfloat[]{\includegraphics[width=0.48\linewidth]{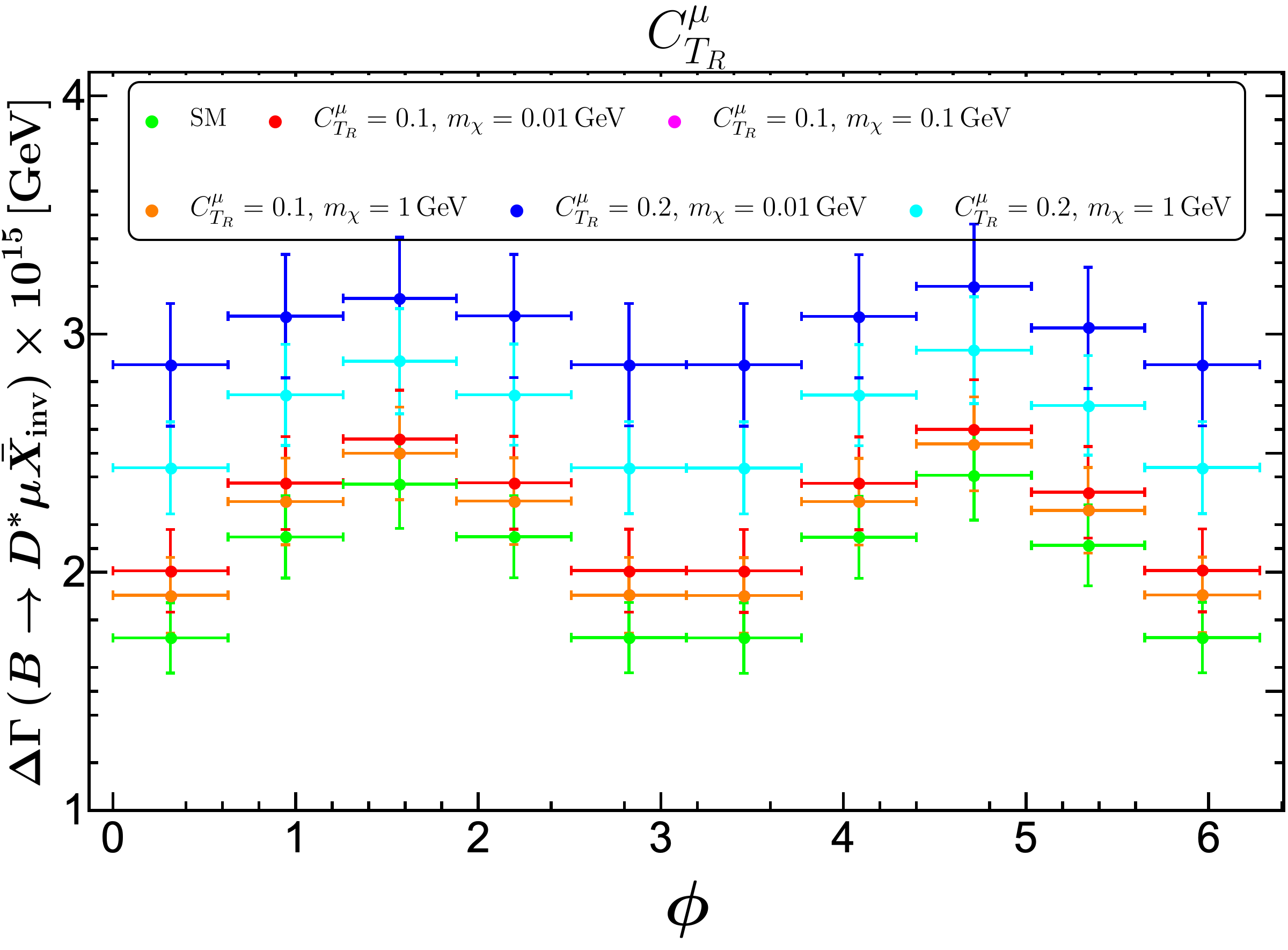}}\vspace{-0.4cm}
	\caption{Variation of $ \Delta \Gamma $ integrated in different bins of $\chi $, for SM and different NP operator scenarios. For two BPs of the WCs, the distributions are shown for the DM mass $ m_{\chi} = (0.01, \, 0.1, \text{ and } 1)$ GeV. The green points in each plot show the SM distribution. } \label{fig:B2Dst_unnorm_chi_SM_NP}
\end{figure}
\begin{figure}[t]
	\centering
	\vspace{-0.4cm}
	\subfloat[]{\includegraphics[width=0.48\linewidth]{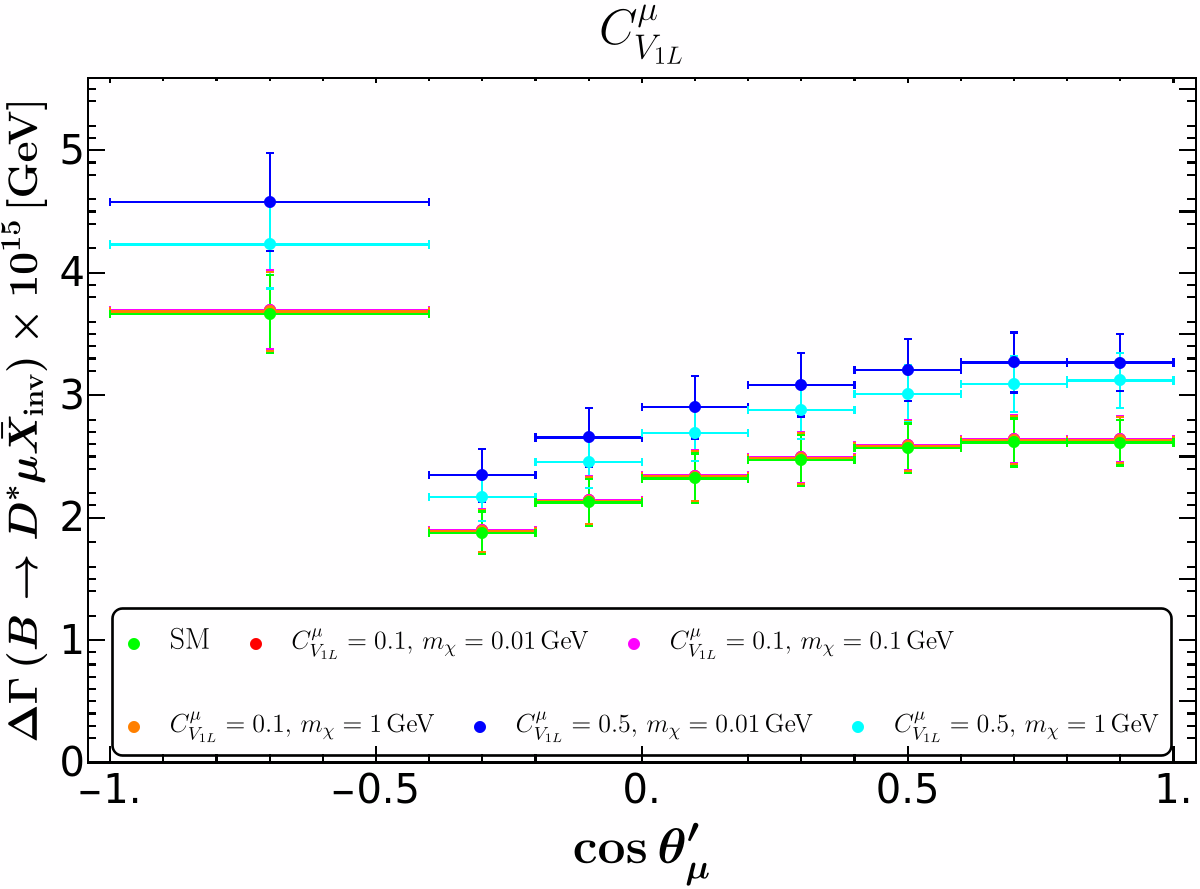}}~~~~~~
	\subfloat[]{\includegraphics[width=0.48\linewidth]{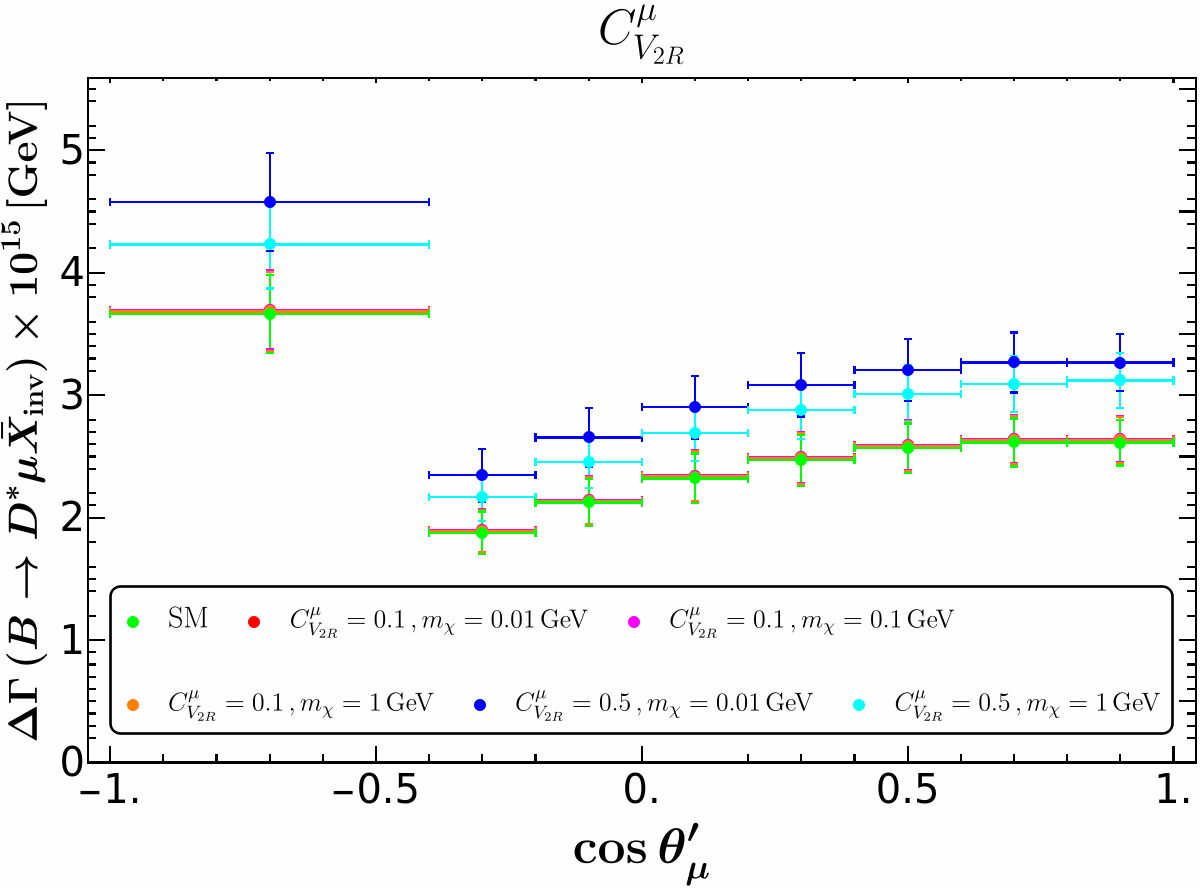}}\vspace{-0.4cm}\hspace{0.0001cm}
	\subfloat[]{\includegraphics[width=0.48\linewidth]{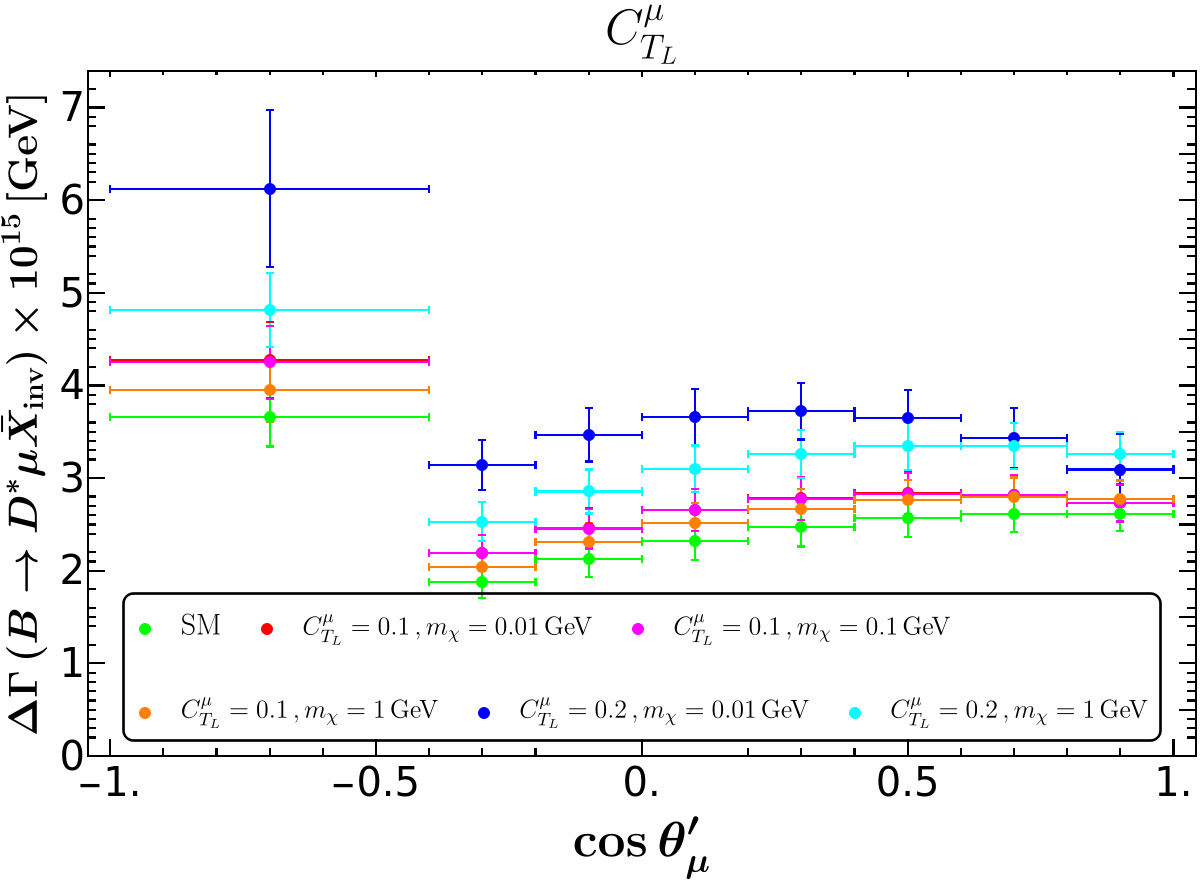}}~~~~~~
	\subfloat[]{\includegraphics[width=0.48\linewidth]{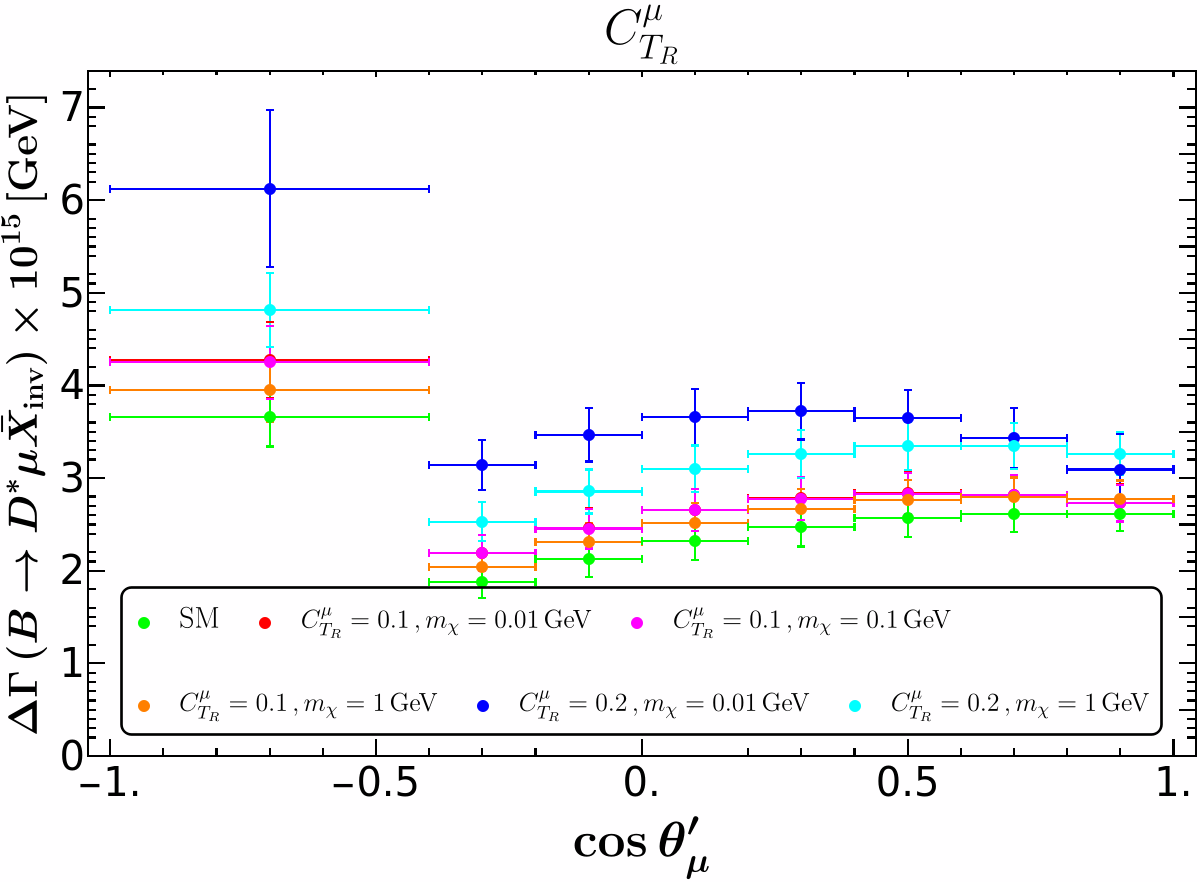}}\vspace{-0.4cm}
	\caption{Variation of $ \Delta \Gamma$ integrated in different bins of the azimuthal angle $\phi $, for different NP operator scenarios. The plots are shown for vector and tensor operators for two BPs of the WC values and for the DM masses $m_{\chi} = (0.01, \, 0.1, \, 1) $ GeV. The green points in each plot are for the SM distribution.} \label{fig:B2Dst_unnorm_cosl_SM_NP}
\end{figure}
\paragraph{\underline{$ \Delta \Gamma$ - integrated in bins of $w$}:} In fig.~\ref{fig:B2Dst_unnorm_omega_SM_NP}, we have shown the distributions of $\Delta\Gamma(B\to D^*\ell^-\bar{X}_{\rm inv})$ and $\Delta\Gamma(B\to D^*\ell^-\bar{\nu})$ with the $w$. Like before, we divide the allowed values of $w$ into small bins. We have shown the plots for two values of the WCs, and for each value of these WCs, we have chosen a few values of $m_{\chi}$. We have shown each bin's respective 1$\sigma$ error bars, which are $\approx 6\%$. The major contributions to these estimated errors come from the errors associated with the respective form factors and the CKM elements. Therefore, the estimated 1$\sigma$ errors are identical in the SM and the scenarios including NP effects. The analysis is performed for the WCs $\mathcal{O}_{V_{1L(R)}}^{\ell}$, $\mathcal{O}_{V_{2R(L))}}^{\ell}$, $\mathcal{O}_{S_{1R(L)}}^{\ell}$, $\mathcal{O}_{S_{2L(R)}}^{\ell}$ and $\mathcal{O}_{T_{L(R)}}^{\ell}$. We have considered the values of the respective WCs at 0.1 and 0.5 (and 0.2 for tensor operators). However, like the case in $B\to D$ decays, we have shown only the plots for the one operator scenarios $\mathcal{O}_{V_{1L}}^{\ell}$, $\mathcal{O}_{V_{2R}}^{\ell}$, $\mathcal{O}_{S_{1R}}^{\ell}$, $\mathcal{O}_{S_{2L}}^{\ell}$ and $\mathcal{O}_{T_{L(R)}}^{\ell}$. The reason for this is as we explained earlier.   

From fig.~\ref{fig:B2Dst_unnorm_omega_SM_NP} we note deviations in a few $w$-bins of the $\Delta\Gamma(B\to D^*\ell^-\bar{X}_{\rm inv})$ distributions as compared to the those in $\Delta\Gamma(B\to D^*\ell^-\bar{\nu})$ for the NP scenarios with operator $\mathcal{O}_{V_{1L(R)}}^{\ell}$,  $\mathcal{O}_{S_{2L(R)}}^{\ell}$ and $\mathcal{O}_{T_{R(L)}}^{\ell}$. 
Like before, we observe most deviations for the $C_{i_{1L(R)}}^{\ell} = 0.5$ and $m_{\chi}=0.01 $ and $1$ GeV. The deviations in the scenarios  $\mathcal{O}_{V_{1L(R)}}^{\ell}$ and $\mathcal{O}_{V_{2L(R)}}^{\ell}$ are larger for lower bins, about 2.5$\sigma$ level and becoming consistent with SM within $1 \sigma$ for the higher bins. 
However, for the scenarios $\mathcal{O}_{T_{R(L)}}^{\ell}$ we observe more deviations. For illustrative purposes, we present results for $C_{T_{L(R)}}^{\mu}=0.1$ and $0.2$. Unlike the other NP scenarios, tensor operators produce considerably larger effects in the observables. Therefore, smaller benchmark values are chosen to ensure a meaningful comparison of the distributions across different NP scenarios. For $m_{\chi} = 0.01$ and $C_{T_{L(R)}}^{\mu} = 0.2$ we see deviation around $\sim 3 \sigma $ and for $C_{T_{L(R)}}^{\mu} = 0.5$ these deviations are larger than 5$\sigma$. 
Unlike the $\Delta\Gamma(B\to D\ell^-\bar{X}_{\rm inv})$, we do not observe any significant deviations in the scenarios $\mathcal{O}_{S_{1R(L)}}^{\ell}$ and $\mathcal{O}_{S_{2L(R)}}^{\ell}$ even for very high WCs.     


\paragraph{\underline{$ \Delta \Gamma  $- integrated in bins of $\cos \theta_{D}$}: }
The plots in fig.~\ref{fig:B2Dst_unnorm_cosD_SM_NP} show the integrated differential decay width, $\Delta\Gamma(B\to D^*\ell^-\bar{X}_{\rm inv})$ and $\Delta\Gamma(B\to D^*\ell^-\bar{\nu})$, in bins of \( \cos \theta_{D} \). Here, we obtain $\Delta\Gamma$ from eq.~\eqref{eq:Obs_Belle_normalised} after integrating the respective $\frac{d\Gamma}{d\cos\theta_{D}}$ over $\cos\theta_{D}$ in a small interval defined by the size of the respective bins. The estimated 1$\sigma$ errors in both the SM and in the scenarios including NP are $\approx 8\%$.  

The overall shape of the distribution remains unchanged under NP effects. However, a shift from the SM prediction is observed, with its magnitude depending on the WC values. Stronger interaction strengths lead to larger deviations. As seen in previous distributions, more significant shifts occur for values of $ m_{\chi} = 0.01$ GeV, and for the values of the WCs around $0.5$. Consistent with the trend in \( d\Gamma \) integrated over \( w \), scalar operators induce only minor shifts in the distribution, and we have not shown those plots separately. 
The largest deviations are caused by the tensor operators (for $C_{T_{L(R)}}^{\mu} = 0.2$), around $(3-5)\sigma$ in the region around $\cos\theta_D  \sim 0$. In the remaining bins, the deviations are around 1.5$\sigma$. For $C_{T_{L(R)}}^{\mu} = 0.5$, the deviations are more than $5 \sigma$.
In the case of $\mathcal{O}_{V_{1L(R)}}^{\ell}$,  $\mathcal{O}_{V_{2L(R)}}^{\ell}$, across all the bins, we note $\gtrsim 1.5~\sigma$ deviations for $m_{\chi} = 0.01$ GeV and the values of the respective WCs $\mathcal{C}_{V_{j_{L(R)}}}^{\ell} \approx 0.5$ (with j = 1,2). 

\paragraph{\underline{$ \Delta \Gamma  $- integrated in bins of $\phi$}: }
The plots in fig.~\ref{fig:B2Dst_unnorm_chi_SM_NP} show the integrated differential decay width for $\Delta\Gamma(B\to D^*\ell^-\bar{X}_{\rm inv})$ decays, binned in the azimuthal angle \( \phi \), for both the SM and various NP scenarios with different values of the WCs and fermion masses \( m_{\chi} \). We obtain $\Delta\Gamma$ from eq.~\eqref{eq:Obs_Belle_normalised} after integrating the respective $\frac{d\Gamma}{d\phi}$ over $\phi$ in a small interval (with bin-length 0.63) defined by the size of the respective bins. The average estimated 1$\sigma$ errors in both the SM and in the scenarios including NP are $\approx 8\%$. The green error bar graphs represent the predictions in the SM for $\Delta\Gamma(B\to D^*\ell^-\bar{\nu})$. The remaining error bars correspond to different NP scenarios. The dark fermion mass is taken to be \( m_{\chi} = (0.01,\, 0.1,\, 1 )\)~GeV. Although the shape of the distribution with respect to the azimuthal angle \( \phi \) remains qualitatively similar, significant shifts in the decay width are observed for the vector and tensor operators. In contrast, we do not observe any deviations in the distributions for $\mathcal{O}_{S_{1R(L)}}^{\ell}$ and $\mathcal{O}_{S_{2L(R)}}^{\ell}$. Hence, we have not shown them separately. 

For the scenarios with $\mathcal{O}_{V_{1L(R)}}^{\ell}$,  $\mathcal{O}_{V_{2L(R)}}^{\ell}$, across all the bins, we note $\sim 2~\sigma$ deviations for $m_{\chi} = 0.01$ GeV and the values of the respective WCs $\mathcal{C}_{V_{j_{L(R)}}}^{\ell} \approx 0.5$ (with j = 1,2). For these operators, at the current level of precision, we could not get sizable deviations for smaller values of the WC, i.e., for $\mathcal{C}_{V_{j_{L(R)}}}^{\ell} \sim 0.1$. However, for the scenario with tensor operators $\mathcal{O}_{T_{L (R)}}^{\ell}$, the average deviations are $\approx 3~\sigma$ for $\mathcal{C}_{T_{L(R)}}^{\ell} \approx 0.2$ and $> 5 \sigma $ for $C_{T_{L(R)}}^{\mu} =0.5$.  These observations are valid for the \( m_{\chi} = (0.01,\, 0.1,\, 1) \)~GeV. However, the distribution deviates more from SM, for \( m_{\chi} = 0.01 \)~GeV. 
 
\begin{figure}[t]
	\centering
	\subfloat[]{\includegraphics[width=0.48\linewidth]{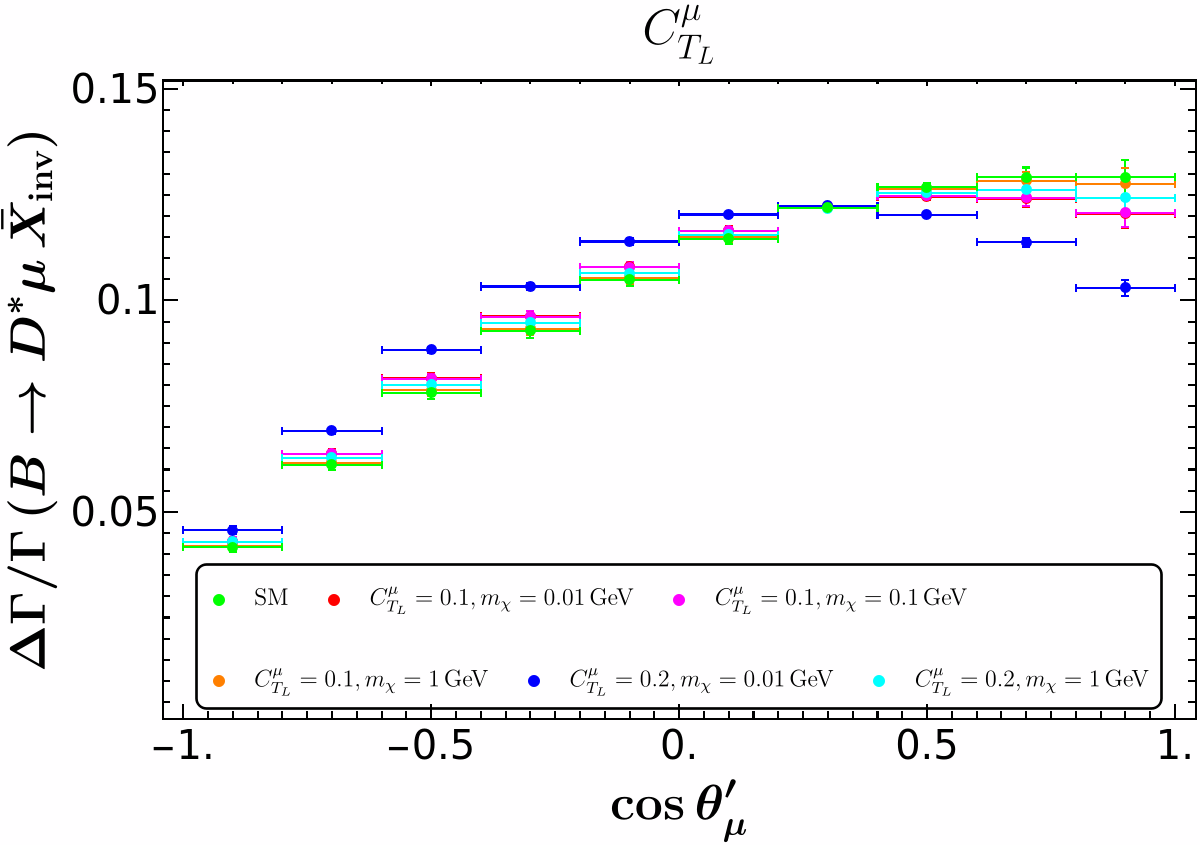}}\hspace{0.0001cm}\vspace{-0.3cm}
	\subfloat[]{\includegraphics[width=0.48\linewidth]{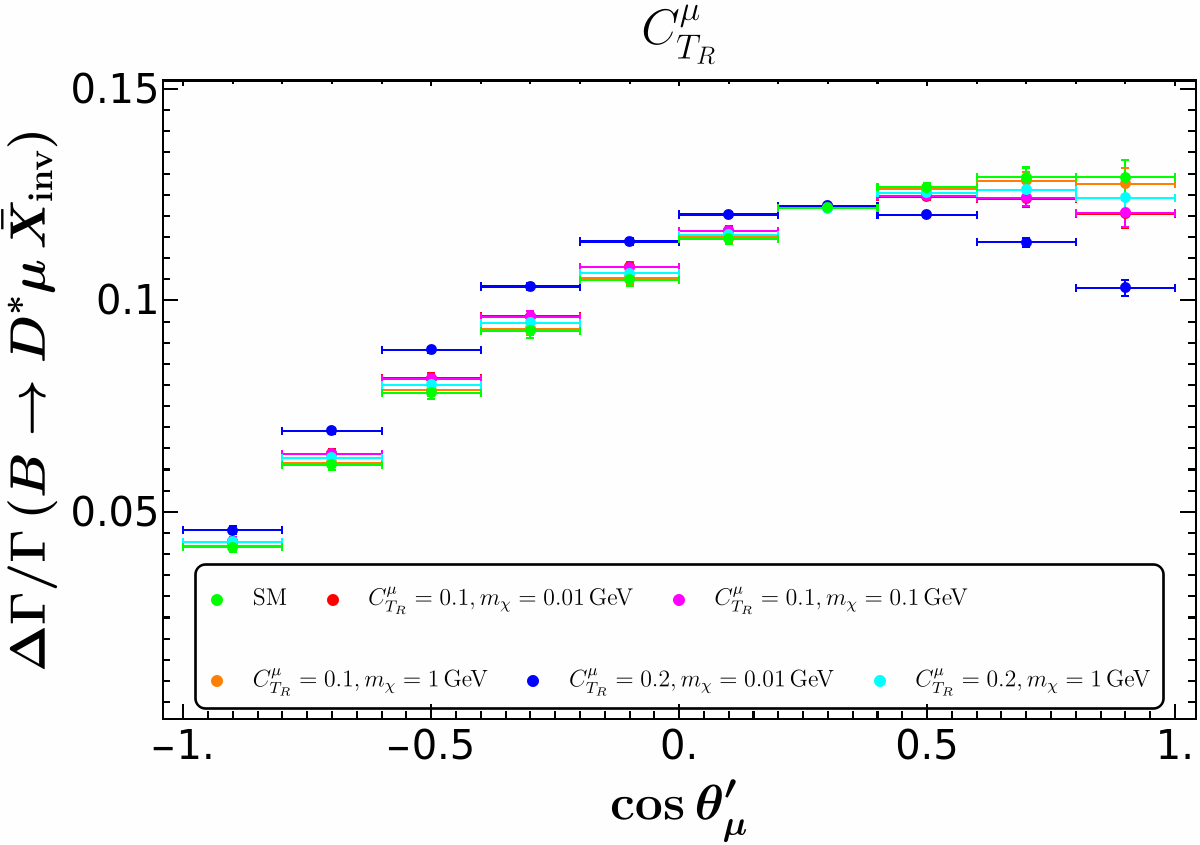}}
	\caption{Variation of $ \Delta \Gamma / \Gamma $ for the decay $B \to D^{*} (\to D \pi) \mu X_{\rm inv}$, integrated in different bins of $ \cos\theta_{\mu}^{\prime} $, for different NP operator scenarios. Each NP plot corresponds to two BPs of the WCs, and for DM mass $ m_{\chi} = (0.01, \, 0.1, \, 1)$ GeV for different NP operators. The green points in each plot correspond to the SM distributions. } \label{fig:B2Dst_normalised_cosl_SM_NP}
\end{figure}

\paragraph{\underline{$ \Delta \Gamma $- integrated in bins of $\cos \theta_{\ell}$}: }
In this paragraph, we discuss the distribution of the decay widths for $\Delta\Gamma(B\to D^*\ell^-\bar{X}_{\rm inv})$ for both SM ($X_{\rm inv} = \nu$) and NP ($X_{\rm inv} = \chi$), integrated in bins of the angle \( \cos \theta_{\ell}^{\prime} \). Note that in our analysis, the angle \( \theta_{\ell}^{\prime} \) is defined as the angle between the lepton and the direction of the \( B \) meson. In experimental analyses \cite{Belle:2023bwv}, however, \( \theta_{\ell} \) is defined as the angle between the lepton and the direction opposite to the \( B \) meson, which corresponds to \( \theta_{\ell}^{\prime} \) in our convention, as illustrated in fig.~\ref{fig:decay_plane}. The relation between these two angles is \( \theta_{\ell} + \theta_{\ell}^{\prime} = \pi \). Since we are comparing our analysis with experimental results, we present the distribution in terms of the angle \( \cos \theta_{\ell}^{\prime} \). 
Therefore, to ensure consistency with the experimental results, we present the distribution in terms of \( \theta_{\ell}^{\prime} \). 
We obtain $\Delta\Gamma$ from eq.~\eqref{eq:Obs_Belle_normalised} after integrating the respective $\frac{d\Gamma}{d\cos\theta_{\ell}^{\prime}}$ over $\cos\theta_{\ell}^{\prime}$ in a small interval defined by the size of the respective bins. The estimated 1$\sigma$ errors in both the SM and in the scenarios including NP are $\approx 8\%$. Fig.~\ref{fig:B2Dst_normalised_cosl_SM_NP} shows the variation of the decay width integrated in bins of \( \cos \theta_{\mu}^{\prime} \) for the above-mentioned decays with $\ell=\mu$.  
Unlike the other decay distributions, the bin widths here are not uniform: the first bin has a width of 0.4, while the remaining bins have widths of 0.2. This is in accordance with the bin sizes defined in the experimental analysis in ref. \cite{Belle:2023bwv}.   

Fig.~\ref{fig:B2Dst_unnorm_cosl_SM_NP} displays the distribution of $\Delta\Gamma$ with $\cos\theta_{\ell}$. The colour coding of the error bar graphs is the same as discussed in the above paragraphs. Also, here we have done the analysis for various NP scenarios corresponding to different benchmark combinations of WCs and the dark fermion mass \( m_{\chi} \). We consider \( m_{\chi} = (0.01,\, 0.1,\, 1) \)~GeV and \( C_{ij} = (0.1,\, 0.5) \) for vector and scalar operators, for tensor we take $C_{ij} = (0.1,\, 0.2) $. Also, in this case, we observe considerable deviations in a few bins for the value of the WCs $C_{i_{jL(R)}}^{\ell} = 0.5 $ (with j=1,2), whereas for tensor operator small deviations can be observed for $C_{T_{L(R)}}^{\ell} = 0.1$. For the vector operators, for $m_{\chi} = 0.01$ GeV and $C_{V_{i L(R)}} = 0.5$, we get $\sim 2 \sigma$ deviation from SM, whereas for $C_{V_{i L(R)}} = 0.1$, the values in each bins, remains consistent within $1 \sigma$. Similar to the other bin-integrated distributions, here also scalar operators do not show deviations with SM and remain consistent within $1 \sigma$. For the tensor operator, as can be seen from the figures, even for $C_{T_{L(R)}}^{\mu} = 0.1$, we get $\gtrsim 1 \sigma$ deviation for $m_{\chi} = 0.01$ GeV, which again decreases for higher mass values of the DM. In this case, for the bins within $-1 < \cos \theta_{\mu}^{\prime} < -0.5$, the deviations between the SM and the NP scenarios are comparatively large. For $C_{T_{L(R)}}^{\mu} = 0.2$, the most deviations are $\sim 3.5 \sigma$, which again increases for higher values of the WC. For $C_{T_{L(R)}}^{\mu} = 0.5$, the deviations are $> 5 \sigma$. 

\begin{figure}[t]
	\centering
	\vspace{-0.5cm}
	\subfloat[]{\includegraphics[scale=0.175]{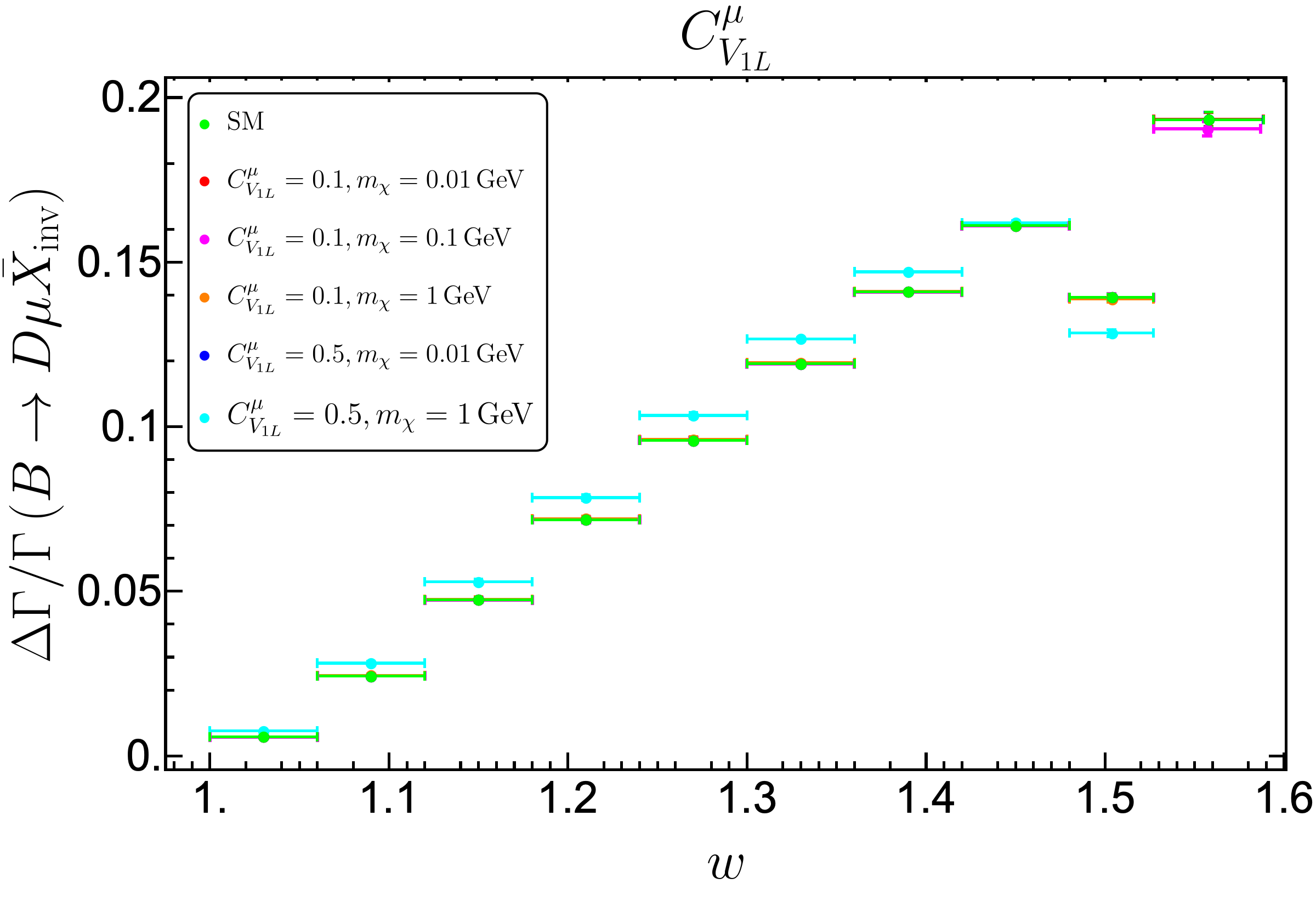}}\hspace{0.0001cm}
	\subfloat[]{\includegraphics[scale=0.175]{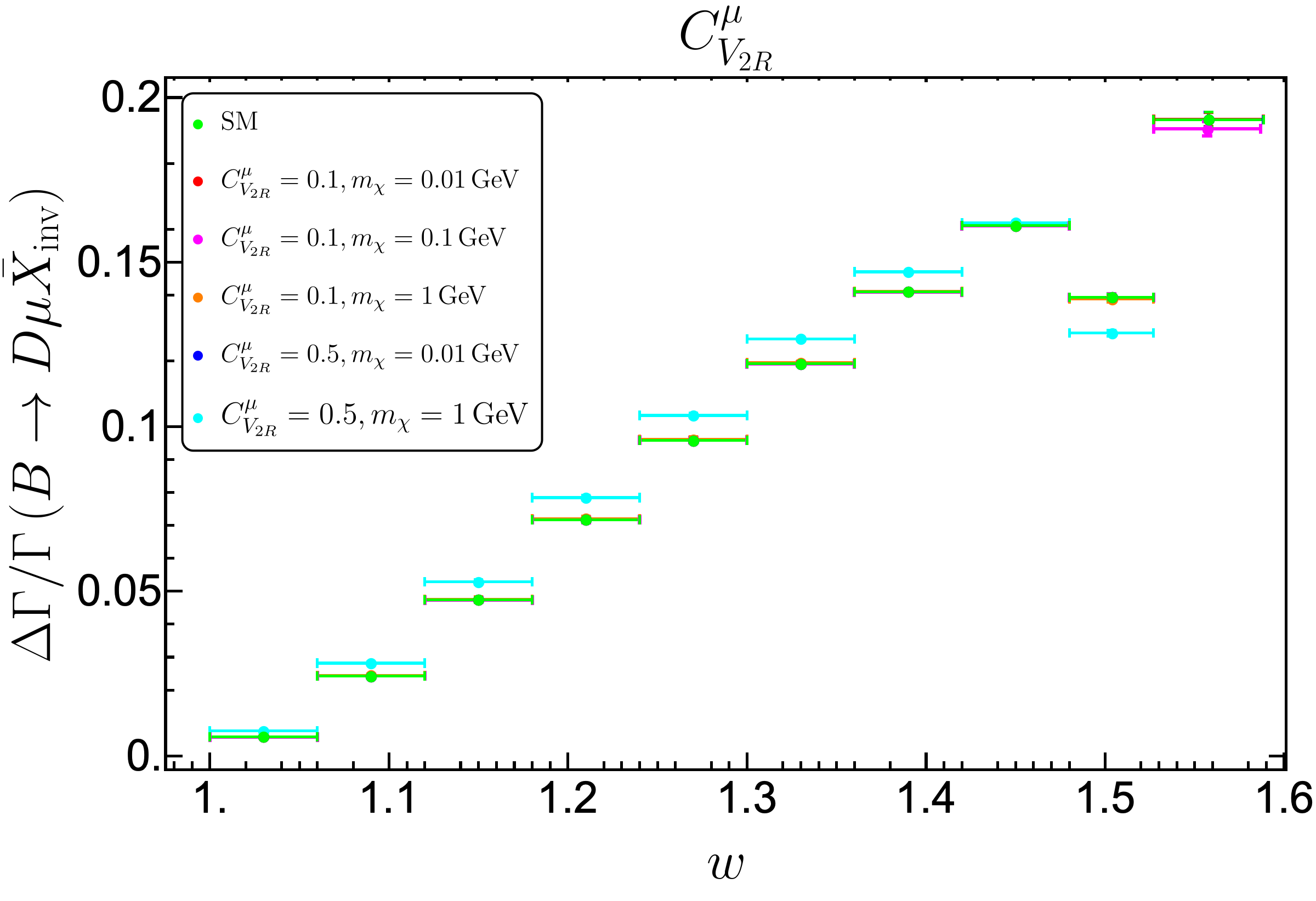}}\hspace{0.0001cm}	
	\subfloat[]{\includegraphics[scale=0.175]{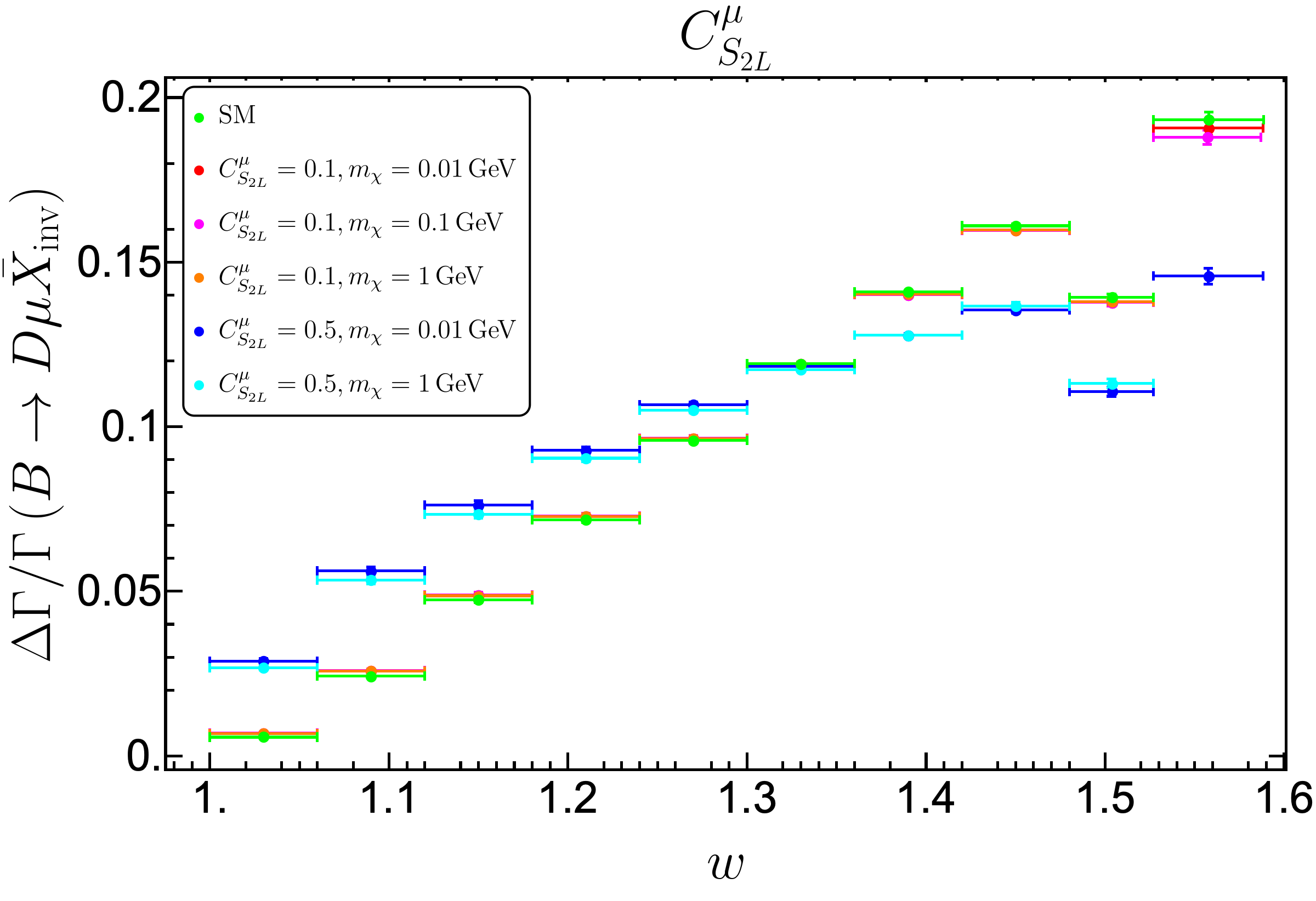}}\hspace{0.0001cm}
	\subfloat[]{\includegraphics[scale=0.175]{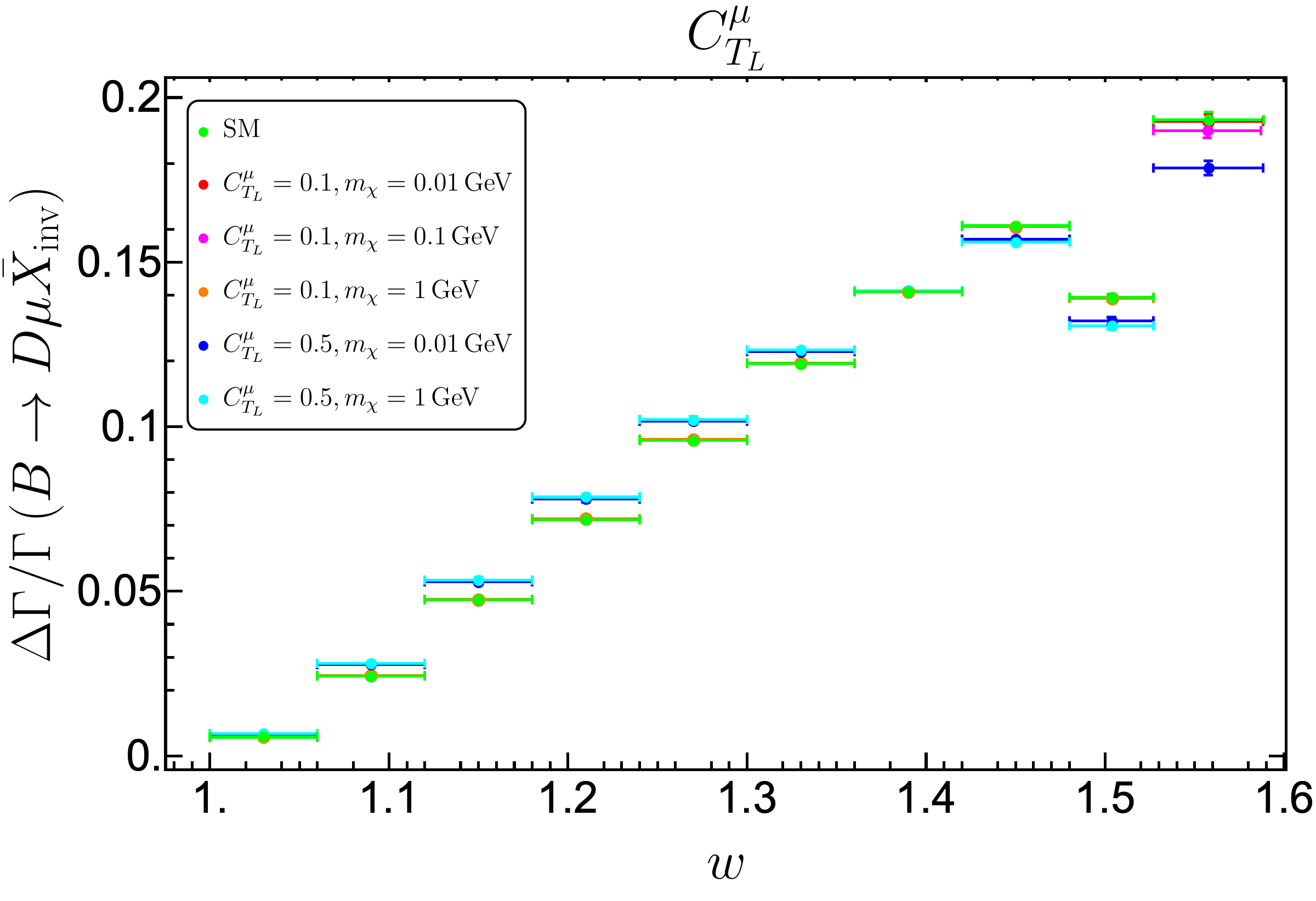}}\hspace{0.0001cm}
	\caption{Variation of the normalised differential decay width, integrated in different bins of $ w $, $ \Delta \Gamma / \Gamma$, for the decay $B \to D\, \mu \, X_{\rm inv}$, for different NP operator scenarios. Each NP plot is done for two BPs of the WCs and $ m_{\chi} =( 0.01\,,0.1\,, 1 ) $ GeV. The green points in each plot are for the SM distribution. }
    \label{fig:B2D_normalised_omega_NP}
\end{figure}

\begin{figure}[t]
	\centering
	\vspace{-0.5cm}
	\subfloat[]{\includegraphics[width=0.48\linewidth]{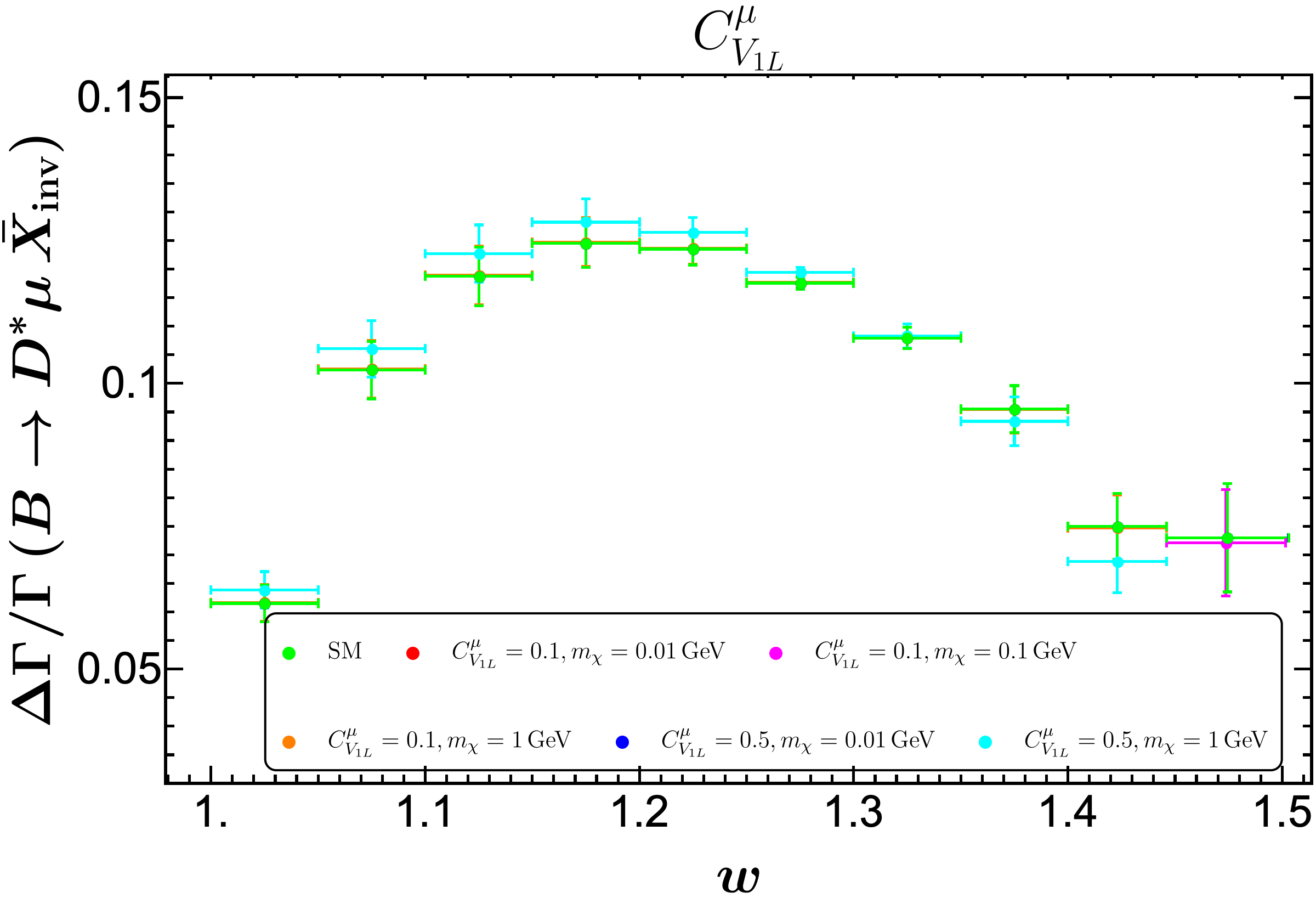}}\hspace{0.0001cm}
	\subfloat[]{\includegraphics[scale=0.175]{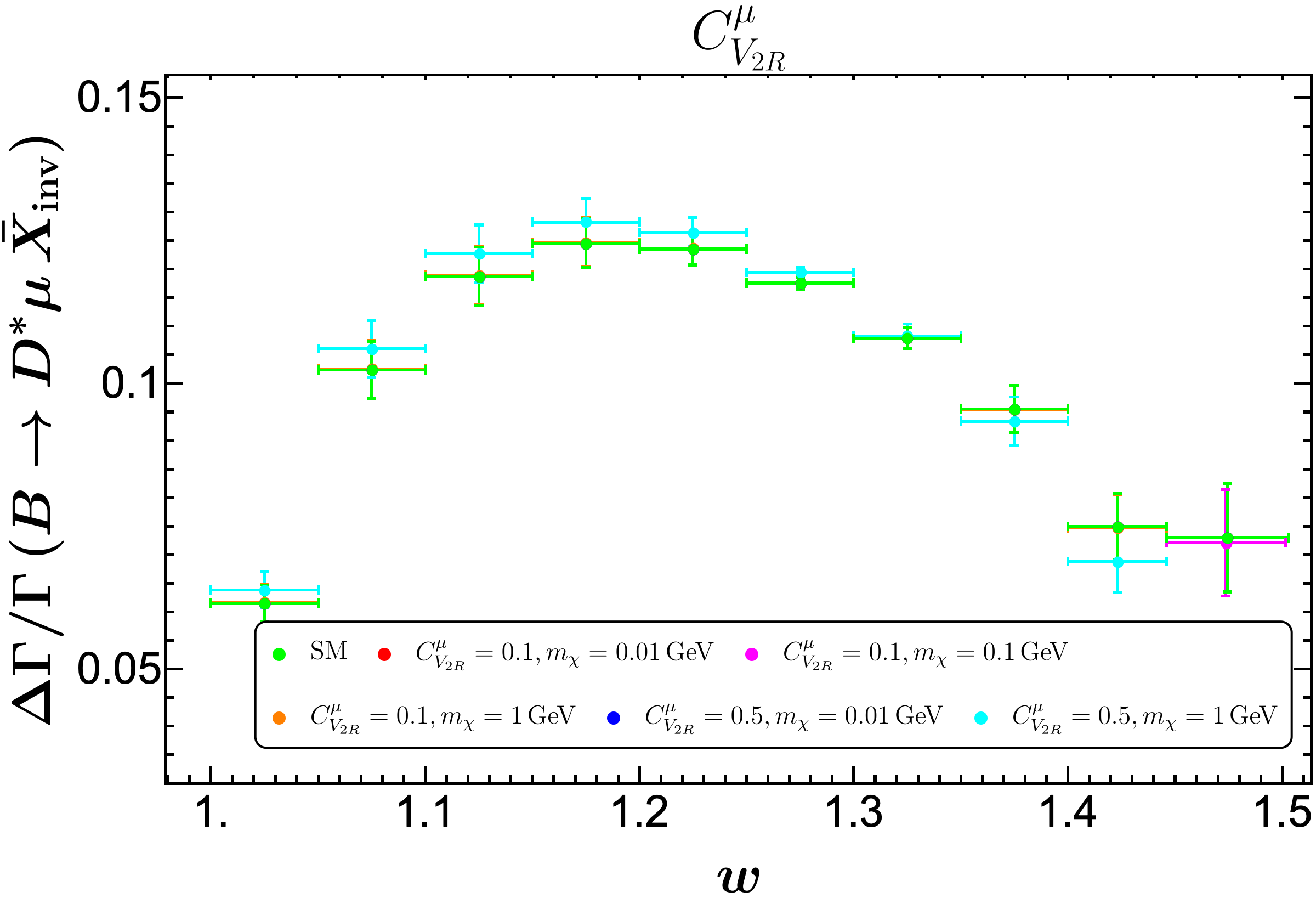}}\hspace{0.0001cm}	
	\subfloat[]{\includegraphics[width=0.48\linewidth]{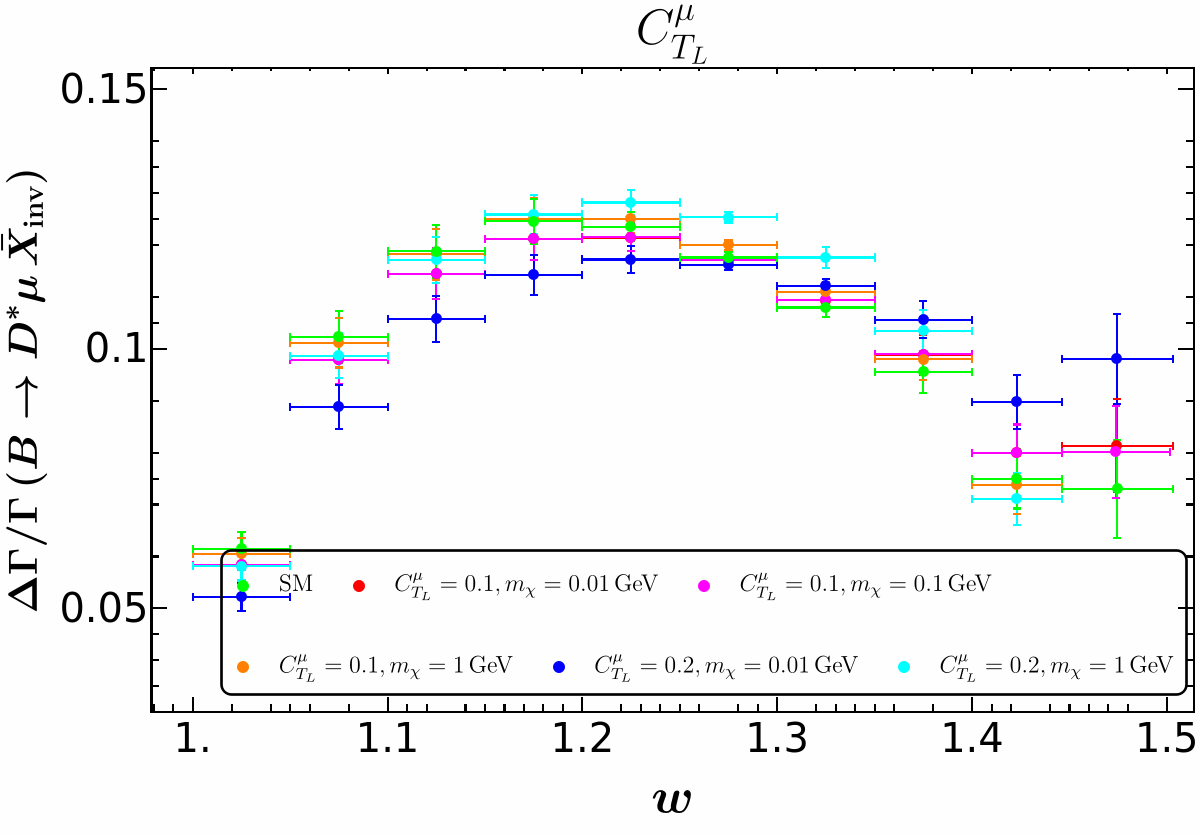}}\hspace{0.0001cm}
	\subfloat[]{\includegraphics[width=0.48\linewidth]{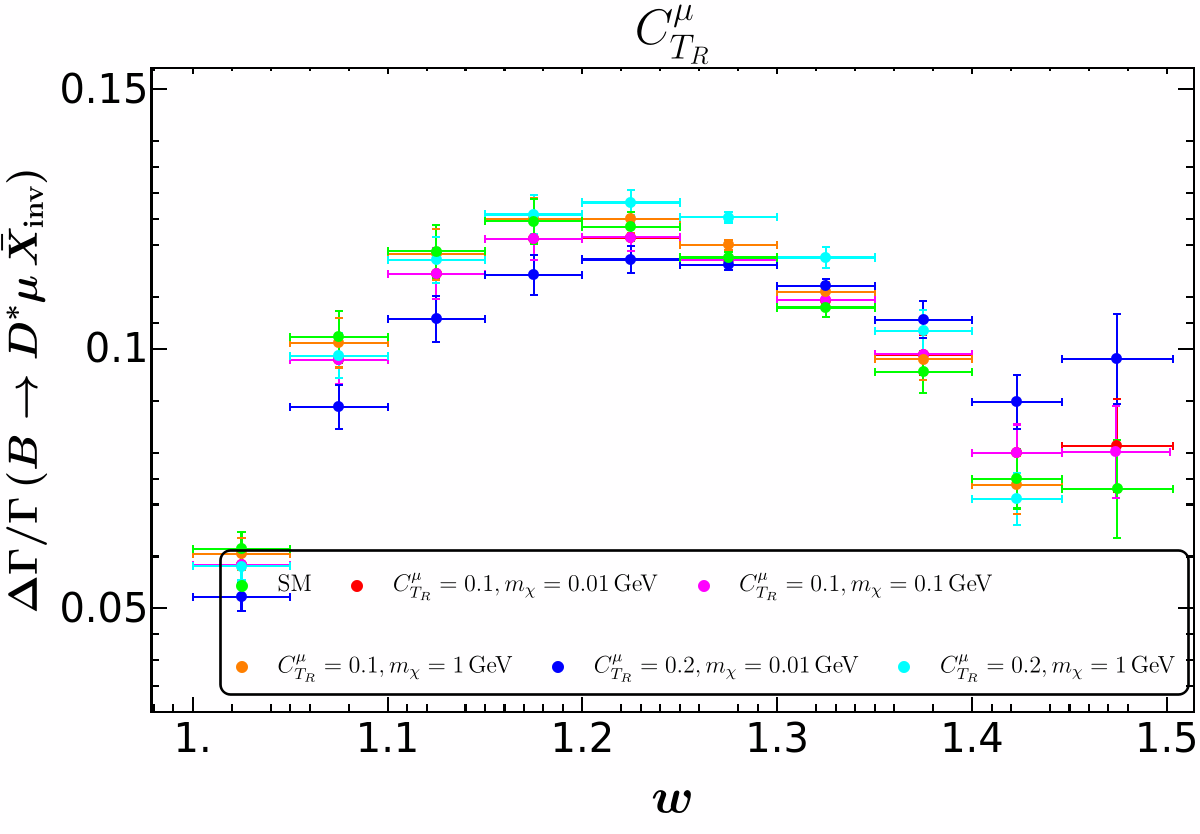}}
	\caption{Variation of $ \Delta \Gamma / \Gamma$, for the decay $B \to D^{*} (\to D \pi)\, \mu \, X_{\rm inv}$ integrated in different bins of $ w $, for different NP operator scenario.  Each NP plot is done for $ m_{\chi} =( 0.01\,,0.1\,, 1 ) $ GeV for two BPs of the NP operators.} \label{fig:B2Dst_normalised_omega_SM_NP}
\end{figure}

\begin{figure}[t]
	\centering
	\subfloat[]{\includegraphics[scale=0.175]{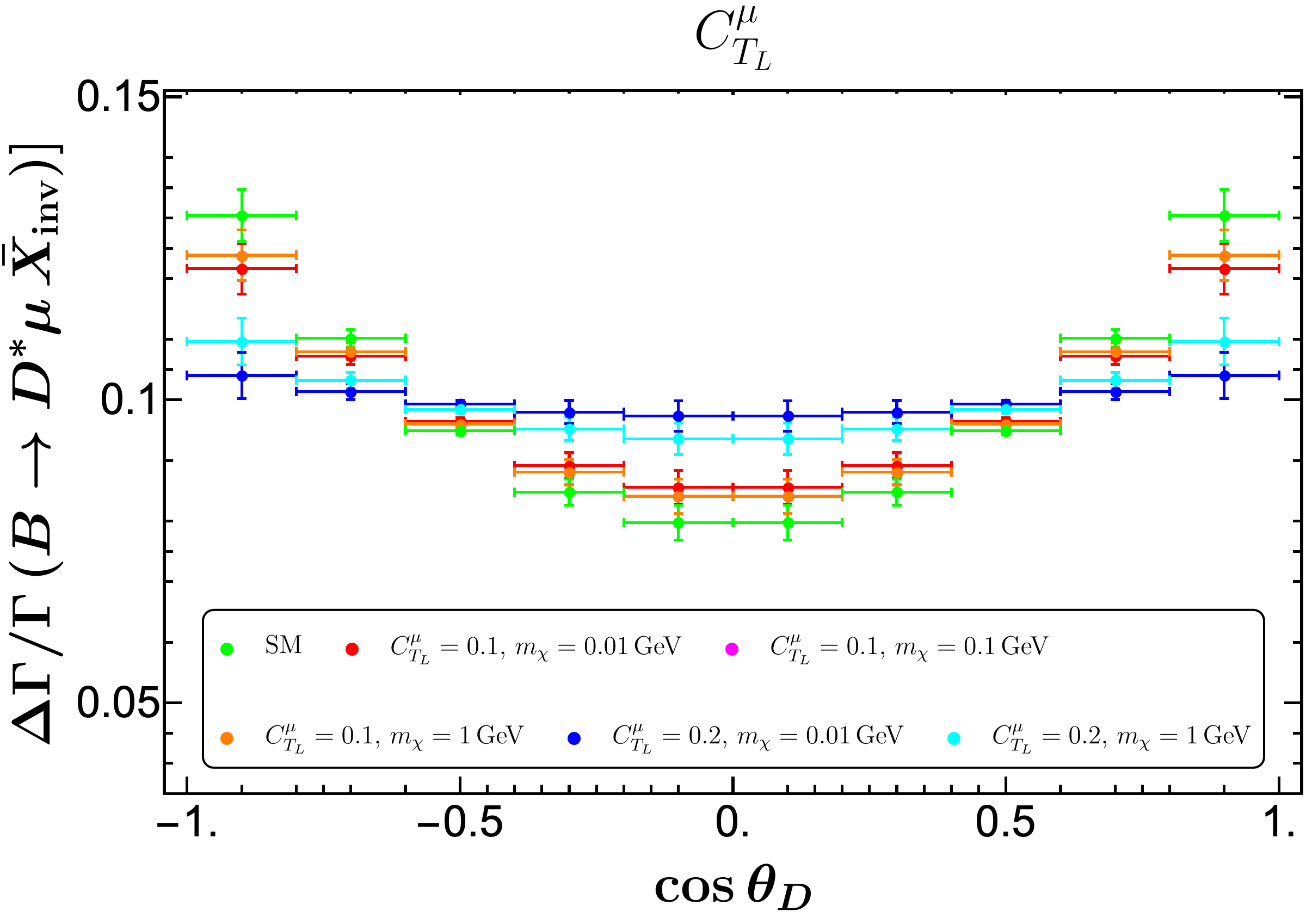}}\hspace{0.0001cm} 
	\subfloat[]{\includegraphics[scale=0.175]{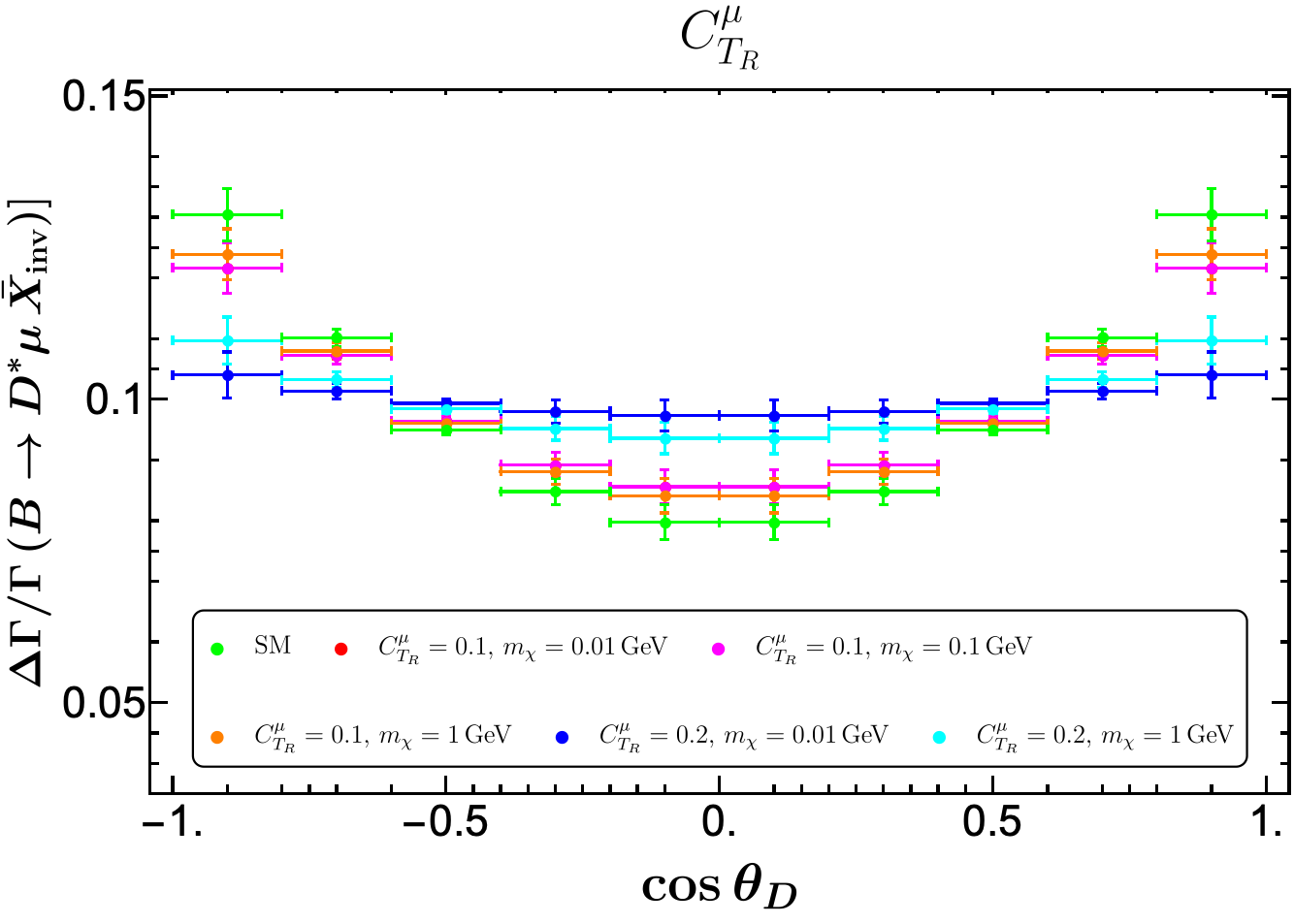}}
	\caption{Variation of the normalised decay width $ \Delta \Gamma / \Gamma $,  for the decay $B \to D^{*} (\to D \pi)\, \mu \, X_{\rm inv}$, integrated in different bins of $ \cos \theta_{D} $. Each plot is shown for two BPs of the WCs, and for the DM masses $m_{\chi} = (0.01, \, 0.1, \, 1)$ GeV. The green points in each plot are for the SM distribution.} \label{fig:B2Dst_unnormalised_cosD_SM_NP}
\end{figure}

\begin{figure}[htbp]
	\centering
	\subfloat[]{\includegraphics[width=0.48\linewidth]{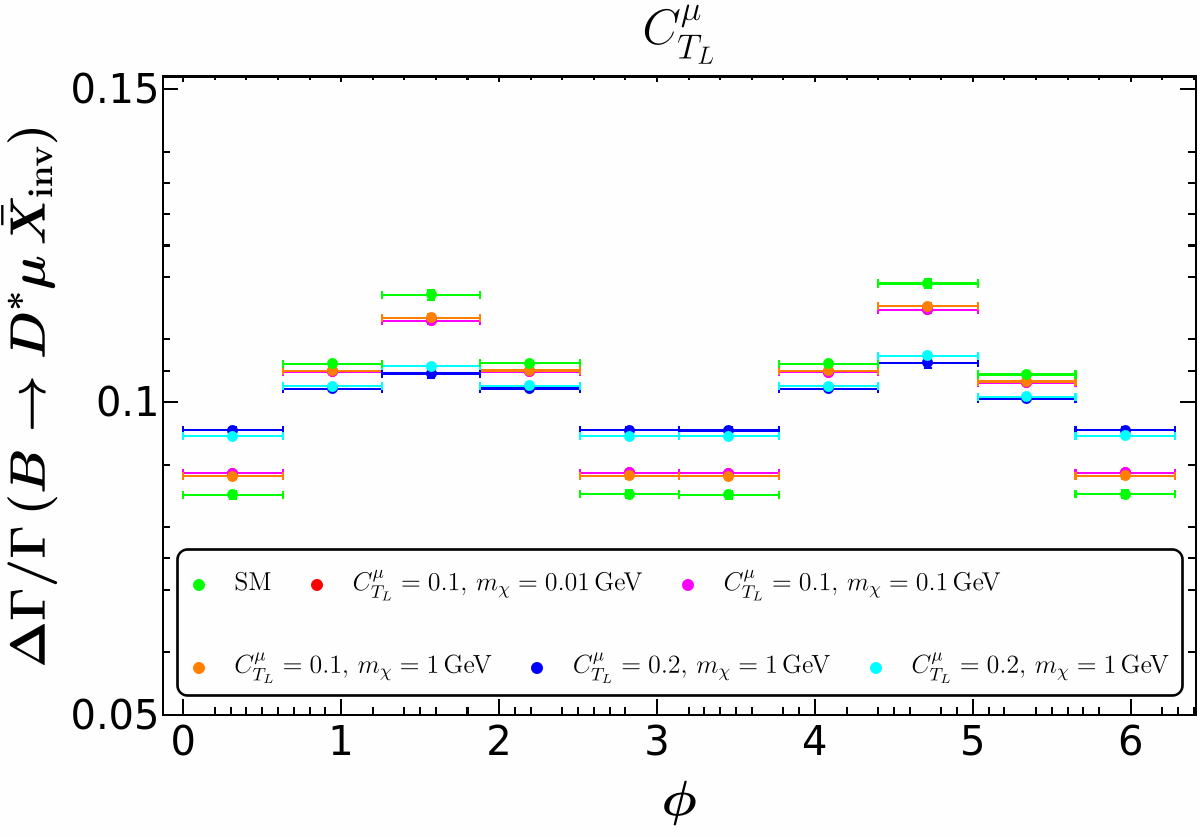}}\hspace{0.0001cm}
	\subfloat[]{\includegraphics[width=0.48\linewidth]{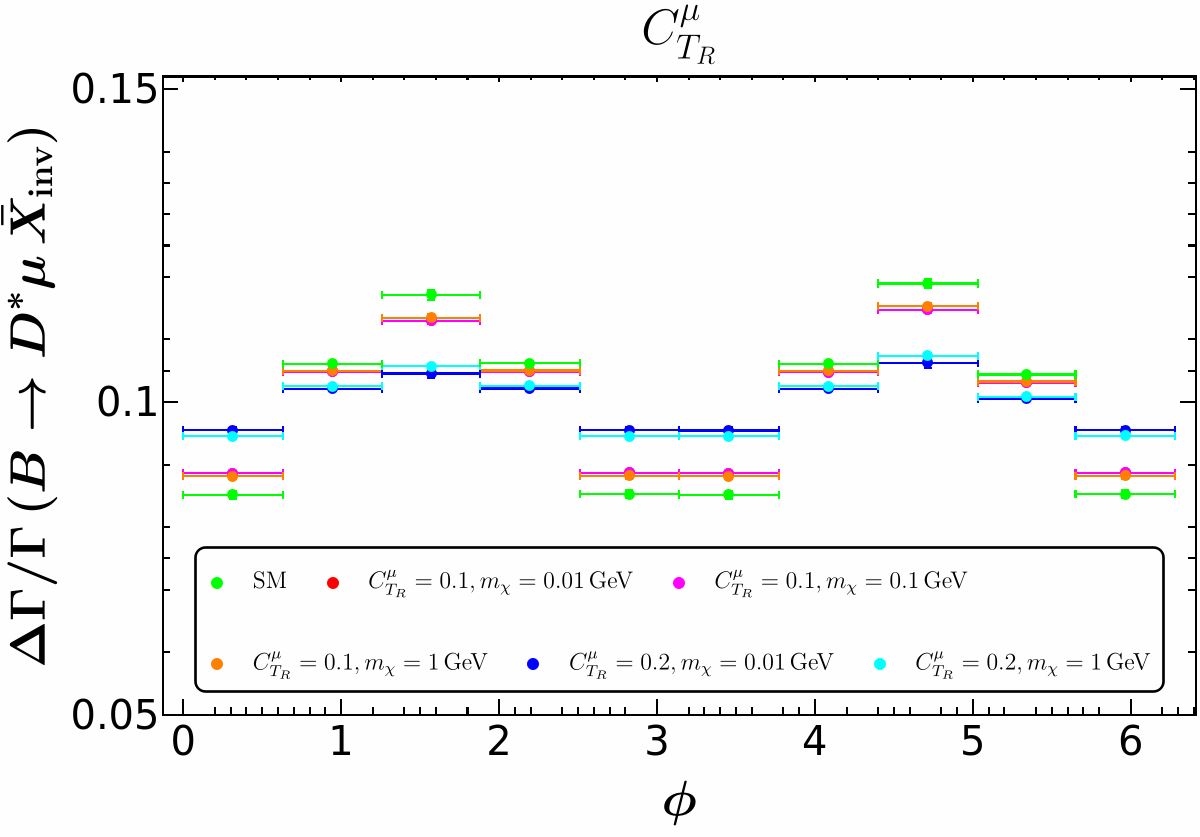}}
	\caption{Variation of the differential decay distribution, $ \Delta \Gamma / \Gamma $,  for the decay $B \to D^{*} (\to D \pi)\, \mu \, X_{\rm inv}$,  integrated in different bins of $ \phi, $ for different NP operator scenarios. Each plot is done for two BPs of the WCs and $ m_{\chi} = (0.01,\,  0.1,\,  1 )$ GeV. The green points in each plot are for the SM distribution. } \label{fig:B2Dst_normalised_chi_SM_NP}
\end{figure}

\subsection{Normalised Decay Rate Distributions}\label{subsec:B2Dst_normalised}
 In a couple of experimental analyses mentioned above, the data on the normalised rates $\Delta\Gamma/\Gamma$ are used to extract $|V_{cb}|$ \cite{Belle:2023bwv, Belle-II:2023okj}. Hence, our analysis defines the normalised rates $\Delta\Gamma/\Gamma$ in small bins of $w$ and other angular variables (wherever applicable). The total rate $\Gamma$ is obtained by integrating the four-fold differential rate over the whole kinematically allowed region in a specific scenario. The allowed ranges of the angular variables will be the same in the SM as well as in all the NP scenarios. However, it is important to note that the allowed ranges of $w$ will be case dependent. The integration ranges will be $w : [1,\,  w_{\rm max}^{\rm SM}]$ for the SM and $[1,\,  w_{\rm max}^{\rm NP}]$ for NP scenarios, with $w_{\rm max}^{\rm SM, NP}$ are defined in eqs.~\eqref{eq:B2Dst_allowed_omega} and \eqref{eq:B2D_allowed_omega}.
 In the remaining region, $w : [w_{\rm max}^{\rm NP},\,  w_{\rm max}^{\rm SM}]$, the contribution is purely from the SM. Following this information, we will obtain the definition for the total decay width in a specific NP scenario as given below: 
 \begin{equation}\label{eq:B2D_full_Br}
 	\Gamma_{\rm tot} = \int_{1}^{w_{\rm max}^{\rm NP}}\frac{d\Gamma}{d w } \bigg|_{\rm tot } d w + \int_{w_{\rm max}^{\rm NP}}^{w_{\rm max}^{\rm SM}} \frac{d\Gamma}{d w } \bigg|_{\rm SM} d w\,.
 \end{equation}
 
We have estimated $\Delta\Gamma/\Gamma$ relevant for $B\to D^{(*)}\ell^-\bar{\nu}$ and $B\to D^{(*)}\ell^-\bar{X}_{\rm inv}$ decays and the total rate $\Gamma$ has been obtained accordingly. For example, we estimate $\Delta\Gamma(B\to D^{(*)}\ell^-\bar{X}_{\rm inv})/\Gamma(B\to D^{(*)}\ell^-\bar{X}_{\rm inv})$ and we obtain the SM scenario from these ratios after setting the new physics WCs to zero. To estimate $\Delta\Gamma(B\to D^{(*)}\ell^-\bar{X}_{\rm inv})$, we have used the same benchmark scenarios which we have discussed in the last subsection. The CKM elements will cancel in these ratios; hence, the only sources of error are the form factors. Due to a strong correlation between the numerator and the denominator, which is coming from the same set of form factors, the errors in these ratios will be less compared to those in the respective $\Delta\Gamma$s. 
 
For $B\to D \ell^-\bar{X}_{\rm inv}$ decays, we have plotted the normalised rates in small $w$ bins in fig.~\ref{fig:B2D_normalised_omega_NP}. We obtain the SM rates by setting the NP WCs to zero, which we plot with green error bars. Each plot corresponds to a different NP scenario we discussed in the last section. Here, we have considered only the scenarios in which we have seen deviations in the unnormalised rates $\Delta\Gamma$ discussed in the last section. Across all the scenarios, our estimate of the errors in the normalised rates is roughly around 2\%. For a normalised distribution, the error entirely comes from the form factors, since no CKM elements are involved. We note sizeable deviation in the one operator scenarios $\mathcal{O}_{S_{1R(L)}}^{\ell}$ and $\mathcal{O}_{S_{2L(R)}}^{\ell}$. In a few bins, the deviations could be much larger than 5$\sigma$ (see fig. \ref{fig:B2D_normalised_omega_NP}). This is primarily due to a considerable reduction in the estimated errors. Also, we observe deviations for the one operator scenarios $\mathcal{O}_{T_{R(L)}}^{\ell}$, $\mathcal{O}_{V_{1L(R)}}^{\ell}$ and  $\mathcal{O}_{V_{2L(R)}}^{\ell}$, respectively. However, in these scenarios, the deviations are relatively smaller around the 3$\sigma$ to 5$\sigma$ level. Like the earlier cases, we observe the maximum deviation for the values of WCs at 0.5.

For $B\to D^* \ell^-\bar{X}_{\rm inv}$ decays in fig.~\ref{fig:B2Dst_normalised_omega_SM_NP}, we have shown the variation of $\Delta \Gamma/\Gamma$ with the recoil parameter $w$. In most of the bins, the estimated 1$\sigma$ errors are about 5\%. We have followed the same colour codes as before to represent the specific scenarios. A comparison of the distributions in different NP scenarios with the SM shows that for the one operator scenarios $\mathcal{O}_{V_{1L(R)}}^{\ell}$ and  $\mathcal{O}_{V_{2L(R)}}^{\ell}$ we do not observe any sizeable deviations. However, for the scenarios $\mathcal{O}_{T_{R(L)}}^{\ell}$, for large values of the WC, deviations are comparatively large, and in a few bins they are even larger than 5$\sigma$. 


Similarly, we have studied the angular distributions of the normalised rates in small bins of $\cos\theta_{D}$, which is shown in fig.~\ref{fig:B2Dst_unnormalised_cosD_SM_NP}, and the estimated 1$\sigma$ errors of the normalised rates are roughly around 2\%. 
Due to a partial cancellation, the effects of NP are not that prominent in these ratios.
In these rates, we only observe significant NP effects for the operators  $\mathcal{O}_{T_{R(L)}}^{\ell}$. For different BPs of the tensor WC, $\mathcal{O}_{T_{R(L)}}^{\ell} = (0.1, 0.2)$, we observe a maximum deviation around $ 1 \sigma, \, 4.5 \sigma$. For increasing the value of the WCs, i.e., for $\mathcal{O}_{T_{R(L)}}^{\ell}=0.5$, the deviation is $>5 \sigma$. In all the rest of the scenarios, the NP effects in these ratios are almost negligible.       
    
Fig.~\ref{fig:B2Dst_normalised_chi_SM_NP} shows the variation of the normalised decay rates in bins of the azimuthal angle $\phi$. In the total rate, the azimuthal angle will vary in the $(0,\pi)$ range. In the normalised decay rate $\Delta\Gamma / \Gamma $, we obtain very small errors in each bin ($\approx (0.5-1)\%$). In the plots, the colour coding in the SM and in all the NP scenarios is similar to that discussed before. We do not observe significant deviations from the SM even for very large vector and scalar WCs. However, tensor operators produce significant deviations, which become more pronounced with increasing values of the corresponding WCs. For $\mathcal{C}_{T_{R(L)}}^{\ell} = 0.1$, the deviation is approximately $\sim 4\sigma$, while for larger values of the Wilson coefficient, the deviation exceeds $5\sigma$. This large deviation arises due to the very small uncertainties associated with the normalised decay width.

Fig.~\ref{fig:B2Dst_normalised_cosl_SM_NP} shows the variation of the integrated differential decay width \( \Delta \Gamma / \Gamma \), binned in \( \cos \theta^{\prime}_{\mu} \), for the muon in the final state of the decay. The uncertainty in the predictions of these normalised rates is also very small ($\approx 2\%$). In the plots, we have followed the same conventions as before. For the vector and scalar operators, the deviation from the SM prediction is very small, even for larger values of the WCs, as the NP effects in the numerator and denominator partially cancel out. Hence, we have not shown those plots. 
In the case of tensor interactions, deviations are observed in most bins. For $C_{T_{L(R)}}^{\mu} = 0.1$, the maximum deviation is around $1.5 \sigma$, and if we increase the WC value, the maximum deviation of $> 5 \sigma$ is observed. 

 
Finally, we note that the allowed ranges of the WCs in NP scenarios with DM signatures can lead to observable deviations in the decay-rate distributions relative to the corresponding SM predictions. Such deviations persist even in the normalised distributions. Therefore, the extraction of $|V_{cb}|$ from these observables should be performed with care, as neglecting such NP effects may bias the determination. In this context, it is worthwhile to consider analyses in which the NP WCs and $|V_{cb}|$ are extracted simultaneously from the decay distributions. In the following subsection, we discuss how the current data can be utilised to constrain the NP WCs and present the corresponding allowed parameter ranges. 

\subsection{Extraction of $|V_{cb}|$ and New Physics Wilson Coefficients}\label{sec:Vcb_fit}
In the previous subsection, we discussed the sensitivity of the decay rate distributions for the processes  
$B \to D^{(*)} \ell X_{\rm inv}$ to NP effects by analysing their behaviour in bins of the kinematic variables $x_i$, where  
$x_i = \{ w,\ \cos\theta_{D^{*}},\ \cos\theta_{\ell},\ \phi \}$. We have also investigated possible deviations arising in different NP scenarios relative to the SM predictions and available experimental data, identifying regions where noticeable discrepancies may occur.
In this section, we proceed to extract the NP WCs using the available and kinematically allowed experimental data. Wherever feasible, we perform a simultaneous extraction of $|V_{cb}|$ alongside the NP WCs. This combined analysis serves a dual purpose: it allows us to constrain the NP parameter space and, at the same time, assess the potential impact of NP contributions on the determination of $|V_{cb}|$.

It is important to note that, in NP scenarios involving a non-zero mass for the invisible dark-sector fermion, $m_\chi$, the kinematically allowed range of the recoil variable $w$ becomes more restricted compared to the SM case. As a consequence, the NP contributions do not populate the highest $w$-bins. Depending on the value of $m_\chi$, the last one or two bins of the $w$-distribution remain unaffected by the presence of NP. In contrast, experimental analyses determine the allowed $w$-range under the assumption of a massless neutrino in the final state~\cite{Belle:2023bwv, Belle-II:2023okj, Belle:2015pkj}. This mismatch implies that the kinematically accessible region in NP scenarios does not always coincide with the experimentally defined binning.

A closely related point concerns the angular observables. The bin-integrated decay rate distributions in the angular variables $(\cos\theta_{D},\ \cos\theta_{\ell},\ \phi)$ are obtained after integrating the differential decay width over the full kinematically allowed range of $w$. As a result, these observables do not retain sensitivity to the NP-induced modifications of the $w$-range. Consequently, they cannot be directly used to extract the NP WCs in the present framework. Therefore, our analysis focuses on the decay rate distributions in $w$-bins, restricted to the kinematic regions allowed within each NP scenario.
For instance, in the process $B \to D^{*} \mu X_{\rm inv}$, the last one and two $w$-bins must be excluded from the fit for $m_\chi \leq 0.1$ and $m_\chi = 1$, respectively. On the other hand, for an electron in the final state, even for $m_\chi = 1$, only the last bin is excluded. This difference originates from the interplay between the bin widths adopted in the experimental analyses and the kinematic limits imposed by the non-zero $m_\chi$.
 
\begin{table}[t]
\centering
\renewcommand{\arraystretch}{1.5}
\setlength{\tabcolsep}{5pt}
\resizebox{0.7\textwidth}{!}{%
\rowcolors{1}{red!10!blue!20}{purple!10!teal!15}
\begin{tabular}{|c c c c|}
        \hline
 \rowcolor{yellow!5!purple!15!blue!10}
  \multicolumn{4}{c}{$B \to D^{(*)}\mu X_{\rm inv}$} \\
\hline
            \rowcolor{blue!10!teal!40!}
\large
Scenario & $ m_{\chi} $ (GeV) & $ |V_{cb}| \times 10^{-3} $ & $ |C_{i}| $ \\
\hline
\hline
\cellcolor{purple!15!blue!15} SM & -  & $ 40.061\pm 0.826 $ & -  \\

\cellcolor{teal!25}&  $0.01$  & $40.062 \pm 0.826 $  &  $0.2567 \pm  0.1236$  \\
\cellcolor{teal!25}&  $0.1$  &  $40.063 \pm 0.826 $  &  $0.2565 \pm  0.1237$  \\
\multirow{-3}{*}{\cellcolor{teal!25} *$C_{V_{1L}}^{\mu}$}  &  $1$  & $40.484 \pm 0.810 $ &  $0.1652 \pm 0.1940 $  \\

\cellcolor{purple!15!blue!15}&  $0.01$  &  $40.062 \pm 0.826 $  &  $0.2567 \pm  0.1236$  \\
\cellcolor{purple!15!blue!15}&  $0.1$  &  $40.063 \pm 0.826 $  &  $0.2565 \pm  0.1237$  \\
\multirow{-3}{*}{\cellcolor{purple!15!blue!15} * $C_{V_{2R}}^{\mu}$}  &  $1$  & $40.484 \pm 0.810 $  &  $0.1652 \pm 0.1940 $  \\

\cellcolor{teal!25} &  $0.01$  &  $41.361 \pm 0.888 $ &  $0.0\pm 0.1093$  \\
\cellcolor{teal!25} &  $0.1$  &  $41.361 \pm 0.888 $   &  $0.0 \pm 0.1093$  \\
\multirow{-3}{*}{\cellcolor{teal!25} $C_{S_{2L}}^{\mu}$}  &  $1$  &  $41.473 \pm 0.901 $  &  $0.0\pm 0.1126$  \\

\cellcolor{purple!15!blue!15} &  $0.01$  &  $41.150 \pm 1.204 $  &  $0.0475\pm 0.0859$  \\
\cellcolor{purple!15!blue!15} &  $0.1$  &  $41.150 \pm 1.204 $  &  $0.0475\pm 0.0859$  \\
\multirow{-3}{*}{\cellcolor{purple!15!blue!15} $C_{T_{L}}^{\mu}$}  &  $1$  &  $41.467 \pm 0.904 $  &  $0.0043 \pm 0.0852$  \\

\cellcolor{teal!25}&  $0.01$  &  $41.161  \pm 1.171 $  &  $0.0467\pm 0.0851$  \\
\cellcolor{teal!25}&  $0.1$  &  $41.158 \pm 1.174$  &  $0.0475\pm 0.0839$  \\
\multirow{-3}{*}{\cellcolor{teal!25} $C_{T_{R}}^{\mu}$}  &  $1$  &  $41.452 \pm 0.958$  &  $0.0171\pm 0.0962$  \\
\hline
\end{tabular}
}
\caption{ The extraction of the CKM element $|V_{cb}|$ and the Wilson coefficients $C_{i}$
is performed in the one-operator scenario from a combined fit of
$B \to D \mu \bar{X}_{\rm inv}$ and $B \to D^{*} \mu \bar{X}_{\rm inv}$,
for several benchmark masses of the NP particle $\chi$. Entries marked with ` *  ', are obtained by treating $|V_{cb}|$ as a nuisance parameter fixed from the SM fit, while the unmarked entries correspond to fits where both $|V_{cb}|$ and $\mathcal{C}_{i}$ are taken as free parameters.
}
\label{tab:fit_results_combined_muon_Ci}
\end{table}

\begin{table}[t]
\centering
\renewcommand{\arraystretch}{1.5}
\setlength{\tabcolsep}{5pt}
\resizebox{0.7\textwidth}{!}{%
\rowcolors{1}{red!10!blue!20}{purple!10!teal!15}
\begin{tabular}{|c c c c|}
\hline
            \rowcolor{yellow!5!purple!15!blue!10}
  \multicolumn{4}{c}{$B \to D^{(*)}\, e \, X_{\rm inv}$} \\
\hline
\rowcolor{blue!10!teal!40!}
\large
Scenario & $ m_{\chi} $ (GeV) & $ |V_{cb}| \times 10^{3} $ & $ |C_{i}| $ \\
\hline \hline

\cellcolor{purple!15!blue!15} SM & -  & $ 41.083\pm 0.515 $ & -  \\

\cellcolor{teal!25} &  $0.01$  &  $41.004 \pm  0.670$  &  $0.10690\pm 0.2515$  \\
\cellcolor{teal!25} &  $0.1$  &  $41.006 \pm  0.670$ &  $0.10580 \pm 0.25410$  \\
\multirow{-3}{*}{\cellcolor{teal!25} *$C_{V_{1L}}^{e}$}  &  $1$  &  $41.071 \pm  0.532$ &  $0.0 \pm  0.1508$  \\

\cellcolor{purple!15!blue!15}  &  $0.01$  & $41.004 \pm  0.670$  &  $0.10690\pm 0.2515$   \\
\cellcolor{purple!15!blue!15} &  $0.1$  &  $41.006 \pm  0.670$ &  $0.10580 \pm 0.25410$  \\
\multirow{-3}{*}{\cellcolor{purple!15!blue!15} *$C_{V_{2R}}^{e}$} &  $1$  & $41.062 \pm  0.532$  &  $0.0 \pm           0.14810$  \\

\cellcolor{teal!25}  &  $0.01$  &  $41.238\pm 0.865$  &  $0.0\pm 0.0627$  \\
\cellcolor{teal!25}  &  $0.1$  &  $41.238\pm 0.865$  &  $0.0\pm 0.0628$  \\
\multirow{-3}{*}{\cellcolor{teal!25} $C_{S_{2L}}^{e}$}  &  $1$  &  $41.185\pm 0.876$  & $0.0\pm 0.06979$  \\

\cellcolor{purple!15!blue!15} &  $0.01$  &  $40.977\pm 1.191$  &  $0.0522\pm 0.0776$  \\
\cellcolor{purple!15!blue!15}  &  $0.1$  &  $40.977\pm 1.191$  &  $0.0522\pm 0.0776$  \\
\multirow{-3}{*}{\cellcolor{purple!15!blue!15}  $C_{T_{L}}^{e}$}  &  $1$  &  $41.185\pm 0.876$  &  $0.0\pm 0.04520$  \\

\cellcolor{teal!25}  &  $0.01$  &  $40.803\pm 1.111$  &  $0.07027\pm 0.0519$  \\
\cellcolor{teal!25}  &  $0.1$  &  $40.885\pm 1.113$  &  $0.06358\pm 0.05715$  \\
\multirow{-3}{*}{\cellcolor{teal!25} $C_{T_{R}}^{e}$} &  $1$  &  $41.174\pm 0.876$  &  $0.0018\pm 0.04524$  \\
\hline
\end{tabular}
}
\caption{Same as Table~\ref{tab:fit_results_combined_muon_Ci}, instead of muon, here electron is the final state lepton. }
\label{tab:fit_results_combined_elec_Ci}
\end{table}
\paragraph{\underline{Inputs for the analysis}:}
The experimental data are provided by Belle \cite{Belle:2015pkj, Belle:2023bwv} and Belle-II \cite{Belle-II:2023okj} collaboration. In this analysis, we have not used the available data on normalised bins available for \( B \to D^{*} \ell \bar{\nu} \) (i.e., \( \Delta \Gamma / \Gamma \)) as these observables are independent of \( |V_{cb}| \). For the decay $B \to D \ell \bar{X}_{\rm inv}$, we have used the bin-integrated observables $ \frac{\Delta\Gamma}{\Delta w} $, divided by the corresponding bin width, i.e.,
\begin{equation}\label{eq:B2D_obs_for_fit}
\frac{\Delta \Gamma }{\Delta w} = \frac{1}{|w_{j} - w_{i}|} \int_{w_{i}}^{w_{j}}\frac{d\Gamma}{d w} d w\,.
\end{equation}
The experimental values of these observables are taken from \cite{Belle:2015pkj}. For the decay $B \to D^{*} \ell \bar{X}_{\rm inv}$, the experimental result is provided for the integrated decay width, without dividing by the bin length:  
\[
\Delta \Gamma = \int_{w_{i}}^{w_{j}} \frac{d \Gamma}{d w} d w\,.
\]  
As mentioned earlier, we consider only the differential decay widths integrated in bins of \( w \), and do not include the rates defined in angular bins in our fit.

 In this analysis, another vital ingrediant is the form factors which we have extracted using the inputs from lattice by Fermilab-MILC~\cite{MILC:2015uhg} and HPQCD~\cite{Na:2015kha} for the decay \( B \to D \ell \bar{X}_{\rm inv} \) and from Fermilab-MILC~\cite{FermilabLattice:2021cdg}, JLQCD~\cite{Aoki:2023qpa} for \( B \to D^{*} \ell \bar{X}_{\rm inv} \) decays. Additionally, we have used the inputs on the form factors from LCSR~\cite{Giacchino:2014moa} given at $q^2=0$. The detail of the analysis is provided in sec.~\ref{subsec:q2shapes}. The extracted values of the BGL coefficients \( a_i \) from the analysis of the $B \to D$  and $B \to D^*$   transitions form factors with the lattice and LCSR inputs are summarised in Tables~\ref{tab:B2D_combined_FF_fit} and~\ref{tab:B2Dst_combined_FF_fit}, respectively. We used these extracted values of the BGL coefficients as nuisance parameters to extract the WCs and $|V_{cb}|$ using experimental data discussed above.

\paragraph{\underline{Results}:}
We have performed both separate and combined fits for the decay modes $B \to D \ell X_{\rm inv}$ and $B \to D^{*} \ell X_{\rm inv}$, considering both muons and electrons in the final state. The results of the combined analysis for the muon channel are presented in Table~\ref{tab:fit_results_combined_muon_Ci}. The fits have been carried out for representative values of the dark fermion mass, $m_{\chi} = (0.01,\,0.1,\,1)\,\text{GeV}$. The corresponding fitted values of $|V_{cb}|$ are displayed in the third column of the table.

For the muon final state, we first perform a fit within the SM hypothesis (i.e., without any dark fermion contribution) using the available data on the differential decay rate in $w$-bins. From this fit, we obtain
\begin{equation}
|V_{cb}| = (40.061 \pm 0.826)\times 10^{-3}.
\end{equation}
In the one-operator NP scenarios $\mathcal{O}_{S_{1R(L)}}^{\ell}$, $\mathcal{O}_{S_{2R(L)}}^{\ell}$, and $\mathcal{O}_{T_{L(R)}}^{\ell}$, we perform a simultaneous extraction of $|V_{cb}|$ and the corresponding WCs. The results are presented in Table~\ref{tab:fit_results_combined_muon_Ci} for representative cases, namely $\mathcal{O}_{S_{2L}}^{\ell}$ and $\mathcal{O}_{T_{L(R)}}^{\ell}$. As discussed earlier, the NP sensitivities of the decay rate distributions for $\mathcal{O}_{S_{1R(L)}}^{\ell}$ and $\mathcal{O}_{S_{2L(R)}}^{\ell}$ are very similar. Therefore, we present results for only one of these scenarios, noting that the fits in the other cases yield comparable outcomes.

Within the current experimental uncertainties, the extracted values of $|V_{cb}|$ in all these one-operator scenarios are consistent with the SM determination at the $1\sigma$ level. We observe that the best-fit values of $|V_{cb}|$ (obtained simultaneously with the NP Wilson coefficients) shifted from the SM fit by at most $\sim 4\%$. The allowed ranges of the scalar and tensor WCs, $C_{S_{1L(R)}}^{\mu}$, $C_{S_{2L(R)}}^{\mu}$, and $C_{T_{L(R)}}^{\mu}$, are typically of order $\sim \mathcal{O}(10^{-1})$.

The situation is qualitatively different for the vector operators $\mathcal{O}_{V_{1L(R)}}^{\mu}$ and $\mathcal{O}_{V_{2L(R)}}^{\mu}$. In these cases, a simultaneous extraction of $|V_{cb}|$ and the corresponding WCs is not feasible. This difficulty arises because their contributions to the decay rate enter approximately as
$|V_{cb}|^2 \left(a_1 + a_2\, C_{V_i}^2 \right)$,
with $a_1 \approx a_2$, leading to a strong degeneracy between $|V_{cb}|$ and the vector Wilson coefficients. Consequently, the fit becomes unstable and results in poorly constrained parameters with large uncertainties. To circumvent this issue, we treat $|V_{cb}|$ as a nuisance parameter, fixing it to the value obtained from the SM fit along with its associated uncertainty. Within this framework, we extract the allowed ranges of $C_{V_{1L(R)}}^{\mu}$ and $C_{V_{2L(R)}}^{\mu}$. We find that, within their $1\sigma$ uncertainties, these coefficients can be as large as $\sim 0.4$ for $m_{\chi} = 0.01$-$0.1~\text{GeV}$.
It is worth noting that, within the $1\sigma$ error bands, non-zero values of these WCs remain allowed. This feature is consistent with the behaviour observed in the decay-rate distributions shown in figs.~\ref{fig:B2D_unnormalised_omega_NP} and~\ref{fig:B2Dst_unnorm_omega_SM_NP}, where certain $w$-bins exhibit noticeable deviations from the corresponding SM predictions, indicating residual sensitivity to NP effects.

The individual fit results for the electron final state are provided in Table~\ref{tab:fit_results_combined_elec_Ci}. In this analysis, we have combined the available data on the decays \( B \to D e X_{\rm inv} \) and \( B \to D^{*} e X_{\rm inv} \). A similar trend in the fitted values of the WCs is observed in this analysis, as in the muon final state. The fitted values of $|V_{cb}|$ in different NP scenarios are consistent with those obtained in the SM within their 1$\sigma$ error bars. However, a shift of about $2\%$ is observed between the best fit values $|V_{cb}|$ extracted in the SM and those of the NP scenarios with a scalar or a tensor operator. Also, the allowed 1$\sigma$ ranges of the new physics WCs could be of order $\mathcal{O}(10^{-1})$. With improved precision, such a large value could contribute significantly to the decay rate distributions of these decays. The above tables of fit results for muon and electron in the final state show that the corresponding WCs, which are extracted from the fit, are more constrained for the case of the electron in the final state.   

\subsection{Prediction of the Decays: $B_{c}^{+} \to \ell^{+} \, X_{\rm inv}$}
Similar to the $B \to D^{(*)} \ell X_{\rm inv}$, one can also study the effect of a low-mass fermionic DM in the decay $B_{c} \to \ell X_{\rm inv}$, with $X_{\rm inv} = \nu $, left handed neutrino for SM and $X_{\rm inv} = \chi$, a BSM dark fermion. The branching ratio of the decay in terms of the Wilson coefficients defined in eq.~\eqref{eq:general_Hamiltonian_b2c}, is given by:
\begin{align}
\mathcal{B}(B_c\to\ell\chi) &= \tau_{B_c}\frac{ G_F^2 |V_{cb}|^2 f_{B_c}^2 }{ 8\pi m_{B_c}^3 } \lambda^{1/2}(m_{B_c}^2,m_\ell^2,m_\chi^2) \nonumber\\[2mm]  &\times \Big[ (m_{B_c}^2-m_\ell^2-m_\chi^2) \left( |A_L|^2+|A_R|^2 \right) - 4m_\ell m_\chi \,{\rm Re}(A_LA_R^\ast) \Big], \end{align}
with 
\begin{align} 
A_L &= m_\ell\, (C_{V_{1L}}^\ell - C_{V_{2L}}^\ell) + \frac{m_{B_c}^2}{m_b+m_c}\, (C_{S_{1L}}^\ell - C_{S_{2L}}^\ell), 
\\[2mm] A_R &= m_\chi\, (C_{V_{1R}}^\ell - C_{V_{2R}}^\ell,) + \frac{m_{B_c}^2}{m_b+m_c}\, (C_{S_{1R}}^\ell - C_{S_{2R}}^\ell). 
\end{align}
The K{\"a}llen $\lambda$ function is defined earlier. The tensor operators do not contribute to this process. By putting $m_{\chi} =0$ and $C_{V_{1L}}^{\ell} =1$, one can obtain the branching for the SM. As in the previous decays, the branching ratios for SM and NP will also be added at the branching level due to differences in available phase space. $f_{B_c}$ is the decay constant of the $B_{c}$ meson, given by HPQCD \cite{Colquhoun:2015oha} as $f_{B_c} = 0.434 (15)$ GeV.  

In this section, we have provided the branching ratio of the decay $B_{c} \to \ell X_{\rm inv}$, for the light leptons in the final state. We have predicted branching ratios using the fitted values of the CKM elements and the Wilson coefficients. The fitted results are given in Tables~\ref{tab:fit_results_combined_muon_Ci} and \ref{tab:fit_results_combined_elec_Ci}. The predicted branching ratios for different values of \(m_\chi\) and for different operator scenarios are presented in Table~\ref{tab:prediction_Bc2lnu}. For the scalar operator scenarios, the fitted Wilson coefficients are consistent with zero, and only their associated uncertainties are obtained from the fit. Consequently, since the branching ratio depends quadratically on the scalar Wilson coefficients and cannot be negative, we quote only the corresponding upper limits on the branching ratios. 

\begin{table}[t]
	\centering
	\renewcommand{\arraystretch}{1.5}
	\setlength{\tabcolsep}{5pt}
	\resizebox{0.7\textwidth}{!}{%
	\rowcolors{1}{lime!7}{teal!15}
		\begin{tabular}{|c | c | c c|}
		      \hline
            \rowcolor{purple!15!blue!10}
		      \multicolumn{4}{c}{Prediction of $\mathcal{B}(B_{c}^{+} \to \ell^{+} X_{\rm inv})$} \\ 
			\hline
			\rowcolor{blue!10!teal!30!}
			\large
			Operators & $m_{\chi}$ (GeV)  & $\mathcal{B}(B_{c} \to \mu \, X_{\rm inv})$ & $\mathcal{B}(B_{c} \to e \, X_{\rm inv})$ \\
			\hline \hline
            SM & - & $ (8.87 \pm 0.73) \times 10^{-5}$ & $(2.18 \pm 0.17) \times 10^{-9}$\\
            \hline
            
            & 0.01 & $(9.45 \pm 0.93) \times 10^{-5}$ & $(5.47 \pm 5.07) \times 10^{-8}$\\
            \cellcolor{teal!15}& 0.1 & $(9.97 \pm 1.29) \times 10^{-5}$ & $(5.23 \pm 5.06) \times 10^{-6}$  \\
            \multirow{-3}{*}{$C_{V_{1(2)L(R)}}^{\mu}$}& 1 & $ < 8.01 \times 10^{-4}$ & $< 7.05 \times 10^{-4}$ \\ \hline

            & 0.01,1  & $<1.07 \times 10^{-2}$ & $<3.45 \times 10^{-3}$ \\
            \multirow{-2}{*}{\cellcolor{lime!7}$C_{S_{1(2)L(R)}}^{\mu}$}& 1 & $<1.08 \times 10^{-2}$ & $<4.06 \times 10^{-3}$ \\
            \hline
            
		\end{tabular}
	}
	\caption{Prediction of the branching ratio for the decay $B_{c}\to \mu (e) X_{\rm inv}$ decay, with $X_{\rm inv} = \nu$ and $\chi$ for SM and NP scenarios, respectively. The predictions are done in one operator scenario with the fit values listed in Tables~\ref{tab:fit_results_combined_muon_Ci} and \ref{tab:fit_results_combined_elec_Ci}.}
\label{tab:prediction_Bc2lnu}
\end{table}

\section{Summary}\label{sec:summary}
In this work, we have studied the charge-current decay process $B \to D^{(*)} \ell X_{\rm inv}$. Here, $X_{\rm inv}$ could represent a dark sector particle in addition to an SM neutrino. We have explored the possibility of getting a light dark fermion, which might be a potential DM candidate, in the final state of the decay. So in addition to $B \to D^{(*)} \ell \bar{\nu}$, we have the decay $B \to D^{(*)} \ell \bar{\chi}$, where $\chi$ can have both right and left chirality. We have considered the most general effective interactions that will contribute to these decays and discussed the probable UV-complete DM models in which such interactions can be generated.

Although the decays $B \to D^{(*)} \ell \bar{\nu}$ and  $B \to D^{(*)} \ell \bar{\chi}$ lead to the same experimental signatures, they represent distinct processes with different kinematically allowed phase space. Hence, the new physics contributions will be added at the decay-width level instead of the amplitude level. We use available lattice inputs from the Fermilab-MILC, HPQCD, and JLQCD collaborations to compute the hadronic form factors. With all these ingrediants in hand, we have predicted the decay rate distributions of $B \to D^{(*)} \ell X_{\rm inv}$ in $w$ and different angular bins (wherever possible) and explored the potential of the resulting observables to probe different NP scenarios, the dark fermion mass $m{\chi}$, and the chirality of the NP operators.
In a couple of one-operator scenarios, we observed deviations from the SM values in several $w$ and angular bins for values of WCs $> 0.1$ (allowed by the UV complete models) and masses $m_{\chi}$ within the range $0.01 \to 1$ (in GeV). In the scenarios with scalar operators, we observe large deviations in $B \to D \, \ell \, X_{\rm inv}$ decays in a couple of $w$ bins. Also, in this mode, we observe sizeable deviations in a few bins for the left-handed tensor current operator and the vector operators. We have found that the integrated rate distribution of $B \to D \,\ell \, X_{\rm inv}$ is unable to distinguish the chiralities of scalar and vector interactions, whereas it can discriminate between left- and right-handed tensor currents, implying that if such a deviation is observed in the decay distribution, it would point towards a dark-sector fermion rather than a right-handed sterile neutrino. The observable $A_{FB}^{\ell}$ related to $B \to D \ell \, X_{\rm inv}$ serves as a sensitive probe of the dark fermion mass and can distinguish vector interactions from scalar and tensor interactions.

Similar observations are obtained for the decay $B \to D^* \, \ell \, X_{\rm inv}$ in both the $w$-binned and angular distributions. The scalar operators showed only mild sensitivity in the rate distributions, whereas the tensor operators induced the largest deviations from the SM predictions. However, unlike the $B \to D \ell X_{\rm inv}$ case, the distributions in $B\to D^* \, \ell \, X_{\rm inv}$ decay failed to significantly discriminate between the left- and right-chiral operators.    

Furthermore, we have done similar studies for the normalised rates $\Delta\Gamma/\Gamma$ for the decay modes mentioned above. We have predicted these rates in different NP scenarios and compared them with the respective SM predictions. These ratios are also sensitive to NP contributions for the values of the WCs $> 0.1$ and the ranges of $m_{\chi}$ as given above.

We have also extracted the CKM element $|V_{cb}|$, along with the new physics WCs, using the available data from Belle and Belle-II measurements on the decay rate distributions $\Delta\Gamma$ of the $B \to D^{(*)} \ell X_{\rm inv}$ decays along with the form-factor inputs from lattice QCD. 
Such a study will guide us to check the impact of any new physics on the extraction of $|V_{cb}|$. At the same time, we will get useful constraints on the new WCs. We have analysed the data available in $B \to D^{(*)} \mu X_{\rm inv}$ and $B \to D^{(*)} e X_{\rm inv}$ separately to probe possible lepton-flavour-dependent effects.
We have found that the allowed $1\sigma$ ranges of the new WCs could be of order $\mathcal{O}(10^{-1})$, which could give a sizeable effect in the decay rate distributions. Given the sizable errors $(1.2\%)$, the extracted values of $|V_{cb}|$ in NP scenarios remain consistent with those in the SM. However, the central values could shift at most by 4\%, which is significant. Therefore, any extraction of $V_{cb}$ from such decay modes must account for possible NP effects to avoid biased determination. Alternatively, a more robust approach would be to extract the NP WCs simultaneously with the CKM element.

\acknowledgments
LK would like to thank Dr.~Ipsita Ray for useful and insightful discussions. LK acknowledges financial support from IIT Gandhinagar under project grant No. OTH/R\&D/13467. SS acknowledges the Anusandhan National Research Foundation (ANRF), Govt.~of India for financial support through a research grant no. ANRF/IRG/2024/000256/PS. RS is supported by the National Natural Science Foundation of China under Grant Nos.~12475094, 12135006, and 12575099, as well as the Science and Technology Innovation Leading Talent Support Program of Henan Province under Grant No.~254000510039.

\appendix
\section{Decay Kinematics}
In the B meson rest frame, the four momenta of $ B(p)\,, D^{(*)}(k) $ and transfer momenta $ q$ are given in \cite{Becirevic:2019tpx}. The polarisation associated with the meson $ D^{*} $  and the virtual boson $ W^{*} $ in $ B$-meson rest frame is given in \cite{Hagiwara:1989cu}.
\subsection{Details of Phase Space Calculation}
The phase space for n-body decay can be written as: 
\begin{equation}\label{eq:phasespace_n-body}
	d \Phi_{n} (P; p_{1},...p_{n})=\delta^{4} (P-\sum_{i=1}^{n}p_{i}) \prod_{i=1}^{n} \frac{d^3 p_{i}}{(2 \pi)^3 2 E_{i}} 
\end{equation}
The three-body phase-space for $ B \to D \ell \nu  $ will be written as: 
\begin{equation}\label{eq:phasespace_3-body}
	d \Phi_{3} = \frac{1}{16 m_{B}  (2 \pi)^3} \, \frac{|\vec{q}|}{m_{B}} \,  \frac{ |p_{\ell}|  }{\sqrt{q^2}} \,  dq^2 \, d\cos \theta
\end{equation}
The four-body phase-space will be written as \cite{Becirevic:2019tpx}: 
\begin{equation}\label{eq:phasespace_4-body}
	d\Phi_{4} = \frac{1}{64 (2 \pi)^8} \, \frac{|\vec{p}_{D^{*}}|}{m_{B}}\,  \frac{|\vec{p}_{D}|}{m_{D\pi}}\, \frac{|\vec{p}_{\ell}|}{\sqrt{q^2}} \, dm^2_{D\pi} dq^2 \,  d\cos \theta_{D\pi} d \phi_{D\pi} \,  d\cos \theta_{D} d\phi_{D} \,  d\cos \theta_{\ell} d\phi_{\ell}\,.
\end{equation}

With the expression for the momenta: 
\begin{equation}
	|\vec{q}|  =-|\vec{p}_{D^{*}}|= -|\vec{p}_{D\pi}|  = -\frac{\lambda^{1/2}(m_{B}^2, m_{D\pi}^2, q^2)}{2 m_{B}}, \quad |\vec{p}_{\ell}| = \frac{\lambda^{1/2} (q^2, m_{\ell}^2, m_{\nu}^2) }{2 \sqrt{q^2}}\, .
\end{equation}
\subsection{Hadronic Helicity Amplitudes}\label{appndxsec:hadronic_amp}
In this section, we present the detailed mathematical expressions of the hadronic and leptonic helicity amplitudes defined in eq.~\eqref{eq:amplitude_gen}.   
The hadronic amplitudes defined in \eqref{eq:hadronic_amplitudes} can be expressed in terms of hadronic form factors. The $ \bar{B} \to D $ transition matrix elements can be written as

\begin{eqnarray}\label{eq:B2D_FF_gen}
		\langle D(k) | \bar{c} \gamma_{\mu } b | \bar{B} (p) \rangle & = & \left[ (p+k)_{\mu} - \frac{m_{B}^2 -m_{D}^2}{q^2} \right]  f_{+}(q^2) + q_{\mu} \frac{m_{B}^2 - m_{D}^2}{q^2} f_{0}(q^2)\,, \\
		\langle D(k) | \bar{c}  b | \bar{B} (p) \rangle & = & \frac{1}{m_{b} - m_{c}} q^{\mu} \langle D(k) | \bar{c} \gamma_{\mu } b | \bar{B} (p) \rangle = \frac{m_{B}^2 - m_{D}^2}{m_{b} - m_{c}} f_{0}(q^2) \,, \\
		\langle D(k) |\bar{c} \sigma_{\mu \nu} b|\bar{B}(p) \rangle & = & - i (p_{\mu} k_{\nu} - k_{\mu} p_{\nu}) \frac{2 f_{T}(q^2)}{m_{B} + m_{D}}\,, \\
		\langle D(k) |\bar{c} \sigma_{\mu \nu} \gamma_5 b|\bar{B}(p) \rangle & = &  - \frac{i}{2} \epsilon_{\mu \nu \alpha \beta} \langle D(k) |\bar{c} \sigma^{\alpha \beta} b|\bar{B}(p) \rangle = - \epsilon_{\mu \nu \alpha \beta} p^{\alpha} k^{\beta} \frac{2 f_{T} (q^2)}{m_{B} + m_{D}}\,.  
\end{eqnarray}

All the other currents will have zero contribution. Here, we get only two independent form factors. $ f_{T} (q^2) $ can be written in terms of $ f_{+}(q^2) $ and $ f_{0}(q^2) $ \cite{Sakaki:2013bfa}.

\begin{eqnarray}\label{eq:B2D_tensor_FF}
	f_{T}(q^2) = \frac{(m_{b} + m_{c})}{2 m_{B} m_{D} (1-r_{D})}&  & \bigg\{ f_{+} (q^2) \left[ m_{D} ( m_{B}^2 - m_{D}^2 + q^2  ) + m_{B}r_{D} (m_{B}^2-m_{D}^2 -q^2 )    \right] \nonumber \\
	& &- f_{0}(q^2) (m_{B}^2 - m_{D}^2)(m_{D} + r_{D} m_{B}) \bigg\} 
\end{eqnarray}
where, $ r_{D}= m_{D}/m_{B} $.

The hadronic matrix elements for $ \bar{B} \to D^{*} $ can be written as follows:
\begin{eqnarray}\label{eq:B2Dst_FF_gen}
	\langle D^{*}(k,\varepsilon ) | \bar{c} \gamma_{\mu } b | \bar{B} (p) \rangle & = & -i \epsilon_{\mu \nu \alpha \beta} \varepsilon^{*\nu} p^{\alpha} k^{\beta} \frac{2 V(q^2)}{m_{B} + m_{D^{*}}}  \,, \\
	\langle D^{*}(k,\varepsilon ) | \bar{c} \gamma_{\mu } \gamma_5 b | \bar{B} (p) \rangle & = & \varepsilon^{*}_{\mu} (m_{B} + m_{D^{*}})A_{1}(q^2) - (p+k)_{\mu} (\varepsilon^{*}.q) \frac{A_{2} (q^2)}{m_{B} + m_{D^{*}}} \nonumber \\ & & 
	- q_{\mu} (\varepsilon^{*}.q) \frac{2 m_{D^{*} } }{q^2} \left[A_{3}(q^2) -A_{0} (q^2)\right]\,, \\
	\langle D^{*}(k,\varepsilon ) | \bar{c} \gamma_5 b | \bar{B} (p) \rangle & =&  -\frac{1}{m_{b} + m_{c}}  q^{\mu} \langle D^{*}(k,\varepsilon ) | \bar{c} \gamma_{\mu } \gamma_5 b | \bar{B} (p) \rangle  \nonumber \\
	&& = -(\varepsilon^{*}.q) \frac{2 m_{D^{*}} }{m_{b} + m_{c}} A_{0} (q^2)\,, \\
	\langle D^{*}(k,\varepsilon ) |\bar{c} \sigma_{\mu \nu} b|\bar{B}(p) \rangle & =&  \epsilon_{\mu \nu \alpha \beta} \bigg[ - \varepsilon^{*\alpha} (p+k)^{\beta}T_{1}(q^2) \nonumber \\
	& & +
	\varepsilon^{*\alpha} q^{\beta} 
	\frac{m_{B}^2 - m_{D^{*}}^2}{q^2} \left(T_{1}(q^2) -T_{2}(q^2) \right) \\
	&& 2 \frac{(\varepsilon^{*}.q)}{q^2} p^{\alpha}k^{\beta} \left( T_{1}(q^2) - T_{2}(q^2) - \frac{q^2}{m_{B}^2 -m_{D^{*}}^2} T_{3}(q^2)\right)
	\bigg] \,, \nonumber \\
	\langle D^{*}(k,\varepsilon ) |\bar{c} \sigma_{\mu \nu} \gamma_5 b|\bar{B}(p) \rangle & = & -\frac{i}{2} \epsilon_{\mu \nu \alpha \beta} \langle D^{*}(k,\varepsilon ) |\bar{c} \sigma^{\alpha \beta} b|\bar{B}(p) \rangle \nonumber \\ 
	&& = i \bigg[ \left(  (p+k)_{\mu} \varepsilon^{*}_{\nu} - \varepsilon^{*}_{\mu} (p+k)_{\nu}  \right) T_{1}(q^2) \nonumber	\\
	&& + \left(\epsilon_{\mu}^{*} q_{\nu} - q_{\mu} \varepsilon_{\nu}^{*} \right) \frac{m_{B}^2 - m_{D^{*}}^2}{q^2} \left[T_{1}(q^2) - T_{2}(q^2)\right] \\
	&& + (\varepsilon^{*}.q) \left[p_{\mu}k_{\nu} - k_{\mu} p_{\nu} \right] \frac{2}{q^2} \left[T_{1}(q^2) - T_{2}(q^2) - \frac{q^2}{m_{B}^2 - m_{D^{*}}^2} T_{3}(q^2)  \right] \Bigg]. \nonumber
\end{eqnarray}

\paragraph{\underline{Helicity Amplitudes for $ B \to D $}: } 
Using the $ \bar{B} \to D $ transition matrix elements defined in eq.~\eqref{eq:B2D_FF_gen}, we obtain the following expressions of the hadronic amplitudes from eq.~\eqref{eq:hadronic_amplitudes}:
	\begin{eqnarray}
	& H_{V,0}^{s} (q^2)& \equiv H_{V_{1},0}^{s} (q^2) = H_{V_{2},0}^{s} (q^2) = \sqrt{\frac{\lambda_{D}(q^2)}{q^2} } f_{+}(q^2) \\
	& H_{V,t}^{s}(q^2) & \equiv H_{V_{1},t}^{s} (q^2) = H_{V_{2},t}^{s} (q^2) = \frac{m_{B}^2 - m_{D}^2}{\sqrt{q^2}} f_{0}(q^2) \\
	& H_{S}^{s}(q^2) & \equiv H_{S_{1}}^{s} (q^2) = H_{S_{2}}^{s} (q^2) \simeq \frac{m_{B}^2 - m_{D}^2}{m_{b}-m_{c}} f_{0}(q^2) \\
	& H_{T}^{s} (q^2) & \equiv H_{T,+-}^{s} (q^2) = H_{T,0t}^{s} (q^2) = -\frac{\sqrt{\lambda_{D}(q^2)}}{m_{B}+m_{D}} f_{T}(q^2)
	\end{eqnarray}

\paragraph{\underline{Helicity Amplitudes for $ B \to D^{*} $}:} 
Using the polarization vector for $D^*$ eq.~\eqref{eq:pol_Dstmeson1} and off-shell w boson polarization vector eq.~\eqref{eq:pol_wboson1}, we derived the hadronic helicity amplitude using definition in eq.~\eqref{eq:hadronic_amplitudes} that we agreeed with~\cite{Hagiwara:1989cu, Sakaki:2013bfa}.
The non-zero hadronic amplitudes are given below: 
\begin{align}
	H_{V,\pm}(q^2)  \equiv H_{V_{1},\pm}^{\pm} (q^2) &= - H_{V_{2},\mp}^{\mp}(q^2)  =  (m_{B}+m_{D^{*}})A_{1}(q^2) \mp \frac{\sqrt{\lambda_{D^{*}}(q^2)}}{m_{B}+m_{D^{*}}}V(q^2)\,,  \\
	H_{V,0}(q^2)  \equiv H_{V_{1},0}^{0}(q^2) & = H_{V_{2,0}}^{0}(q^2)  = \frac{m_{B}+m_{D^{*}}}{2 m_{D^{*}} \sqrt{q^2}} \big[-(m_{B}^2 - m_{D^{*}}^2 - q^2)A_{1}(q^2) \nonumber \\
	&{}\ \ \ \ \ \ \ \ \ \quad\quad\quad\quad\quad\quad\quad\quad\quad + \frac{\lambda_{D^{*}}}{(m_{B} + m_{D^{*}})^2}A_{2}(q^2) \big]\,, \\
	H_{V,t} (q^2) \equiv H_{V_{1},t}^{0}(q^2)& =  - H_{V_{2},t}^{0}(q^2)  \simeq  -  \frac{\lambda_{D^{*}} (q^2)}{q^2} A_{0}(q^2)\,, \\
	H_{S}(q^2) \equiv H_{S_{1}}^{0}(q^2) & = - H_{S_{2}}^{0}(q^2)  \simeq - \frac{\sqrt{ \lambda_{D^{*}}(q^2)}}{m_{b}+m_{c}}A_{0}(q^2),\\
	H_{T,\pm}(q^2) \equiv \pm H_{T,\pm t}^{\pm }(q^2)& =  \frac{1}{\sqrt{2}} \left[ \pm (m_{B}^2 - m_{D^{*}}^2)T_{2}(q^2) + \sqrt{\lambda_{D^{*}}(q^2) } T_{1} (q^2) \right], 
	\end{align}
	\begin{align}	
	H_{T,0}(q^2) \equiv H_{T,+-}^{0}(q^2) = H_{T,0t}^{0}(q^2) & = \frac{1}{2 m_{D^{*}}} \big[  -(m_{B}^2 + 3 m_{D^{*}}^2 - q^2) T_{2}(q^2)   \nonumber  \\
	&{}\quad\quad\ \ \ \ \ \ \ + \frac{\lambda_{D^{*}}(q^2)}{m_{B}^2 - m_{D^{*}}^2} T_{3}(q^2)	 \big] .
	\end{align}
\subsection{Leptonic Helicity Amplitudes}\label{appndxsec:leptonic_helicity}
In this section, we present the detailed mathematical expressions of the leptonic helicity amplitudes defined in eq.~\eqref{eq:amplitude_gen}. We will write the expression of leptonic amplitude for including the dark-fermions, from which we can get the amplitudes for massless neutrinos by setting $ m_{\chi} \to 0$. The Dirac spinors used to calculate leptonic helicity amplitudes in \eqref{eq:leptonic_helicity_amp} are given as : 
\begin{equation}\begin{split}
		u(p)^+ = \begin{pmatrix}
			\sqrt{E_{\ell} + m_{\ell}}~ \cos \frac{\theta_{\ell}}{2} \\
			\sqrt{E_{\ell} + m_{\ell}}~ \sin \frac{\theta_{\ell}}{2}~ e^{i\phi} \\
			\sqrt{E_{\ell} - m_{\ell}}~ \cos \frac{\theta_{\ell}}{2} \\
			\sqrt{E_{\ell} - m_{\ell}} ~\sin \frac{\theta_{\ell}}{2}~ e^{i\phi} 
		\end{pmatrix}, \quad 
		u(p)^- = \begin{pmatrix}
			-\sqrt{E_{\ell} + m_{\ell}}~  \sin \frac{ \theta_{\ell}}{2}~ e^{-i\phi} \\
			\sqrt{E_{\ell} + m_{\ell}}~ \cos \frac{\theta_{\ell}}{2} \\
			\sqrt{E_{\ell} - m_{\ell}}~ \sin \frac{\theta_{\ell}}{2} ~e^{-i\phi} \\
			-\sqrt{E_{\ell} - m_{\ell}}~\cos\frac{\theta_{\ell}}{2} 
		\end{pmatrix},  \\
		v(p)^+ = \begin{pmatrix} 
			\sqrt{E_{X} - m_{X}}~  \cos \frac{\theta_{\ell}}{2}~ e^{-i \phi} \\
			\sqrt{E_{X} - m_{X}}~ \sin \frac{\theta_{\ell}}{2} \\
			-\sqrt{E_{X} + m_{X}}~ \cos \frac{ \theta_{ \ell}}{2} ~e^{-i \phi } \\
			-\sqrt{E_{X} + m_{X}}~\sin \frac{\theta_{\ell}}{2} 
		\end{pmatrix}, \quad 
		v(p)^- = \begin{pmatrix} 
			\sqrt{E_{X} - m_{X}}~ \sin\frac{ \theta_{\ell}}{2} \\
			- \sqrt{E_{X} - m_{X}}~ \cos \frac{\theta_{\ell}}{2}~ e^{i \phi} \\
			\sqrt{E_{X} + m_{X}}~ \sin \frac{ \theta_{\ell}}{2} \\
			-\sqrt{E_{X} + m_{X}} ~\cos \frac{\theta_{\ell}}{2}~ e^{i \phi )} 
		\end{pmatrix}
	\end{split}
\end{equation}
Here, $X= \{\nu, \,\chi \}$ for SM and NP respectively. 
Using the above leptonic helicity spinors and polarization vectors for the \text{$w^*$} boson eq.~\eqref{eq:pol_wboson2} and $g_{\mu\nu}$, we obtained the leptonic helicity amplitudes using eq.~\eqref{eq:leptonic_helicity_amp} which are presented in the following items:
\begin{itemize}
	\item The vector current with left-handed massive neutrino (dark-fermion) is given by:
	\begin{subequations}
		\begin{eqnarray}\label{eq:leptonic_vector_left}
			L^{V,L}_{\frac{1}{2},\frac{1}{2}} & = & \frac{1}{2} \, e^{-i \phi} \left\{ \cos \theta_{\ell} \,,  e^{-i\phi} \frac{\sin \theta_{\ell}}{\sqrt{2}}\,, -e^{i\phi}\frac{\sin \theta_{\ell}}{\sqrt{2}}\,, -1 \right\} X_{+-} \,, \\
			L^{V,L}_{\frac{1}{2},-\frac{1}{2}} & = & \frac{1}{2}\left\{ \sin \theta_{\ell} \,, e^{-i\phi}  \frac{(1-\cos \theta_{\ell} )}{\sqrt{2}} , e^{i \phi} \frac{(1+\cos \theta_{\ell} )}{\sqrt{2}}   \,, 0 \right\} X_{--} \,, \\
			L^{V,L}_{-\frac{1}{2}, \frac{1}{2}} & = & \frac{1}{2}\, \left\{ \sin \theta_{\ell} \,, -e^{-i\phi} \frac{(1+\cos \theta_{\ell})}{\sqrt{2}}\,, -e^{i\phi} \frac{(1-\cos \theta_{\ell})}{\sqrt{2}} \,, 0 \right\} X_{++} \,, \\
			L^{V,L}_{-\frac{1}{2}, -\frac{1}{2}} & = & \frac{1}{2} e^{i\phi} \left\{ -\cos \theta_{\ell} \,,- e^{- i\phi} \frac{\sin \theta_{\ell}}{\sqrt{2}}  \,,  e^{i\phi}  \frac{\sin \theta_{\ell}}{\sqrt{2}} \,, -1  \right\}X_{-+}\,.
		\end{eqnarray}
	\end{subequations}
\item The vector current with right-handed massive neutrino (dark-fermion) is given by:
\begin{subequations}
	\begin{eqnarray}\label{eq:leptonic_vector_right}
		L^{V,R}_{\frac{1}{2},\frac{1}{2}} & = & \frac{1}{2} \, e^{-i \phi} \left\{ \cos \theta_{\ell} \,,  e^{-i\phi} \frac{\sin \theta_{\ell}}{\sqrt{2}}\,, -e^{i\phi}\frac{\sin \theta_{\ell}}{\sqrt{2}}\,, 1 \right\} X_{-+}\,, \\
		L^{V,R}_{\frac{1}{2},-\frac{1}{2}} & = & \frac{1}{2}\left\{ \sin \theta_{\ell} \,, e^{-i\phi}  \frac{(1-\cos \theta_{\ell} )}{\sqrt{2}} \,,e^{i \phi} \frac{(1+\cos \theta_{\ell} )}{\sqrt{2}}   \,, 0 \right\} X_{++}\,, \\
		L^{V,R}_{-\frac{1}{2}, \frac{1}{2}} & = & \frac{1}{2}\, \left\{ \sin \theta_{\ell} \,, -e^{-i\phi} \frac{(1+\cos \theta_{\ell})}{\sqrt{2}}\,, -e^{i\phi} \frac{(1-\cos \theta_{\ell})}{\sqrt{2}} \,, 0 \right\} X_{--} \,,\\
		L^{V,R}_{-\frac{1}{2}, -\frac{1}{2}} & = & \frac{1}{2} e^{i\phi} \left\{ -\cos \theta_{\ell} \,, -e^{- i\phi} \frac{\sin \theta_{\ell}}{\sqrt{2}}  \,,  e^{i\phi}  \frac{\sin \theta_{\ell}}{\sqrt{2}} \,, 1  \right\}X_{+-}\,.
	\end{eqnarray}
\end{subequations}

\item The scalar current for a left-handed massive neutrino can be given by: 
\begin{subequations}
	\begin{eqnarray}
		&L^{S,L}_{\frac{1}{2},\frac{1}{2}}  = & \frac{1}{2}\,e^{-i\phi}\, X_{++} \,, \\
		& L^{S,L}_{\frac{1}{2},-\frac{1}{2}} = & 0\,,  \\
		& L^{S,L}_{-\frac{1}{2},\frac{1}{2}} = & 0 \,, \\
		& L^{S,L}_{-\frac{1}{2},-\frac{1}{2}}  = & \frac{1}{2}\, e^{i\phi}\, X_{--} \,.
	\end{eqnarray}
\end{subequations}
\item The scalar current for a right-handed massive neutrino can be given by: 
\begin{subequations}
	\begin{eqnarray}
		& L^{S,R}_{\frac{1}{2},\frac{1}{2}}  = & - \frac{1}{2}\,e^{-i\phi}\, X_{--} \,, \\
		& L^{S,R}_{\frac{1}{2},-\frac{1}{2}} = & 0 \,,  \\
		& L^{S,R}_{-\frac{1}{2},\frac{1}{2}} = & 0 \,,  \\
		& L^{S,R}_{-\frac{1}{2},-\frac{1}{2}}  = & -\frac{1}{2}\, e^{i\phi}\, X_{++} \,.
	\end{eqnarray}
\end{subequations}
\item The tensor current for a left-handed massive neutrino can be given by: 
\begin{subequations}
	\begin{eqnarray}
		L_{\frac{1}{2},\frac{1}{2},t\lambda'}^{T,L} & = & -\frac{1}{2} e^{-i\phi} \left\{  \cos \theta_{\ell} \,, e^{-i\phi} \frac{\sin \theta_{\ell}}{\sqrt{2}} \,, -e^{i\phi} \frac{\sin \theta_{\ell}}{\sqrt{2}} \,, 0 \right\} X_{++} \,,\\
		L_{\frac{1}{2},-\frac{1}{2},t\lambda'}^{T,L} & = & -\frac{1}{2}  \left\{   \sin \theta_{\ell} \,, e^{-i\phi}\, \frac{(1-\cos \theta_{\ell})}{\sqrt{2}} \,, e^{i\phi} \frac{(1+\cos \theta_{\ell})}{\sqrt{2}} \,,0 \right\} X_{-+} \,,\\
		L_{-\frac{1}{2},\frac{1}{2},t\lambda'}^{T,L} & = & -\frac{1}{2}  \left\{  \sin \theta_{\ell} \,, -e^{-i\phi}\, \frac{(1+\cos \theta_{\ell})}{\sqrt{2}} \,, -e^{i\phi} \frac{(1-\cos \theta_{\ell})}{\sqrt{2}} \,, 0 \right\} X_{+-} \,,\\
		L_{-\frac{1}{2},-\frac{1}{2},t\lambda'}^{T,L} & = & -\frac{1}{2} e^{i\phi} \left\{  -\cos\theta_{\ell} \,,    -e^{-i\phi} \frac{\sin \theta}{\sqrt{2}} \,,  e^{i\phi} \frac{\sin \theta_{\ell} }{\sqrt{2}} \,, 0 \right\} X_{--} \,,
	\end{eqnarray}
\end{subequations}
\begin{subequations}
	\begin{eqnarray}
		\nonumber \\
		L_{\frac{1}{2},\frac{1}{2},0\lambda'}^{T,L} & = & -\frac{1} {2} e^{-i\phi} \left\{ 0, -\, e^{-i\phi} \frac{\sin \theta_{\ell}}{\sqrt{2}} \,, -e^{i\phi} \frac{ \sin \theta_{\ell}}{\sqrt{2}}, -\cos \theta_{\ell} \,, \right\} X_{++} \,,\\
		L_{\frac{1}{2},-\frac{1}{2},0\lambda'}^{T,L} & = & -\frac{1}{2}  \left\{0, -e^{-i\phi} \frac{1-\cos \theta_{\ell}}{\sqrt{2}} \,, e^{i\phi} \frac{(1+\cos \theta_{\ell})}{\sqrt{2}}, -\sin \theta_{\ell} \,, \right\} X_{-+} \,,\\
		L_{-\frac{1}{2},\frac{1}{2},0\lambda'}^{T,L} & = & -\frac{1}{2}  \left\{ 0, e^{-i \phi} \frac{(1+\cos \theta_{\ell}) }{\sqrt{2}} \,, -e^{i\phi} \frac{(1-\cos \theta_{\ell})}{\sqrt{2}}\,, - \sin \theta_{\ell} \right\} X_{+-} \,,\\
		L_{-\frac{1}{2},-\frac{1}{2},0\lambda'}^{T,L} & = & - \frac{1}{2} e^{i\phi} \left\{0 \,, e^{-i\phi} \frac{\sin \theta_{\ell} }{\sqrt{2}} \,, e^{i\phi} \frac{\sin \theta_{\ell}}{\sqrt{2}}, \, \cos \theta_{\ell}  \right\} X_{--} \,,
	\end{eqnarray}
\end{subequations}
\begin{subequations}
	\begin{eqnarray}
		L_{\frac{1}{2},\frac{1}{2},+\lambda'}^{T,L} & = & -\frac{1}{2} e^{-i\phi} \left\{ e^{-i\phi} \frac{\sin \theta_{\ell}}{\sqrt{2}} \,, 0 \,,  \cos\theta_{\ell} \,, -e^{-i\phi} \frac{\sin \theta_{\ell} }{\sqrt{2}}\right\} X_{++} \,,\\
		L_{\frac{1}{2},-\frac{1}{2},+\lambda'}^{T,L} & = & -\frac{1}{2} \left\{ e^{-i\phi} \frac{ (1-\cos \theta_{\ell})}{\sqrt{2}} \,, 0 \,,  \sin \theta_{\ell} \,,- e^{-i\phi} \frac{(1-\cos \theta_{\ell} )}{\sqrt{2}} \right\} X_{-+} \,,\\
		L_{-\frac{1}{2},\frac{1}{2},+\lambda'}^{T,L} & = & -\frac{1}{2} \left\{   -e^{-i\phi}\frac{(1+\cos \theta_{\ell})}{\sqrt{2}} \,, 0\,, \sin \theta_{\ell} \,,   e^{-i\phi}\frac{(1+\cos \theta_{\ell})}{\sqrt{2}}   \right\} X_{+-} \,,\\
		L_{-\frac{1}{2},-\frac{1}{2},+\lambda'}^{T,L} & = & -\frac{1}{2} e^{i\phi} \left\{- e^{-i\phi}\frac{\sin \theta_{\ell}}{\sqrt{2}} \,, 0 \,, - \cos \theta_{\ell} \,, e^{-i\phi} \frac{\sin \theta_{\ell}}{\sqrt{2}} \right\} X_{--} \,,
	\end{eqnarray}
\end{subequations}
\begin{subequations}
	\begin{eqnarray}
		L_{\frac{1}{2},\frac{1}{2},-\lambda'}^{T,L} & = & -\frac{1}{2} e^{-i\phi} \left\{ e^{i\phi} \frac{\sin \theta_{\ell}}{\sqrt{2}} \,, - \cos \theta_{\ell} \,, 0 \,, e^{i\phi} \frac{\sin \theta_{\ell}}{\sqrt{2}}   \right\} X_{++} \,,\\
		L_{\frac{1}{2},-\frac{1}{2},-\lambda'}^{T,L} & = & -\frac{1}{2}  \left\{  - e^{i\phi} \frac{(1+\cos \theta_{\ell})}{\sqrt{2}} \,, -\sin \theta_{\ell} \,, 0 \,,  -e^{i\phi} \frac{(1+\cos \theta_{\ell})}{\sqrt{2}}  \right\} X_{-+} \,,\\
		L_{-\frac{1}{2},\frac{1}{2},-\lambda'}^{T,L} & = & -\frac{1}{2} \left\{ e^{i\phi}\frac{(1-\cos \theta_{\ell} )}{\sqrt{2}} \,, -\sin \theta_{\ell} \,, 0 \,, e^{i\phi} \frac{(1-\cos \theta_{\ell})}{\sqrt{{2}}} \right\} X_{+-} \,,\\
		L_{-\frac{1}{2},-\frac{1}{2},-\lambda'}^{T,L} & = & -\frac{1}{2} e^{i\phi} \left\{ -e^{i\phi} \frac{\sin \theta_{\ell}}{\sqrt{2}} \,, \cos \theta_{\ell} \,, 0 \,, - e^{i\phi} \frac{\sin \theta_{\ell}}{\sqrt{2}}  \right\} X_{--} \,.
	\end{eqnarray}
\end{subequations}
The tensor current for a right-handed massive neutrino is given by: 
\begin{subequations}
	\begin{eqnarray}
		L_{\frac{1}{2},\frac{1}{2},t\lambda'}^{T,R} & = & -\frac{1}{2}  e^{-i \phi}\left\{ \cos \theta_{\ell} \,, e^{-i\phi} \frac{\sin \theta_{\ell}}{\sqrt{2}} \,,  -e^{i\phi} \frac{\sin \theta_{\ell} }{\sqrt{2}}\,, 0 \right\} X_{--} \,,\\
		L_{\frac{1}{2},-\frac{1}{2},t\lambda'}^{T,R} & = & -\frac{1}{2}  \left\{  \sin \theta_{\ell} \,, e^{-i \phi} \frac{(1-\cos \theta_{\ell})}{\sqrt{2}} \,, e^{i\phi} \frac{(1+\cos \theta_{\ell} )}{\sqrt{2}} \,, 0  \right\} X_{+-} \,,\\
		L_{-\frac{1}{2},\frac{1}{2},t\lambda'}^{T,R} & = & -\frac{1}{2}  \left\{ \sin \theta_{\ell} \,, -e^{-i\phi} \frac{(1 + \cos \theta_{\ell} )}{\sqrt{2}}\,, -e^{i\phi} \frac{(1-\cos \theta_{\ell})}{\sqrt{2}} \,, 0    \right\} X_{-+} \,,\\
		L_{-\frac{1}{2},-\frac{1}{2},t\lambda'}^{T,R} & = & -\frac{1}{2} e^{i \phi}\left\{  -\cos\theta_{\ell} \,,    -e^{-i\phi} \frac{\sin \theta_{\ell}}{\sqrt{2}} \,,  e^{i\phi} \frac{\sin \theta_{\ell} }{\sqrt{2}} \,, 0  \right\} X_{++} \,,
	\end{eqnarray}
\end{subequations}
\begin{subequations}
	\begin{eqnarray}
		L_{\frac{1}{2},\frac{1}{2},0\lambda'}^{T,R} & = & -\frac{1}{2} e^{-i\phi} \left\{ 0 \,,e^{-i\phi} \frac{\sin \theta_{\ell}}{\sqrt{2}} \,, e^{i\phi} \frac{\sin \theta_{\ell}}{\sqrt{2}} \,, -\cos \theta_{\ell}  \right\} X_{--} \,,\\
		L_{\frac{1}{2},-\frac{1}{2},0\lambda'}^{T,R} & = & -\frac{1}{2}  \left\{  0 \,,  e^{-i\phi} \frac{(1-\cos \theta_{\ell})}{\sqrt{2}} \,, -e^{i\phi} \frac{(1+\cos \theta)}{\sqrt{2}} \,, -\sin \theta_{\ell} \right\} X_{+-} \,,\\
		L_{-\frac{1}{2},\frac{1}{2},0\lambda'}^{T,R} & = & -\frac{1}{2}  \left\{  0 \,, -e^{-i\phi} \frac{(1+\cos \theta_{\ell})}{\sqrt{2}} \,,  e^{i\phi} \frac{(1-\cos \theta_{\ell} )}{\sqrt{2}} \,, -\sin \theta_{\ell}   \right\} X_{-+} \,,\\
		L_{-\frac{1}{2},-\frac{1}{2},0\lambda'}^{T,R} & = & -\frac{1}{2} e^{i\phi} \left\{  0 \,, -e^{-i\phi} \frac{\sin \theta_{\ell} }{\sqrt{2}} \,, -e^{i\phi} \frac{\sin \theta_{\ell}}{\sqrt{2}} \,, \cos \theta_{\ell} \right\} X_{++} \,,
	\end{eqnarray}
\end{subequations}
\begin{subequations}
	\begin{eqnarray}
		L_{\frac{1}{2},\frac{1}{2},+\lambda'}^{T,R} & = & -\frac{1}{2} e^{-i\phi} \left\{ -e^{-i\phi}\, \frac{\sin \theta_{\ell}}{\sqrt{2}} \,, 0\,, -\cos \theta_{\ell} \,, -e^{-i\phi} \,\frac{\sin \theta_{\ell}}{\sqrt{2}} \right\} X_{--} \,,\\
		L_{\frac{1}{2},-\frac{1}{2},+\lambda'}^{T,R} & = & -\frac{1}{2}  \left\{ -e^{-i\phi} \frac{(1-\cos \theta_{\ell})}{\sqrt{2}} \,, 0 \,,  -\sin \theta_{\ell}  \,,  -e^{-i\phi} \frac{(1-\cos \theta_{\ell})}{\sqrt{2}}  \right\} X_{+-} \,,\\
		L_{-\frac{1}{2},\frac{1}{2},+\lambda'}^{T,R} & = & - \frac{1}{2}   \left\{ e^{-i\phi} \frac{(1+\cos \theta_{\ell})}{\sqrt{2}} \,, 0 \,, - \sin \theta_{\ell}  \,,  e^{-i\phi} \frac{(1+\cos \theta_{\ell})}{\sqrt{2}} \right\} X_{-+} \,,\\
		L_{-\frac{1}{2},-\frac{1}{2},+\lambda'}^{T,R} & = & -\frac{1}{2} e^{i \phi} \left\{ e^{-i \phi} \frac{\sin \theta_{\ell}}{\sqrt{2}} \,, 0 \,, \cos \theta_{\ell}  \,, e^{-i \phi} \frac{\sin \theta_{\ell}}{\sqrt{2}} \right\} X_{++} \,,
	\end{eqnarray}
\end{subequations}
\begin{subequations}
	\begin{eqnarray}
		L_{\frac{1}{2},\frac{1}{2},-\lambda'}^{T,R} & = & -\frac{1}{2} e^{-i \phi} \left\{ -e^{i \phi} \frac{\sin \theta_{\ell}}{\sqrt{2}} \,,  \cos \theta_{\ell} \,, 0 \,,  e^{i \phi} \frac{\sin \theta_{\ell}}{\sqrt{2}}  \right\} X_{--} \,,\\
		L_{\frac{1}{2},-\frac{1}{2},-\lambda'}^{T,R} & = & - \frac{1}{2} \left\{ e^{i\phi}  \frac{(1+\cos \theta_{\ell})}{\sqrt{2}} \,,   \sin \theta_{\ell} \,, 0 \,, - e^{i\phi} \frac{(1+\cos \theta_{\ell})}{\sqrt{2}}   \right\} X_{+-} \,,\\
		L_{-\frac{1}{2},\frac{1}{2},-\lambda'}^{T,R} & = & - \frac{1}{2}   \left\{ -e^{i\phi} \frac{(1-\cos \theta_{\ell})}{\sqrt{2}} \,,  \sin \theta_{\ell} \,, 0 \,, e^{i\phi} \frac{(1-\cos \theta_{\ell})}{\sqrt{2}}\right\} X_{-+} \,,\\
		L_{-\frac{1}{2},-\frac{1}{2},-\lambda'}^{T,R} & = & -\frac{1}{2} e^{i\phi}  \left\{ e^{i\phi} \frac{\sin \theta_{\ell} }{\sqrt{2}} \,, -\cos \theta_{\ell} \,, 0 \,, - e^{i\phi} \frac{\sin \theta_{\ell}}{\sqrt{2}} \right\} X_{++} \,.
	\end{eqnarray}
\end{subequations}
\end{itemize}
The leptonic tensor current will change sign under the exchange of polarisation index, i.e.,
\begin{equation}\label{eq:leptonic_tensor_relation}
	L_{\lambda_{\ell}, \lambda_{\nu}, \lambda \lambda'}^{T,L(R)} = - L_{\lambda_{\ell}, \lambda_{\nu}, \lambda' \lambda}^{T,L(R)} \,,
\end{equation}
where, $ \lambda_{\epsilon} = \{0,+,-,t\} $, and we have taken
{\scriptsize 
	\begin{eqnarray}\label{eq:kinetic_func}
		X_{\pm \pm } & = & \left( \sqrt{\frac{(\sqrt{q^2}-m_{\chi}^2)^2-m_{\ell}^2}{\sqrt{q^2}} } \pm  \sqrt{\frac{(\sqrt{q^2}+m_{\chi}^2)^2-m_{\ell}^2}{\sqrt{q^2}} }\right) \left(  \sqrt{\frac{(\sqrt{q^2}-m_{\ell}^2)^2-m_{\chi}^2}{\sqrt{q^2}} } \pm  \sqrt{\frac{(\sqrt{q^2}+m_{\ell}^2)^2-m_{\chi}^2}{\sqrt{q^2}} } \right)  \,. \nonumber \\ 
\end{eqnarray}}
Our findings for leptonic helicity amplitude differ from ref.~\cite{Mandal:2020htr, Datta:2022czw} for right-handed cases with a relative $-\text{ve}$ sign. They took a different $g_{\mu\nu}$ convention for leptonic helicity amplitude within the SM and took a different convention for the NP amplitude calculation. So, this is the primary reason why our results differ with a relative $-\text{ve}$ sign. But final results for total decay distribution evaluated to be the same in the massless limit for neutrinos ($m_\chi \to 0$) as the total amplitude in eq.~\eqref{eq:amplitude_gen} is the convolution of hadronic and leptonic helicity with matric tensor. So, the extra minus sign absorbs accordingly while calculating the decay amplitude, resulting in a consistent result.
\section{Details on Multi-Higgs Doublet Model} \label{appndx:model_3HDM}
Lets consider we have three scalar doublets $\Phi_{1}$, with $i = 1, 2 , 3$. The field $\Phi_{1}$ and $\Phi_{2}$ are even under the $\mathbb{Z}_{2}$ symmetry while $\Phi_{3}$ is even under it. The soft breaking of $\mathbb{Z}_{2}$ symmetry allow the fields $\Phi_{2}$ and $\Phi_3$ mix. The general scalar potential will be given by:
\begin{align*}
V (\Phi_1, \Phi_2, \Phi_3) &= -\mu_1^2(\Phi_1^\dagger\Phi_1)-\mu_2^2(\Phi_2^\dagger\Phi_2)-\mu_3^2(\Phi_3^\dagger\Phi_3) \\
& \quad + \Big[ M_{12}^2 (\Phi_{1}^{\dagger}\Phi_{2}) + \mu_{23}^{2}(\Phi_{2}^{\dagger}\Phi_{3}) + \text{h.c.} \Big] \\
&\quad + \lambda_{1}(\Phi_{1}^{\dagger}\Phi_{1})^{2} + \lambda_{2}(\Phi_{2}^{\dagger}\Phi_{2})^{2} + \lambda_{3}(\Phi_{3}^{\dagger}\Phi_{3})^{2} \nonumber \\
&\qquad + \lambda_1(\Phi_1^\dagger\Phi_1)^2+\lambda_2(\Phi_2^\dagger\Phi_2)^2+\lambda_3(\Phi_3^\dagger\Phi_3)^2 \\
&\qquad + \lambda_4(\Phi_1^\dagger\Phi_2)(\Phi_2^\dagger\Phi_1)
+ \lambda_5(\Phi_1^\dagger\Phi_3)(\Phi_3^\dagger\Phi_1)
+ \lambda_6(\Phi_2^\dagger\Phi_3)(\Phi_3^\dagger\Phi_2) \\
&\qquad + \big[\,\lambda_7(\Phi_1^\dagger\Phi_2)^2
+ \lambda_8(\Phi_1^\dagger\Phi_3)^2
+ \lambda_9(\Phi_2^\dagger\Phi_3)^2 + \text{h.c.}\,\big].
\end{align*}
Here, the fields $\Phi_{2}$ is odd and $\Phi_{1}$ is even under a discrete symmetry $\mathbb{Z}_{2}$. We allow soft-breaking for the terms for $\Phi_2$ and $\Phi_3$, this explicitly breaks $\mathbb{Z}_2$ but only by a dimension-2 operator, and is therefore called a \emph{soft} $\mathbb{Z}_2$-breaking term; it produces a tree-level $\Phi_2$--$\Phi_3$ mixing proportional to \(\mu_{13}^2\).
The fields $\Phi_{1}$ and $\Phi_2$ can also mix. We assume the mixing angle to be very small. 
The mixing matrix will read as:
\begin{equation}
\mathcal{M}^2 = \begin{pmatrix}
\mu_{1}^2 & M_{12}^2 & 0 \\
M_{12}^2 & \mu_{2}^2 &\mu_{23}^2 \\
0 & \mu_{23}^2 & \mu_{3}^2
\end{pmatrix},
\end{equation}
We assume $\Phi_1$ is the SM-like doublet and carries the dominant vev $v_1\simeq v\approx 246\,$GeV. The vevs for other doublets are negligibly small. We work with three $SU(2)_L$ scalar doublets with hypercharge $+1/2$, has the form:
\begin{equation}
\Phi_i = \begin{pmatrix} \phi_i^+ \\
\tfrac{1}{\sqrt2}\left(v_i + \rho_i + i\eta_i\right)\end{pmatrix},\qquad i=1,2,3.
\end{equation}
The hierarchy in vev is given by: $v_{1} >> v_{2,3} \sim 0$. 
Now, different scalars (charges, CP even, odd) will have different rotation matrices. Since here, we are only interested in the charged scalar interaction, we will write those explicitly. Note that our mixing prescription is:
$\Phi_{1}$ and $\Phi_2$ has very feeble mixing. $\Phi_2, \Phi_3$ have small but moderate mixing. $\Phi_1$ does not mix with $\Phi_3$. The mass eigenstate charged scalars can be written as:
\begin{align}
\begin{pmatrix}
G^+ \\[4pt] H_1^+ \\[4pt] H_2^+
\end{pmatrix}
&= R_\pm
\begin{pmatrix}
\phi_1^+ \\[4pt] \phi_2^+ \\[4pt] \phi_3^+
\end{pmatrix},
\\[6pt]
R_\pm &= R_{23}(\theta_{23})\,R_{12}(\theta_{12}),
\end{align}
with the forms of the rotational matrix:
\begin{align}
R_{12}(\theta_{12}) =
\begin{pmatrix}
\cos\theta_{12} & \sin\theta_{12} & 0\\[6pt]
-\sin\theta_{12} & \cos\theta_{12} & 0\\[6pt]
0 & 0 & 1
\end{pmatrix},\ \ \ \ 
R_{23}(\theta_{23}) =
\begin{pmatrix}
1 & 0 & 0\\[6pt]
0 & \cos\theta_{23} & \sin\theta_{23}\\[6pt]
0 & -\sin\theta_{23} & \cos\theta_{23}
\end{pmatrix}.
\end{align}

\begin{align}
R_\pm \;=\; R_{23}(\theta_{23})\,R_{12}(\theta_{12})
&=
\begin{pmatrix}
\cos\theta_{12} & \sin\theta_{12} & 0\\[6pt]
-\cos\theta_{23}\sin\theta_{12} & \cos\theta_{23}\cos\theta_{12} & \sin\theta_{23}\\[6pt]
\sin\theta_{23}\sin\theta_{12} & -\sin\theta_{23}\cos\theta_{12} & \cos\theta_{23}
\end{pmatrix}.
\end{align}
In the small mixing approximation, the rotation matrix will be:
\begin{align}
R_\pm &\simeq
\begin{pmatrix}
1 & \theta_{12} & 0\\[6pt]
-\theta_{12} & 1 & \theta_{23}\\[6pt]
\theta_{12} \, \theta_{23} & -\theta_{23} & 1
\end{pmatrix} \,.
\end{align}
Again in the small mixing limit of $\Phi_2$ and $\Phi_3$, we can write $\cos\theta_{23} \to 1$ and $\sin \theta_{23} \to \theta_{23}$. 
Depending on the charge assignment on the quark and lepton singlets and doublets, we get different types of three-Higgs-doublet models. 
\begin{table}[h!]
\centering
\begin{tabular}{c|ccccccc}
  & $\Phi_{2}$ & $\Phi_{3}$ & $q_{L}/ \ell_{L}$ & $u_{R}$ & $d_{R}$ & $e_{R}$ & $\Psi_{R}$ \\
\hline
 1st & $-$ & $+$ & $+$ & $-$ & $-$ & $+$ & $+$ \\
2nd & $-$ & $+$ & $+$ & $-$ & $+$ & $+$ & $+$ \\
\hline
\end{tabular}
\caption{$\mathbb{Z}_{2}$ charge assignment to the scalar doublets and the SM fermion doublets and singlets. }\label{tab:3HDM_types}
\end{table}
In our case, we consider the doublet $\Phi_{1}$ interacts with all lepton and quark singlets and doublets. Depending on the charge of the $\Phi_2$ and $\Phi_3$, there will be different interaction Lagrangians. For example, if the doublets have a charge assignment of the 1st column of Table~\ref{tab:3HDM_types}, we get the following interaction:
The corresponding Yukawa interaction will be:
\begin{align}
\mathcal{L}_{Y} = \;&
- Y_{U}^{(1)} \overline{q}_{L}  \tilde{\Phi}_{1}' u_{R} - Y_{D}^{(1)} \overline{q}_{L}  \Phi_{1}' d_{R} -  Y_{e}^{(1)} \overline{\ell}_{L}  \Phi_{1}' e_{R}
 \nonumber \\
&- Y_{U}^{(2)}  \overline{q}_{L} \tilde{\Phi}_{2}' u_{R}
-  Y_{D}^{(2)}  \overline{q}_{L} \Phi_{2}' d_{R} \nonumber \\
& -  Y_{e}^{(3)} \overline{\ell}_{L} \Phi_{3} e_{R} -  Y_{\nu}^{(3)} \overline{\ell}_{L} \Phi_{3} \Psi_{R} 
+ \text{h.c.}
\end{align}
Final interaction of the quarks with the (new) charged scalars can be written as:
\begin{align}
    \mathcal{L}_{q} & \supset \bar{u} \left[ ( Y_{D}^{(2)} - \theta_{12} Y_{D}^{(1)} ) P_{R} + (- Y_{U}^{(2)} + \theta_{12}  Y_{U}^{(1)})P_{L} \right] d \, H_{1}^{+} \nonumber  \\
    & \bar{u} \left[ (- \theta_{23} Y_{D}^{(2)} + \theta_{12} \theta_{23} Y_{D}^{(1)} ) P_{R} + ( \theta_{23} Y_{U}^{(2)} - \theta_{12} \theta_{23} Y_{U}^{(1)} ) P_{L}\right] d \, H_{2}^{+} 
\end{align}
The interactions with the DM can be written as:
\begin{align}
    \mathcal{L}_{\ell} & \supset - \bar{\ell} \left[ \theta_{23} \, Y_{\nu}^{(3)} P_{R}  \right] \Psi \, H_{1}^{+} \nonumber  \\
    & - \bar{\ell} \left[ Y_{\nu}^{(3)} P_{R} \right] \Psi \, H_{2}^{+} \,. 
\end{align}
Similar way, the interaction for 2nd charge assignment (can be found from Table~\ref{tab:3HDM_types}) can be written as:
The corresponding Yukawa interaction (for type-II) will be:
\begin{align}
\mathcal{L}_{Y} = \;&
- Y_{U}^{(1)} \overline{q}_{L}  \tilde{\Phi}_{1} u_{R} - Y_{D}^{(1)} \overline{q}_{L}  \Phi_{1} d_{R} -  Y_{e}^{(1)} \overline{\ell}_{L}  \Phi_{1}' e_{R} \nonumber \\
&- Y_{U}^{(2)}  \overline{q}_{L} \tilde{\Phi}_{2} u_{R} \nonumber \\
& -  Y_{D}^{(3)}  \overline{q}_{L}  \Phi_{3} d_{R} -  Y_{e}^{(3)} \overline{\ell}_{L} \Phi_{3} e_{R} -  Y_{\nu}^{(3)} \overline{\ell}_{L} \Phi_{3} \Psi_{R} 
+ \text{h.c.}
\end{align}
Final interaction of the quarks with the (new) charged scalars can be written as:
\begin{align}
\mathcal{L}_{q} & \supset \bar{u} \left[ - Y_{U}^{(2)} P_{L} + \theta_{23} Y_{D}^{(3) P_{R}}  \right] d \, H_{1}^{+} \nonumber \\
& + \bar{u} \left[ \theta_{23} Y_{U}^{(2)} P_{L} + Y_{D}^{(3)} P_{R} \right] d \,  H_{2}^{+}
\end{align}

The interactions with the DM can be written as:
\begin{align}
    \mathcal{L}_{\ell} & \supset \bar{\ell} \left[ -  \theta_{23} Y_{\nu} P_{R} \right] \Psi_{R} H_{1}^{+} \nonumber \\
    &+ \bar{\ell} \left[ - Y_{\nu} P_{R} \right] \Psi H_{2}^{+}
\end{align}
We can take $H_{1}^{+}$ to be very heavy or equivalently $\theta_{23}$ to be very small, then we will get non-$\theta_c$ suppressed contribution to the $\mathcal{C}_{S_{1R}}^{\ell}$.

Similarly, for another charge assignment, if the Yukawa interaction is like the following:
\begin{align}
\mathcal{L}_{Y} = \;&
- Y_{U}^{(1)} \overline{q}_{L}  \tilde{\Phi}_{1} u_{R} - Y_{D}^{(1)} \overline{q}_{L}  \Phi_{1} d_{R} -  Y_{e}^{(1)} \overline{\ell}_{L}  \Phi_{1}' e_{R} \nonumber \\
&- Y_{D}^{(2)}  \overline{q}_{L} \Phi_{2} d_{R} \nonumber \\
& -  Y_{U}^{(3)}  \overline{q}_{L}  \tilde{\Phi}_{3} u_{R} -  Y_{e}^{(3)} \overline{\ell}_{L} \Phi_{3} e_{R} -  Y_{\nu}^{(3)} \overline{\ell}_{L} \Phi_{3} \Psi_{R} 
+ \text{h.c.}
\end{align}
This will contribute similarly to the $\mathcal{C}_{S_{2R}}^{\ell}$ Wilson.

\bibliographystyle{JHEP}
\bibliography{refs}
\end{document}